\documentclass[a4paper,12pt]{article}
\usepackage{amssymb,amsmath,amsthm}
\usepackage{graphicx}
\usepackage{subfigure}
\usepackage{natbib}
\newtheorem{theorem}{Theorem}[section]
\newtheorem{corollary}{Corollary}[section]

\numberwithin{equation}{section}

\title{\textbf{Grouped Variable Selection via Nested Spike and Slab Priors}}
\author{Tso-Jung Yen\footnote{Postdoctoral Fellow, Institute of Statistical Science, Academia Sinica. E-mail: tjyen@stat.sinica.edu.tw}
\\Institute of Statistical Science\\Academia Sinica
\and 
Yu-Min Yen\footnote{PhD Candidate, Department of Finance, London School of Economics and Political Science, Houghton Street, London WC2A 2AE, UK. E-mail: Y.YEN@lse.ac.uk.}
\\Department of Finance\\London School of Economics and Political Science
}

\begin{document}
\maketitle
\begin{abstract}\textbf{Abstract.} In this paper we study grouped variable selection problems by proposing a specified prior, called the nested spike and slab prior, to model collective behavior of regression coefficients. At the group level, the nested spike and slab prior puts positive mass on the event that the $l_{2}$-norm of the grouped coefficients is equal to zero. At the individual level, each coefficient is assumed to follow a spike and slab prior. We carry out maximum a posteriori estimation for the model by applying blockwise coordinate descent algorithms to solve an optimization problem involving an approximate objective modified by majorization-minimization techniques. Simulation studies show that the proposed estimator performs relatively well in the situations in which the true and redundant covariates are both covered by the same group. Asymptotic analysis under a frequentist's framework further shows that the $l_{2}$ estimation error of the proposed estimator can have a better upper bound if the group that covers the true covariates does not cover too many redundant covariates. In addition, given some regular conditions hold, the proposed estimator is asymptotically invariant to group structures, and its model selection consistency can be established without imposing irrepresentable-type conditions.
\end{abstract}

\section{Introduction}
Variable selection has long been an important issue in regression-based statistical analysis. Recently, many efficient methods have been developed to tackle the problems in the situation when the number of covariates is large. At the same time, many efforts have also been made in understanding the statistical properties of these methods. In this paper we focus on grouped variable selection problems. More specifically, we study variable selection in the following regression model:
\begin{eqnarray}
\label{intro01}
y_{i} = \Bigg(\sum_{j\in G_{1}}x_{ij}\beta_{j}\Bigg) + \Bigg(\sum_{j\in G_{2}}x_{ij}\beta_{j}\Bigg)+\cdots+\Bigg(\sum_{j\in G_{m}}x_{ij}\beta_{j}\Bigg)+\epsilon_{i}, 
\end{eqnarray}
where $y_{i}$ is the response variable for subject $i$, $G_{k}\subseteq\{1,2,\cdots,p\}$ is the index set associated to the $k$th group, and $\epsilon_{i}$ is the corresponding error term following some specified distribution. Throughout the paper, we focus on non-overlapping cases, i.e. for two index sets $G_{k}$ and $G_{k^{\prime}}$ with $k,k^{\prime}\in\{1,2,\cdots,m\}$, we assume $G_{k}\cap G_{k^{\prime}}=\emptyset$ for $k\neq k^{\prime}$. Now let $\beta_{G_{k}}$ denote the regression vector with entries indexed by $G_{k}$. Grouped variable selection aims to select covariates groupwisely, that is, entries in $\beta_{G_{k}}$ are either estimated with non-zero values or they are all estimated with zero values. In grouped variable selection, one benchmark method for estimating $\beta=(\beta_{G_{1}},\beta_{G_{2}},\cdots,\beta_{G_{m}})$ is the group lasso \cite{yuanandlin06}:
\begin{eqnarray}
\label{group_lasso_estimator}
\widehat{\beta}_{\text{GL}} &=&\arg\min_{\beta}\bigg\{\frac{1}{2}\bigg|\bigg|y-\sum_{k=1}^{m}X_{G_{k}}\beta_{G_{k}}\bigg|\bigg|_{2}^{2}+\lambda\sum_{k=1}^{m}w_{k}||\beta_{G_{k}}||_{2}\bigg\},
\end{eqnarray}
where $X_{G_{k}}$ is an $n\times |G_{k}|$ matrix representing the covariates indexed by $G_{k}$, $\lambda\geq 0$ is the tuning parameter, and $w_{k}$ is a specified weight corresponding to the $k$th group. 

The group lasso estimator (\ref{group_lasso_estimator}) has several advantages over the lasso in dealing with the variable selection problem associated with model (\ref{intro01}). First, since the $l_{2}$-norm $||\beta_{G_{k}}||_{2}$ is not separable in $\beta_{G_{k}}$, the group lasso provides a more suitable way for regression coefficient estimation when either covariates have meaningful interpretations as a whole \cite{meieretal08,obozinskietal11,chiquetetal11}, or they can be expressed as a group of dummy variables \cite{yuanandlin06}, or they are represented as linear combinations of basis functions \cite{bach08,ravikumaretal09,huangetal10}. In addition, as shown in \cite{huangandzhang10, lounicietal11}, given some regular conditions hold, the $l_{2}$ estimation error of (\ref{group_lasso_estimator}) can have an order of magnitude similar or even smaller than that of the lasso estimator. Moreover, like the lasso, (\ref{group_lasso_estimator}) can also enjoy model selection consistency if some irrepresentable-type conditions \cite{zhaoandyu06} are satisfied \cite{bach08,ravikumaretal09,obozinskietal11,lounicietal11}.

Note that the group lasso estimator (\ref{group_lasso_estimator}) is only able to produce between-group-sparsity, that is, once the $l_{2}$-norm $||\beta_{G_{k}}||_{2}$ is estimated with a non-zero value, all entries in $\beta_{G_{k}}$ will be estimated with non-zero values. However, sometimes the pre-specified group structure may not exactly cover the true covariates. As a result of that, redundant covariates may be wrongly selected in the model, along with the true covariates. To correct this, one need to consider within-group-sparsity. Friedman et al. \cite{friedmanetal10a} proposed the sparse group lasso estimation by adding an $l_{1}$ penalty to the objective function stated in (\ref{group_lasso_estimator}). Under the sparse group lasso estimation, within-group-sparsity can be reached, since with the $l_{1}$ penalty the regression coefficients in the active groups are allowed to have zero-valued estimates. 

In this paper we will study the grouped variable selection problem by developing a specified spike and slab prior \cite{mitchellandbeauchamp88}, called the nested spike and slab prior, to model the group regression coefficient vector $\beta_{G_{k}}$. The nested spike and slab prior assigns positive mass on events $\{||\beta_{G_{k}}||_{2}=0\}$ and $\{||\beta_{G_{k}}||_{2}\neq 0\}$ to represent the sparsity between group coefficient vectors $\beta_{G_{1}},\beta_{G{2}},\cdots,\beta_{G_{m}}$. Given that $||\beta_{G_{k}}||_{2}\neq 0$, it further assigns each entry in $\beta_{G_{k}}$ with a spike and slab prior \cite{mitchellandbeauchamp88}. Under the nested spike and slab prior, sparsity between groups and sparsity within a group can be achieved simultaneously with a positive probability. 

We then develop a method to carry out maximum a posteriori (MAP) estimation for the model. More specifically, we formulate the estimation problem as an optimization problem in which the objective function is approximated by the majorization-minimization algorithms \cite{hunterandlange05, wuandlange10}. We then solve the optimization problem by proposing blockwise coordinate descent algorithms based on the ideas developed in \cite{friedmanetal10a,foygelanddrton10}. Simulation studies show that the proposed estimator performs relatively well in the situations in which the within-group-sparsity is present. However, its performance may get deteriorated if the true covariates are scattered over a large number of groups that contain many redundant covariates. 

Further we will show that under a frequentist's framework, the proposed MAP estimator can have a better $l_{2}$ estimation error bound if the number of groups that cover the true covariates and the numbers of redundant covariates in such groups are small. In addition, if some regular conditions on tuning parameters hold, the values of the proposed estimates will be asymptotically invariant to group structures. We will also establish model selection consistency for the proposed estimator. The result does not require one to impose the irrepresentable-type conditions.   

The paper is organized as follows. In Section \ref{nss_prior} we develop the nested spike and slab prior and construct a Bayesian hierarchical model based on the proposed prior. We then present a method to carry out maximum a posteriori estimation for the model. In Section \ref{simulation_study} we conduct a simulation study to demonstrate finite sample properties of the proposed estimator. In Section \ref{asymptotic_analysis} we establish asymptotic results for the proposed estimator under a frequentist's framework. Section \ref{real_data_examples} contains two real data examples. Section \ref{discussion} is the discussion.

\section{Notation}
For the $k$th index set $G_{k}$, we let $q_{k}$ denote the number of elements in it, i.e. $q_{k}=|G_{k}|$. For the $j$th covariate, we let $k_{j}$ denote the index for the group that $j$ belongs to, that is, if $j\in G_{k^{\prime}}$, then $k_{j}=k^{\prime}$. For a $p$-dimensional vector $b=(b_{1},b_{2},\cdots,b_{p})$, we define $(b)_{j}=b_{j}$ and $b_{G_{k}}$ be the vector whose entries are those indexed by $G_{k}$ in $b$. For the vector $b$, we define the associated $l_{1}$-norm by $||b||_{1}=\sum_{j=1}^{p}|b_{j}|$ and $l_{2}$-norm by $||b||_{2}=(\sum_{j=1}^{p}|b_{j}|^{2})^{1/2}$. We define the sign function of $z$ by $\text{sign}(z)=1$ if $z> 0$; $\text{sign}(z)=-1$ if $z<0$; $\text{sign}(z)=0$ if $z=0$. Finally, we define the soft-thresholding operator $ST_{\lambda}$ by
\begin{eqnarray}
\label{softthresholding}
ST_{\lambda}(z) = \text{sign}(z)(|z|-\lambda)_{+}.
\end{eqnarray}

\section{Nested spike and slab prior}\label{nss_prior}
Since our aim is to jointly select covariates indexed by $G_{k}$, therefore the information about whether $\beta_{G_{k}}$ is a zero vector or not is crucial. Practically, we assign probability mass on event $\{||\beta_{G_{k}}||_{2}\neq 0\}$ to express our belief that $\beta_{G_{k}}$ is not a zero vector. Let $\theta_{k}$ denote the probability. With $\theta_{k}$, we further assume $\beta_{G_{k}}$ follows a distribution which has a density given by 
\begin{eqnarray}
\label{nss02}
f(\beta_{G_{k}})&=&\theta_{k}\Bigg\{\prod_{j\in G_{k}}\bigg[\omega_{j}g(\beta_{j})+(1-\omega_{j})\delta_{(-\xi,\xi)}(\beta_{j})\bigg]\Bigg\}\nonumber\\
& & + (1-\theta_{k})\delta_{0}(||\beta_{G_{k}}||_{2}),
\end{eqnarray}
where $\omega_{j}\in[0,1]$, $g(\beta_{j})$ is some specified density defined on $\mathbb{R}\setminus(-\xi,\xi)$, and $\delta_{0}(||\beta_{G_{k}}||_{2})$ is the Dirac delta function centered at event $\{||\beta_{G_{k}}||_{2}=0\}$. The density (\ref{nss02}) is called the nested spike and slab prior, since the joint spike and slab prior assigned on entries in $\beta_{G_{k}}$ at the individual level is wrapped by a spike and slab prior assigned at the group level. The nested spike and slab prior (\ref{nss02}) implies that $\beta_{G_{k}}$ has probability $\theta_{k}$ to be a non-zero vector. In addition, given that $\beta_{G_{k}}$ is not a zero vector, the entries in $\beta_{G_{k}}$ are independently distributed, and each entry will have probability $\omega_{j}$ to follow a distribution with density $g(\beta_{j})$ and probability $1-\omega_{j}$ to fall uniformly in the region $(-\xi,\xi)$. 

For practical purposes, we introduce two sets of Bernoulli variables $\gamma=(\gamma_{1},\gamma_{2},\cdots,\gamma_{m})$ and $\alpha=(\alpha_{1},\alpha_{2},\cdots,\alpha_{p})$. The former will be used to model regression coefficients at the group level while the latter will be used to model regression coefficients at the individual level. Below we reformulate the nested spike and slab prior (\ref{nss02}) in terms of $\alpha$ and $\gamma$. For group $k$, we let $\gamma_{k}\sim$ Bernoulli$(\theta_{k})$. For $j\in G_{k}$, we assume $\alpha_{j}|\ \gamma_{k}=1$ $\sim$ Bernoulli$(\omega_{j})$. Here $\alpha_{j}$ is defined conditional on $\gamma_{k}=1$, reflecting the nested structure of (\ref{nss02}). Now conditional on $\gamma_{k}$ and $\alpha_{G_{k}}$, the density $f(\beta_{G_{k}}|\ \gamma_{k},\alpha_{G_{k}})$ has the same format as the nested spike and slab prior (\ref{nss02}) with $\theta_{k}$ replaced by $\gamma_{k}$ and $\omega_{j}$ replaced by $\alpha_{j}$. Further it can be shown that the expectation $\mathbb{E}_{\gamma_{k},\alpha_{G_{k}}}[f(\beta_{G_{k}}|\gamma_{k},\alpha_{G_{k}})]$ is the nested spike and slab prior (\ref{nss02}). In addition, given $\gamma_{k}$ and $\alpha_{G_{k}}$ are known, the prior density $f(\beta_{G_{k}}|\ \gamma_{k},\alpha_{G_{k}})$ has an equivalent representation:
\begin{eqnarray}
\label{nss03}
f(\beta_{G_{k}}|\ \gamma_{k},\alpha_{G_{k}})=\Bigg\{\prod_{j\in G_{k}} g(\beta_{j})^{\alpha_{j}}\delta_{(-\xi,\xi)}(\beta_{j})^{1-\alpha_{j}}\Bigg\}^{\gamma_{k}} \delta_{0}(||\beta_{G_{k}}||_{2})^{1-\gamma_{k}}.
\end{eqnarray}
Below we will use the augmented form (\ref{nss03}) to derive the joint posterior density of $\beta,\alpha$ and $\gamma$. 

\subsection{Model}\label{the_model}
We now turn back to regression model (\ref{intro01}). With the prior setting given above, we can construct a hierarchical Bayesian model and carry out inference on parameters in (\ref{intro01}). For practical purposes, we will only focuses on a situation in which the region $(-\xi,\xi)$ is a small region concentrating around 0, that is, $\xi\rightarrow 0$. Under this situation, we can represent (\ref{intro01}) in terms of Bernoulli variables $\alpha$ and $\gamma$ by $y_{i}=\sum_{k=1}^{m}\gamma_{k}(\sum_{j\in G_{k}}x_{ij}\alpha_{j}\beta_{j})+\epsilon_{i}$. Given that there are $n$ subjects, we assume
\begin{eqnarray}
\label{model02}
y_{i}|\ X, \beta, \alpha, \gamma, \sigma^{2} &\sim&\text{ Normal}\Bigg\{\sum_{k=1}^{m}\gamma_{k}\bigg(\sum_{j\in G_{k}}x_{ij}\alpha_{j}\beta_{j}\bigg),\sigma^{2}\Bigg\},\text{ for }i=1,2,\cdots,n,\nonumber\\
\beta_{G_{k}}|\ \alpha_{G_{k}},\gamma_{k}, \sigma^{2},\lambda &\sim&\gamma_{k}\Bigg[\prod_{j\in G_{k}}\bigg\{\alpha_{j}\text{Normal}(0,\sigma^{2}\lambda^{-1})\mathbb{I}\{\mathbb{R}\setminus(-\xi,\xi)\}\nonumber\\
& &+(1-\alpha_{j})\delta_{(-\xi,\xi)}(\beta_{j})\bigg\}\Bigg]\nonumber\\
& &+(1-\gamma_{k})\delta_{0}(||\beta_{G_{k}}||_{2}),\text{ for }k=1,2,\cdots,m,\nonumber\\
\alpha_{G_{k}}|\ \gamma_{k},\omega_{G_{k}} &\sim&\bigg\{\prod_{j\in G_{k}}\text{Bernoulli}(\omega_{j})\bigg\}^{\gamma_{k}}\delta_{0}(\alpha_{G_{k}})^{1-\gamma_{k}},\text{ for }k=1,2,\cdots,m,\nonumber\\
\gamma_{k}|\ \theta_{k}&\sim&\text{ Bernoulli}(\theta_{k}),\text{ for }k=1,2,\cdots,m.
\end{eqnarray}
Under hierarchical Bayesian model (\ref{model02}), the joint posterior density of $\beta,\alpha$ and $\gamma$ is given by
\begin{eqnarray}
\label{model03}
& &f(\beta,\alpha,\gamma|\ y, X, \lambda, \sigma^{2},\omega, \theta)\nonumber\\
& &\ \ \ \ \ \ \ \ \ \ \ \
\propto f(y|\ X,\beta,\alpha,\gamma,\sigma^{2})f(\beta|\ \alpha,\gamma,\sigma^{2},\lambda)f(\alpha|\ \gamma,\omega)f(\gamma|\ \theta),
\end{eqnarray}
where $y=(y_{1},y_{2},\cdots,y_{n})$, and for notational simplicity, similar definitions are applied to $\omega$ and $\theta$. With the joint posterior density (\ref{model03}), various methods can be proposed to make inference on the parameters. Here we adopt the maximum a posteriori (MAP) approach to carrying out the parameter estimation. We define the maximum a posteriori estimator for $\beta$, $\alpha$ and $\gamma$ by
\begin{eqnarray}
(\widehat{\beta},\widehat{\alpha},\widehat{\gamma})=\arg\min_{\beta,\alpha,\gamma}{-2\log f(\beta,\alpha,\gamma|\ y, X, \lambda, \sigma^{2},\omega, \theta)},\nonumber
\end{eqnarray}
where
\begin{eqnarray}
\label{model04}
& &-2\log f(\beta,\alpha,\gamma|\ y, X, \lambda, \sigma^{2},\omega,\theta)\nonumber\\
& &\ \ \ \ \ \ \ \ \ \ \ \ \ \ \ 
=-2\log f(y|\ X,\beta,\alpha,\gamma,\sigma^{2})\nonumber\\
& &\ \ \ \ \ \ \ \ \ \ \ \ \ \ \ \ \ \ 
-2\log\{f(\beta|\ \alpha, \gamma, \sigma^{2},\lambda)f(\alpha|\ \gamma,\omega)f(\gamma|\ \theta)\}\nonumber\\
& &\ \ \ \ \ \ \ \ \ \ \ \ \ \ \ \ \ \ 
-2\log\{\text{normalizing constant}\}.
\end{eqnarray}
\subsection{Parameter estimation}\label{parameter_estimation}
By definition, we can write $\gamma_{k}=\mathbb{I}\{||\beta_{G_{k}}||_{2}\neq 0\}$ and $\alpha_{j}=\mathbb{I}\{\beta_{j}\notin(-\xi,\xi)|\ ||\beta_{G_{k_{j}}}||_{2}\neq 0\}$, where $k_{j}$ is the index for the group that $j$ belongs to. With argumented representation (\ref{nss03}), the second term on the right hand side of (\ref{model04}) can be expressed as
\begin{eqnarray}
\label{model05}
& &-2\log\{f(\beta|\ \alpha,\gamma,\sigma^{2},\lambda)f(\alpha|\ \gamma,\omega)f(\gamma|\ \theta)\}\nonumber\\
& &\ \ \ \ \ \ \ \ \ \ \ \ \ \ \ \ \ \ \ \ \ \
=\frac{\lambda}{\sigma^{2}}\sum_{k=1}^{m}\gamma_{k}\bigg(\sum_{j\in G_{k}}\alpha_{j}\beta_{j}^{2}\bigg)+\log\bigg(\frac{2\pi\sigma^{2}}{\lambda}\bigg)\sum_{k=1}^{m}\sum_{j\in G_{k}}\gamma_{k}\alpha_{j}\nonumber\\
& &\ \ \ \ \ \ \ \ \ \ \ \ \ \ \ \ \ \ \ \ \ \
+\sum_{k=1}^{m}\sum_{j\in G_{k}}\bigg[\log\bigg(\frac{1-\omega_{j}}{\omega_{j}}\bigg)^{2}\bigg]\gamma_{k}\alpha_{j}\nonumber\\
& &\ \ \ \ \ \ \ \ \ \ \ \ \ \ \ \ \ \ \ \ \ \ +\sum_{k=1}^{m}\bigg\{\log\bigg[\bigg(\frac{1-\theta_{k}}{\theta_{k}}\bigg)^{2}\prod_{j\in G_{k}}\bigg(\frac{1}{1-\omega_{j}}\bigg)^{2}\bigg]\bigg\}\gamma_{k}.
\end{eqnarray}
Here we have used the facts that $(1-\alpha_{j})\log\delta_{(-\xi,\xi)}(\beta_{j})=0$, $(1-\gamma_{k})\log\delta_{0}(||\beta_{G_{k}}||_{2})=0$, and $(1-\gamma_{k})\log\delta_{0}(\alpha_{G_{k}}) = 0$ in deriving (\ref{model05}). 

In addition, given that $\xi\rightarrow 0$, we have $\alpha_{j}\approx\mathbb{I}\{\beta_{j}\neq 0|\ ||\beta_{G_{k_{j}}}||_{2}\neq 0\}$. Further by a direct calculation, we have $\gamma_{k_{j}}\alpha_{j}=\mathbb{I}\{\beta_{j}\neq 0\cap ||\beta_{G_{k_{j}}}||_{2}\neq 0\}=\mathbb{I}\{ ||\beta_{G_{k_{j}}}||_{2}\neq 0|\ \beta_{j}\neq 0\}\mathbb{I}\{\beta_{j}\neq 0\}$. Note that the expectation of the index $\mathbb{I}\{||\beta_{G_{k_{j}}}||_{2}\neq 0|\ \beta_{j}\neq 0\}$ is $\mathbb{P}(||\beta_{G_{k_{j}}}||_{2}\neq 0|\ \beta_{j}\neq 0)$, which is obviously equal to $1$ since $j\in G_{k_{j}}$ and $\beta_{j}\neq 0$ implies $||\beta_{G_{k_{j}}}||_{2}\neq 0$ almost surely. This further implies that $\mathbb{I}\{ ||\beta_{G_{k_{j}}}||_{2}\neq 0|\ \beta_{j}\neq 0\}$ is equal to 1 almost surely. Therefore we have
\begin{eqnarray}
\label{model06}
\gamma_{k_{j}}\alpha_{j}=\mathbb{I}\{\beta_{j}\neq 0\}.
\end{eqnarray}
Now consider the hyperparameters $\lambda$, $\sigma^{2}$, $\theta$, $\omega$. Since there is no easy way to determine values of these hyperparameters, therefore for practical purposes, we will impose some constraints on these hyperparameters. We assume $\omega_{j}=\omega_{1}$ for all $j$. Further we define $\rho_{1}=\sigma^{2}\log\{[(2\pi\sigma^{2})/\lambda][(1-\omega_{1})/\omega_{1}]^{2}\}$ and assume $\rho_{1}\geq 0$. For the fourth term on the right hand side of (\ref{model05}) that involves $\theta_{k}$'s, we adopt the following parametrization. We will assume all $\gamma_{k}$'s in the fourth term on the right hand side of (\ref{model05}) have an equal weight. Given that $\omega_{j}=\omega_{1}$ for all $j$, we can choose appropriate $\theta_{k}$'s from interval $[0,1]$ to make the weights of $\gamma_{k}$'s the same for all $k$. Let $\theta_{k}^{*}$ be such appropriate value of $\theta_{k}$. With values of $\theta_{k}^{*}$'s, we define $\rho_{2} = \sigma^{2}\log\{[(1-\theta_{k}^{*})/\theta_{k}^{*}]^{2/\sqrt{q_{k}}}(1-\omega_{1})^{-2\sqrt{q_{k}}}\}$, where $q_{k}=|G_{k}|$. We assume $\rho_{2}\geq 0$. 

With (\ref{model06}) and the definitions of $\rho_{1}$ and $\rho_{2}$, minimizing (\ref{model04}) with respect to $\beta$, $\alpha$ and $\gamma$ is equivalent to minimizing the function
\begin{eqnarray}
\label{model07}
V(\beta)&=&\bigg|\bigg|y-\sum_{k=1}^{m}X_{G_{k}}\beta_{G_{k}}\bigg|\bigg|_{2}^{2} + \lambda\sum_{k=1}^{m}||\beta_{G_{k}}||_{2}^{2}\nonumber\\
& &+\rho_{1}\sum_{k=1}^{m}\sum_{j\in G_{k}}\mathbb{I}\{\beta_{j}\neq 0\}+\rho_{2}\sum_{k=1}^{m}\sqrt{q_{k}}\mathbb{I}\big\{||\beta_{G_{k}}||_{2}\neq 0\big\}
\end{eqnarray}
with respect to $\beta$. Here we define the gvsnss estimator (\textbf{G}rouped \textbf{V}ariable \textbf{S}election via \textbf{N}ested \textbf{S}pike and \textbf{S}lab Priors) as the one that minimizes (\ref{model07}). Below we provide a numerical procedure to calculate the gvsnss estimator.
\subsubsection{Majorization-minimization algorithms}\label{mm_algorithm}
\begin{figure}[h]
\begin{center}
\includegraphics[width=0.45\textwidth]{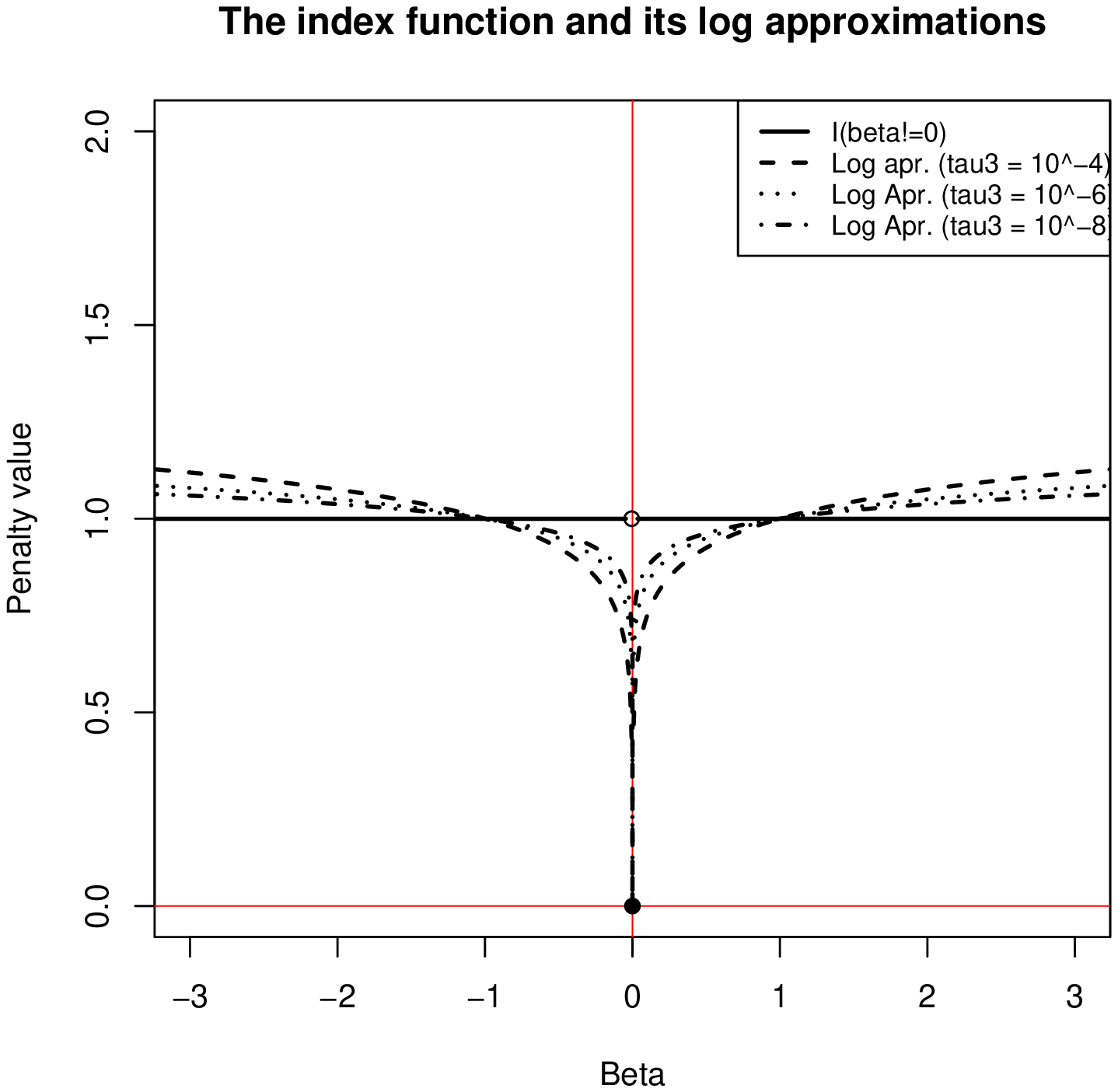}
\includegraphics[width=0.45\textwidth]{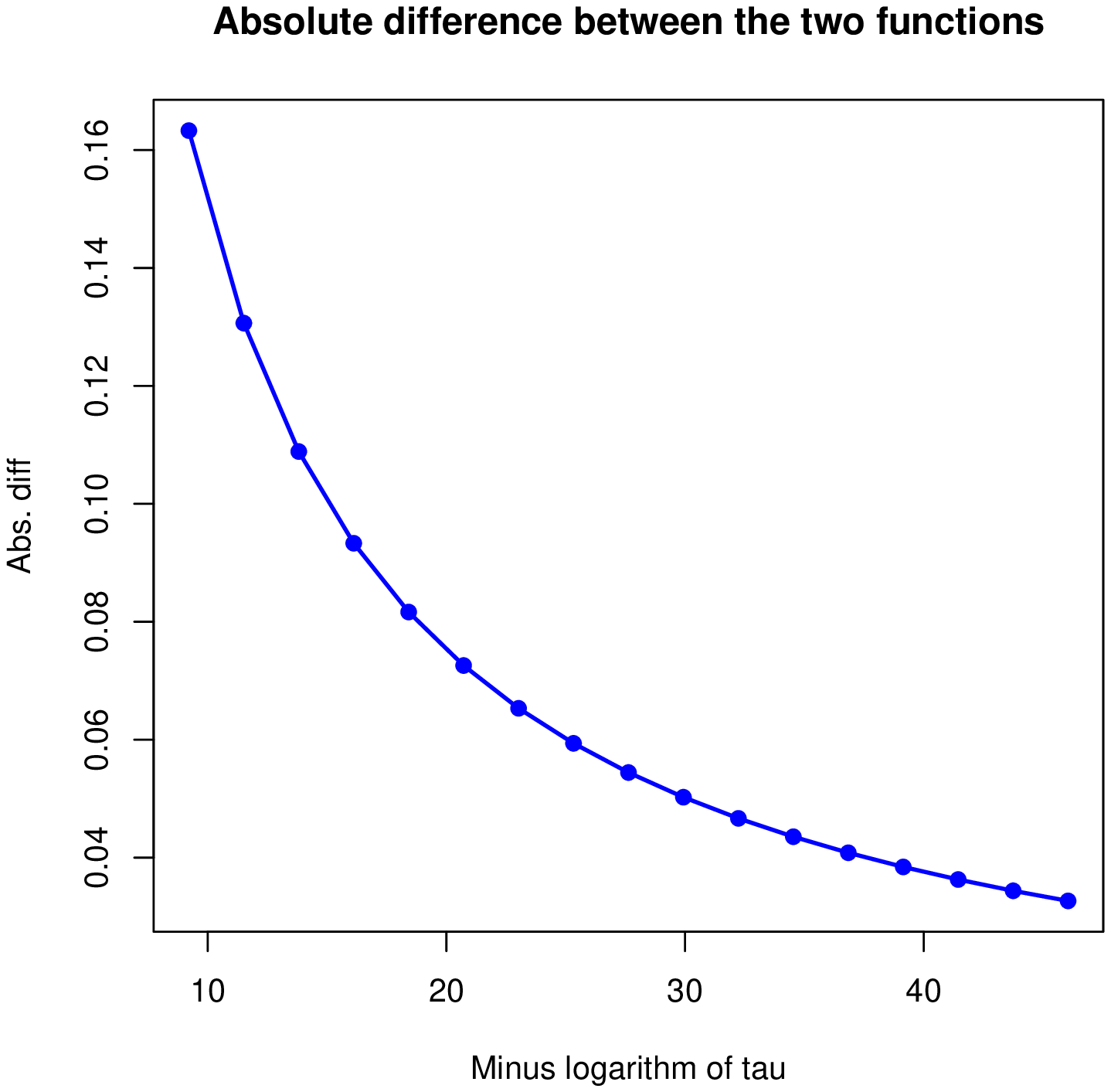}
\end{center}\caption{Left: The index function and its log approximations; Right: The mean absolute difference between the index function and its log approximation as a function of $-\log\tau$. Each point is an average over absolute differences with input values from $[-10,10]$.} 
\label{figure_sect3_01}
\end{figure}

Since the last two terms in (\ref{model07}) are discrete in their domain, the minimization problem involving (\ref{model07}) is combinatorial and in general is considered to be difficult. Here we adopt a continuous relaxation procedure to modify (\ref{model07}). More specifically, we use the function
\begin{eqnarray}
\label{model08}
g_{\tau}(a)=\frac{\log(1 + \tau^{-1}|a|)}{\log(1+\tau^{-1})}
\end{eqnarray} 
to approximate index function $\mathbb{I}\{a\neq 0\}$. It can be shown that $g_{\tau}(a)\rightarrow\mathbb{I}\{a\neq 0\}$ as $\tau\rightarrow 0$ \cite{sriperumbuduretal09,yen11}. Figure \ref{figure_sect3_01} shows $\mathbb{I}\{a\neq 0\}$ and $g_{\tau}(a)$ and the absolute difference between the two functions as a function of $-\log\tau$. Since (\ref{model08}) is continuous on $\mathbb{R}$, the combinatorial nature of $\mathbb{I}\{a\neq 0\}$ is relaxed. However, (\ref{model08}) is not convex in $a$, and using (\ref{model08}) for continuous relaxation on (\ref{model07}) still makes (\ref{model07}) remain non-convex. We adopt a majorization-minimization approach to tackling this problem. Majorization-minimization (MM) algorithms \cite{hunterandlange05,wuandlange10} aim to solve difficult minimization problems by modifying the corresponding objective functions so that solution spaces of the modified ones are easier to explore. For an objective function $V^{*}(a)$, the modification procedure relies on finding a function $V^{**}(a;a^{(d)})$ that satisfies the following properties: 
\begin{eqnarray}
\label{mm01}
V^{**}(a; a^{(d)})&\geq& V^{*}(a)  \ \ \ \text{ for all }a,\nonumber\\
V^{**}(a^{(d)}; a^{(d)})&=&V^{*}(a^{(d)}).
\end{eqnarray}
In (\ref{mm01}), the objective function $V^{*}(a)$ is said to be majorized by $V^{**}(a;a^{(d)})$. In this sense, $V^{**}(a;a^{(d)})$ is called the majorization function. In addition, (\ref{mm01}) implies that $V^{**}(a; a^{(d)})$ is tangent to $V^{*}(a)$ at $a^{(d)}$. Moreover if $a^{(d+1)}$ is a minimizer of $V^{**}(a; a^{(d)})$, then (\ref{mm01}) further implies that $V^{*}(a^{(d)})=V^{**}(a^{(d)}; a^{(d)})\geq V^{**}(a^{(d+1)}; a^{(d)})\geq V^{*}(a^{(d+1)})$, which means that the iteration procedure $a^{(d)}$ pushes $V^{*}(a)$ toward its minimum.  

Now we turn back to function (\ref{model08}). Note that, since $\log(a)$ is a concave function of $a$ for $a>0$, therefore the inequality
\begin{eqnarray}
\label{mm02}
\log(a^{\prime}) + \frac{a}{a^{\prime}}-1\geq\log(a)
\end{eqnarray}
holds for all $a>0$ and $a^{\prime}>0$. Note that the left hand side of (\ref{mm02}) is convex in $a$. In addition, if we let $a=a^{\prime}$, then (\ref{mm02}) becomes an equality, which implies that the left hand side of (\ref{mm02}) satisfies the properties stated in (\ref{mm01}), therefore is a valid function for majorizing $\log(a)$.

Now by applying (\ref{model08}) and the left hand side of (\ref{mm02}) to $\sum_{k=1}^{m}\sum_{j\in G_{k}}\mathbb{I}\{\beta_{j}\neq 0\}$, we can establish the following inequality:
\begin{eqnarray}
\label{model09}
& &\sum_{k=1}^{m}\sum_{j\in G_{k}}\mathbb{I}\{\beta_{j}\neq 0\}\nonumber\\
&=&\lim_{\tau\rightarrow 0}\sum_{k=1}^{m}\sum_{j\in G_{k}}\frac{\log(1 + \tau^{-1}|\beta_{j}|)}{\log(1+\tau^{-1})}\nonumber\\
&\leq&\lim_{\tau\rightarrow 0}\frac{1}{\log(1+\tau^{-1})}\sum_{j=1}^{p}\Bigg(\log\big(1 + \tau^{-1}|\beta_{j}^{\prime}|\big) + \frac{\tau+|\beta_{j}|}{\tau+|\beta_{j}^{\prime}|}-1\Bigg).
\end{eqnarray}
Similarly for $\sum_{k=1}^{m}\{||\beta_{G_{k}}||_{2}\neq 0\}$, we have
\begin{eqnarray}
\label{model10}
\sum_{k=1}^{m}\sqrt{q_{k}}\mathbb{I}\{||\beta_{G_{k}}||_{2}\neq 0\}&\leq&\lim_{\tau\rightarrow 0}\frac{1}{\log(1+\tau^{-1})}\nonumber\\
& &\times\sum_{k=1}^{m}\sqrt{q_{k}}\Bigg(\log\big(1 + \tau^{-1}||\beta_{G_{k}}^{\prime}||_{2}\big) + \frac{\tau+||\beta_{G_{k}}||_{2}}{\tau+||\beta_{G_{k}}^{\prime}||_{2}}-1\Bigg).\nonumber\\
\end{eqnarray}
\subsubsection{Blockwise coordinate descent algorithms}\label{blockwise_cd_algorithms}
With the majorization-minimization results (\ref{model09}) and (\ref{model10}), we can establish an iterative scheme to find the minimizer of (\ref{model07}). In practice, we use the blockwise iterative scheme 
\begin{eqnarray}
\label{model11}
\widehat{\beta}_{G_{k}}^{(d+1)}&=&\arg\min_{\beta_{G_{k}}}\bigg\{\big|\big|r_{-G_{k}}-X_{G_{k}}\beta_{G_{k}}\big|\big|_{2}^{2} +\lambda||\beta_{G_{k}}||_{2}^{2}\nonumber\\
& &\ \ \ \ \ \ \ \ \ \ \ \ \ \ \ \ \ \ \ \ \ +\lambda_{1}||\widehat{\nu}_{G_{k}}^{(d)}\beta_{G_{k}}||_{1}+\lambda_{2}\widehat{\phi}_{k}^{(d)}||\beta_{G_{k}}||_{2}\bigg\}
\end{eqnarray}
to find the solution that minimizes (\ref{model07}), where $\lambda_{1}=\rho_{1}\lim_{\tau\rightarrow 0}[\log(1+\tau^{-1})]^{-1}$, $\lambda_{2}=\rho_{2}\lim_{\tau\rightarrow 0}[\log(1+\tau^{-1})]^{-1}$, and $r_{-G_{k}}=y - \sum_{k^{\prime}\neq k}X_{G_{k^{\prime}}}\beta_{G_{k^{\prime}}}$. In addition, for $j\in G_{k}$, $\widehat{\nu}_{j}^{(d)}=\lim_{\tau\rightarrow0}(\tau+|\widehat{\beta}_{j}^{(d)}|)^{-1}$, and $\widehat{\phi}_{k}^{(d)}=\lim_{\tau\rightarrow0}\sqrt{q_{k}}(\tau+||\widehat{\beta}_{G_{k}}^{(d)}||_{2})^{-1}$. 

With the objective function stated in (\ref{model11}), one can derive associated KKT conditions and solve them for the minimizer $\widehat{\beta}_{G_{k}}^{(d+1)}$. However, the third and fourth terms on the right hand side of (\ref{model11}) are not smooth, therefore special attention is needed to obtain a gradient-like vector for (\ref{model11}). Here we adopt a subgradient-based approach to tackling this problem. For the idea of subgradients and related theoretical properties, please see Section B.5 of \cite{bertsekas99}. By applying the subgradient calculus to the objective function in (\ref{model11}) with respect to $\beta_{G_{k}}$, we can obtain a gradient-like vector for the objective function. Then by setting the vector to zero, we obtain the subgradient equations
\begin{eqnarray}
\label{sub_equations}
2X_{G_{k}}^{T}r_{-G_{k}}-2X_{G_{k}}^{T}X_{G_{k}}\beta_{G_{k}}-2\lambda\beta_{G_{k}}-\lambda_{1}\widehat{\nu}_{G_{k}}^{(d)}h_{G_{k}} - \lambda_{2}\widehat{\phi}_{k}^{(d)}v_{G_{k}}=0,
\end{eqnarray}
where $h_{G_{k}}$ is a subgradient vector of the $l_{1}$-norm $||\beta_{G_{k}}||_{1}$, and its entry is defined as that, for $j\in G_{k}$ $h_{j} = 1$ if $\beta_{j} > 0$; $h_{j}=h_{j}^{*}\in[-1,1]$ if $\beta_{j}=0$; and $h_{j}=-1$ if $\beta_{j} < 0$. In addition, $v_{G_{k}}$ is a subgradient vector of the $l_{2}$-norm $||\beta_{G_{k}}||_{2}$ and is defined as
\begin{eqnarray}
\label{def_vgk}
v_{G_{k}} =
\begin{cases}
\beta_{G_{k}}/||\beta_{G_{k}}||_{2}\ \ \ \ \ \ \ \ \ \ \ \ \ \ \ \ \ \ \ \ \text{if }||\beta_{G_{k}}||_{2}\neq 0,\\
v_{G_{k}}^{*}\text{ such that }||v_{G_{k}}^{*}||_{2}^{2}\leq 1\ \ \ \ \text{if }||\beta_{G_{k}}||_{2}=0.
\end{cases} 
\end{eqnarray}
Below we adopt a method provided by Friedman et al. \cite{friedmanetal10a} to solve the subgradient equations (\ref{sub_equations}). The method uses a testing procedure to identify whether $\beta_{G_{k}}$ is a zero vector or not. First note that, if $\beta_{G_{k}}=0$, then the subgradient equations (\ref{sub_equations}) becomes
\begin{eqnarray}
\label{model12}
2X_{G_{k}}^{T}r_{-G_{k}} - \lambda_{1}\widehat{\nu}_{G_{k}}^{(d)}h_{G_{k}} = \lambda_{2}\widehat{\phi}_{k}^{(d)}v_{G_{k}}.
\end{eqnarray}
Now by definition (\ref{def_vgk}), if $||\beta_{G_{k}}||_{2}=0$, i.e. $\beta_{G_{k}}$ is a zero vector, then $||v_{G_{k}}||_{2}\leq 1$, therefore (\ref{model12}) implies that 
\begin{eqnarray}
\label{model13}
||2X_{G_{k}}^{T}r_{-G_{k}}-\lambda_{1}\widehat{\nu}_{G_{k}}^{(d)}h_{G_{k}}||_{2}\leq\lambda_{2}\widehat{\phi}_{k}^{(d)}.
\end{eqnarray}
To numerically verify the condition (\ref{model13}), we need to know $h_{G_{k}}$. Friedman et al. \cite{friedmanetal10a} provided a practical way to estimate $h_{G_{k}}$ by solving the least squares problem $\min_{h_{G_{k}}}||2X_{G_{k}}^{T}r_{-G_{k}}-\lambda_{1}\widehat{\nu}_{G_{k}}^{(d)}h_{G_{k}}||_{2}^{2}$ subject to $-1\leq h_{j}\leq 1$ for $j\in G_{k}$. The resulting estimate takes a soft-thresholding form, and by plugging it into (\ref{model13}), one obtains
\begin{eqnarray}
\label{model14}
\Big|\Big|ST_{\lambda_{1}\widehat{\nu}_{G_{k}}^{(d)}}(2X_{G_{k}}^{T}r_{-G_{k}})\Big|\Big|_{2}\leq\lambda_{2}\widehat{\phi}_{k}^{(d)},
\end{eqnarray}
Note that if condition (\ref{model14}) holds, we let $\widehat{\beta}_{G_{k}}^{(d+1)}=0$, otherwise we go further to estimate entries in $\beta_{G_{k}}$ with other values. 

Below we describe a numerical procedure for estimating non-zero entries in $\beta_{G_{k}}$. First note that, as shown in \cite{wuandlange08}, the $l_{2}$-norm $||\beta_{G_{k}}||_{2}$ on the right hand side of (\ref{model11}) can be bounded in a way such that
\begin{eqnarray}
\label{mm03}
||\beta_{G_{k}}^{\prime}||_{2} + \frac{1}{2||\beta_{G_{k}}^{\prime}||_{2}}(||\beta_{G_{k}}||_{2}^{2}-||\beta_{G_{k}}^{\prime}||_{2}^{2})\geq ||\beta_{G_{k}}||_{2}.
\end{eqnarray}
Here the function on the left hand side is convex in $\beta_{G_{k}}$. Now if we let $\beta_{G_{k}}=\beta_{G_{k}}^{\prime}$, then the equality will hold between the two sides of (\ref{mm03}). Therefore the function on the left hand side of (\ref{mm03}) majorizes $||\beta_{G_{k}}||_{2}$.  With the majorization result (\ref{mm03}), we construct the following iterative scheme:
\begin{eqnarray}
\label{model15}
\widehat{\beta}_{G_{k}}^{(d_{1}+1,d_{2}+1)}&=&\arg\min_{\beta_{j}}\bigg\{\big|\big|r_{-G_{k}}-X_{G_{k}}\beta_{G_{k}}\big|\big|_{2}^{2}\nonumber\\
& &\ \ \ \ \ \ \ \ \ \ \ \ \ \ \ \ \  +\lambda_{1}||\widehat{\nu}_{G_{k}}^{(d_{1})}\beta_{G_{k}}||_{1}+\Bigg(\lambda+\frac{\lambda_{2}\widehat{\phi}_{k}^{(d_{1})}}{2||\widehat{\beta}_{k}^{(d_{1}+1,d_{2})}||_{2}}\Bigg)||\beta_{G_{k}}||_{2}^{2}\bigg\}\nonumber\\
\end{eqnarray}
to obtain $\widehat{\beta}_{G_{k}}^{(d_{1}+1)}$. The scheme (\ref{model15}) can be approximated by the following iterative least squares procedure: 
\begin{eqnarray}
\label{model17}
\widehat{\beta}_{G_{k}}^{(d_{1}+1,d_{2}+1)} = \bigg[X_{G_{k}}^{T}X_{G_{k}}+\bigg(\lambda+\frac{\lambda_{2}\widehat{\phi}_{k}^{(d_{1})}}{2||\widehat{\beta}_{G_{k}}^{(d_{1}+1,d_{2})}||_{2}}\bigg)I_{q_{k}\times q_{k}}\bigg]^{-1}ST_{\lambda_{1}\widehat{\nu}_{G_{k}}^{(d_{1})}/2}\big(X_{G_{k}}^{T}r_{-G_{k}}\big),
\end{eqnarray}
where $ST_{\lambda_{1}\widehat{\nu}_{G_{k}}^{(d_{1})}/2}(X_{G_{k}}^{T}r_{-G_{k}})$ is the soft thresholding operator defined in (\ref{softthresholding}). A least squares result similar to (\ref{model17}) can be found in \cite{foygelanddrton10}. For large-scale problems, we construct a one dimensional soft thresholding scheme to approximate (\ref{model15}). The soft-thresholding scheme is given by
\begin{eqnarray}
\label{model18}
\widehat{\beta}_{j}^{(d_{1}+1,d_{2}+1)}=\Bigg(\sum_{i=1}^{n}x_{ij}^{2}+\lambda+\frac{\lambda_{2}\widehat{\phi}_{k_{j}}^{(d_{1})}}{2||\widehat{\beta}_{k_{j}}^{(d_{1}+1,d_{2})}||_{2}}\Bigg)^{-1}ST_{\lambda_{1}\widehat{\nu}_{j}^{(d_{1})}/2}\Bigg(\sum_{i=1}^{n}x_{ij}r_{i,-j}^{(*)}\Bigg),
\end{eqnarray}
where $r_{i,-j}^{(*)}=r_{i,-G_{k}}- \sum_{j^{\prime}\neq j; j^{\prime},j\in G_{k}}x_{ij^{\prime}}\beta_{j^{\prime}}^{(*)}$ with $\beta_{j^{\prime}}^{(*)}=\beta_{j^{\prime}}^{(d_{1}+1,d_{2}+1)}$ for $j^{\prime}<j$ and $\beta_{j^{\prime}}^{(*)}=\beta_{j^{\prime}}^{(d_{1}+1,d_{2})}$ for $j^{\prime}>j$.
\subsection{Determining tuning parameter values}\label{determining_hyper_values}
For $\lambda_{1}$, $\lambda_{2}$ and $\lambda$, we adopt a grid search strategy to find their optimal values. Here we assume that each column of design matrix $X$ is standardized. To find optimal $\lambda_{1}$, we search along a grid of candidate values in the interval $[0, \lambda_{1}^{*}]$, where $\lambda_{1}^{*}$ is defined as $
\lambda_{1}^{*}=2.05\tau\times\max_{j\in\{1,2,\cdots,p\}}|x_{j}^{T}y|$. To find optimal $\lambda_{2}$, we search along a grid of candidate values in the interval $[0, \lambda_{2}^{*}]$, where $\lambda_{2}^{*}=1.1\tau\times\max_{k\in\{1,2,\cdots,m\}}||2X_{G_{k}}^{T}y||_{2}/\sqrt{q_{k}}$. For $\lambda$, we assume it decreases with sample size $n$ and is proportional to $\lambda_{2}$. More specifically, we let $\lambda = \lambda_{2}/(10n)$. With the reparametrization on $\lambda$ given above, we only need to do grid searches for $\lambda_{1}$ and $\lambda_{2}$. For parameter $\tau$, we let $\tau = 5\times 10^{-4}$.  
\subsection{Connection with other approaches}\label{connection_with_other_approaches}
Recent research on variable selection using maximum a posteriori estimation includes \cite{genkinetal07,yen11}. Armagan et al. \cite{armaganetal11} developed a shrinkage-based method for variable selection based on the generalized double Pareto priors. The idea of using spike and slab priors in grouped variable selection has also been adopted by Scheipl et al. \cite{scheipletal11}, who developed an MCMC-based approach to carrying out posterior inference on additive regression models. The idea of using (\ref{model08}) in approximating an index function has been mentioned in \cite{candesetal08, sriperumbuduretal09,mazumderetal11,yen11}. Tipping \cite{tipping01} has pointed out a connection between the log function (\ref{model08}) and the improper Student's $t$ density.

\section{Simulation study}\label{simulation_study}
In this section we study finite sample properties of the gvsnss estimator by fitting regression models with simulated data. In the simulation study, we assume the covariates are randomly divided into $m$ groups, and the true covariates, i.e. the covariates with non-zero coefficients, are covered by $r\leq m$ groups. We will focus on the following two situations: 
\begin{itemize}
\item[\textbf{i.}]The true covariates are covered by the $r$ groups, but at the same time, some redundant covariates, i.e. the covariates with zero coefficients, are also covered by the $r$ groups. 
\item[\textbf{ii.}]The true covariates are re-assigned with different group labels. In this situation, $r$, the number of groups that covers the true covariates, will change. 
\end{itemize}
To create the first situation, we focuses on varying the level of sparsity in the groups that contain the true covariates. To create the second situation, we focuses on re-assigning covariates to other groups according to some group switching probabilities. Under the two situations, each simulation experiment is characterized by the pair $($spr, mis-labeled$)$, where "spr" denotes the level of within-group-sparsity and "mis-labeled" denotes the group switching probability. For a covariate in an active group, spr $=0.3$ means that the value of its coefficient will have probability $0.3$ to be coerced to zero, and mis-labeled $=0.3$ means that it will be re-assigned with a different group label with probability $0.3$. 

Below we introduce the basic simulation scheme. For the $n\times p$ design matrix $X$, we generate its rows i.i.d. from MVN$(0,I_{p\times p})$. For regression coefficients $\beta=(\beta_{1},\beta_{2},\cdots,\beta_{p})$, we first randomly assign the corresponding covariates into $m$ groups. We then choose $r\leq m$ groups of covariates and generate their coefficients i.i.d. from Normal$(0,1)$. We further set coefficients of the covariates in the rest of $m-r$ groups to zero. We then re-proceed each coefficient by either coercing its value to zero or re-assigning its covariate with a different group label according to the pre-specified values in $($spr, mis-labeled$)$. For the error vector $\epsilon$, we generate its entries i.i.d. from Normal$(0,1)$. Finally, we compute the response vector $y = X\beta + \epsilon$.

\subsection{Methods for comparisons}
We conducted two gvsnss estimations for the regression model. The first one used five fold cross validation for tuning parameter selection. The second one used the following logarithm of the Bayes factor:
\begin{eqnarray}
\label{bayes_factor}
\log \text{BF}(\widehat{S},\text{null}; y)= \frac{n}{2}\log\bigg\{\frac{y^{T}y}{y^{T}(\lambda^{-1}X_{\widehat{S}}X_{\widehat{S}}^{T}+I_{n\times n})^{-1}y}\bigg\}-\frac{1}{2}\log \big|\lambda^{-1}X_{\widehat{S}}X_{\widehat{S}}^{T}+I_{n\times n}\big|\nonumber\\
\end{eqnarray}
for tuning parameter selection, where $\widehat{S}=\{j:\widehat{\beta}_{\text{gvsnss},j}\neq 0\}$. The logarithm Bayes factor (\ref{bayes_factor}) corresponds to the model that assigns Normal$(0,\sigma^{2}/\lambda)$ on $\beta_{j}$ and Inverse-Gamma$(\tau_{1},\tau_{2})$ on $\sigma^{2}$ with $\tau_{1}$ and $\tau_{2}$ both approaching to zero. For tuning parameter selection, we searched optimal $\lambda_{1}$ along a grid of $20$ candidate values and optimal $\lambda_{2}$ along a grid of another $20$ candidate values.   

We also conducted three other estimations for the regression model. The first one is the group lasso using five fold cross validation for tuning parameter selection. The second one is also the group lasso but using a naive AIC for tuning parameter selection. The naive AIC is given by nAIC $= ||y-X\widehat{\beta}_{\text{GL}}||_{2}^{2}/\widehat{\sigma}^{2}+2\widehat{s}_{\text{GL}}$, where $\widehat{\sigma}^{2}$ is estimated from the null model and $\widehat{s}_{\text{GL}}$ is the number of non-zero entries in $\widehat{\beta}_{\text{GL}}$. Numerical calculations for the two group lasso estimations were done by using R package grplasso \cite{meieretal08}. The third one is the lasso using ten fold cross validation for tuning parameter estimation. We used R package glmnet \cite{friedmanetal09} to carry out numerical computations for the lasso estimation. For all the three estimations, we searched optimal tuning parameters along a grid of $100$ candidate values. 

We collected three performance measures at each simulation run. The first one is the sign-adjusted false positive rate, which is defined as
\begin{eqnarray}
\label{sfpr}
\text{SFPR}=\frac{\#\{j\in \widehat{S}:\text{sign}(\widehat{\beta}_{j})\neq\text{sign}(\beta_{\text{true},j})\}}{|\widehat{S}|}.\nonumber
\end{eqnarray}
The second one is the squared $l_{2}$ estimation error, which is defined as 
\begin{eqnarray}
\label{l2_dis}
l_{2}\text{-dis}=\frac{\sum_{j=1}^{p}(\widehat{\beta}_{j}-\beta_{\text{true},j})^{2}}{p}.\nonumber
\end{eqnarray}
The third one is the predictive mean squared error, which is defined as 
\begin{eqnarray}
\label{pmse}
\text{PMSE}=\frac{\sum_{i=1}^{n^{\prime}}(y_{i,\text{new}}-x_{i,\text{new}}^{T}\widehat{\beta})^{2}}{n^{\prime}},\nonumber
\end{eqnarray}
where $n^{\prime}=10\times n$, $y_{i,\text{new}}$ and $x_{i,\text{new}}$ are new data points generated under the same simulation scheme.
\subsection{Results}
In practice, we let $p = 200$, $m = 10$, and $r = 2$. We considered different values of sample size $n$ and the pair $($spr, mis-labeled$)$ in generating data points.

\begin{figure}[t]
\begin{center}
\includegraphics[width=0.32\textwidth]{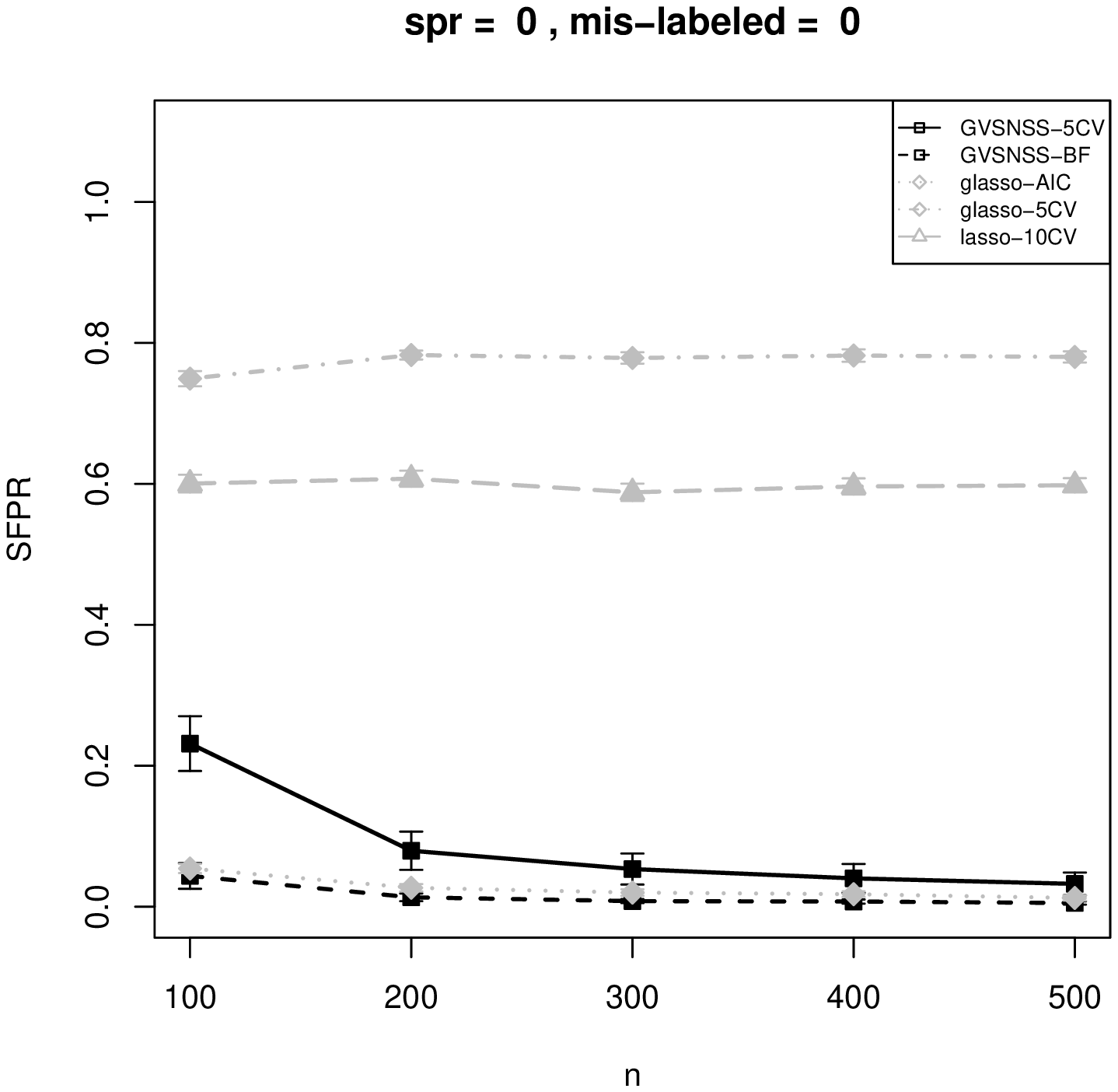}
\includegraphics[width=0.32\textwidth]{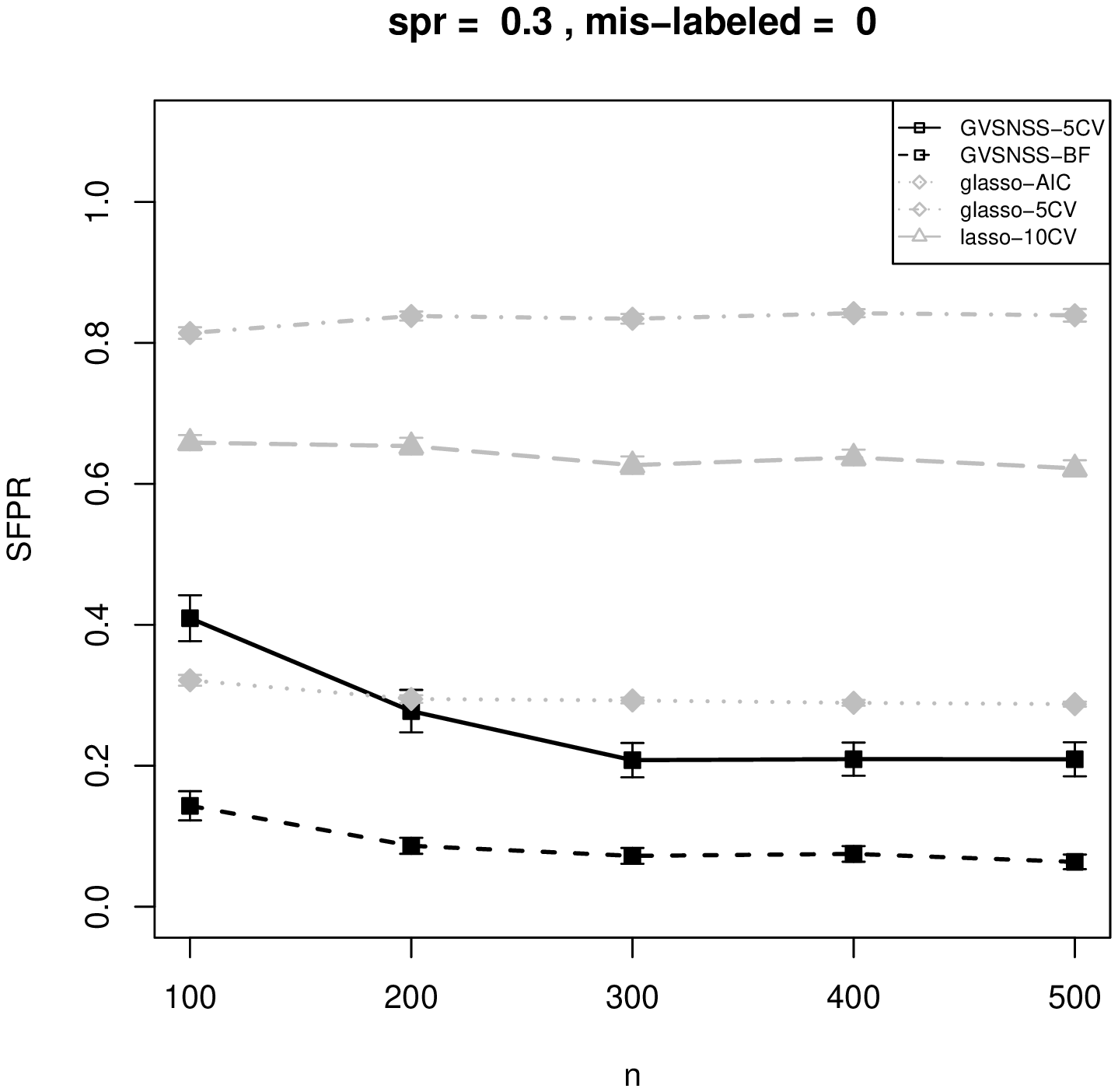}
\includegraphics[width=0.32\textwidth]{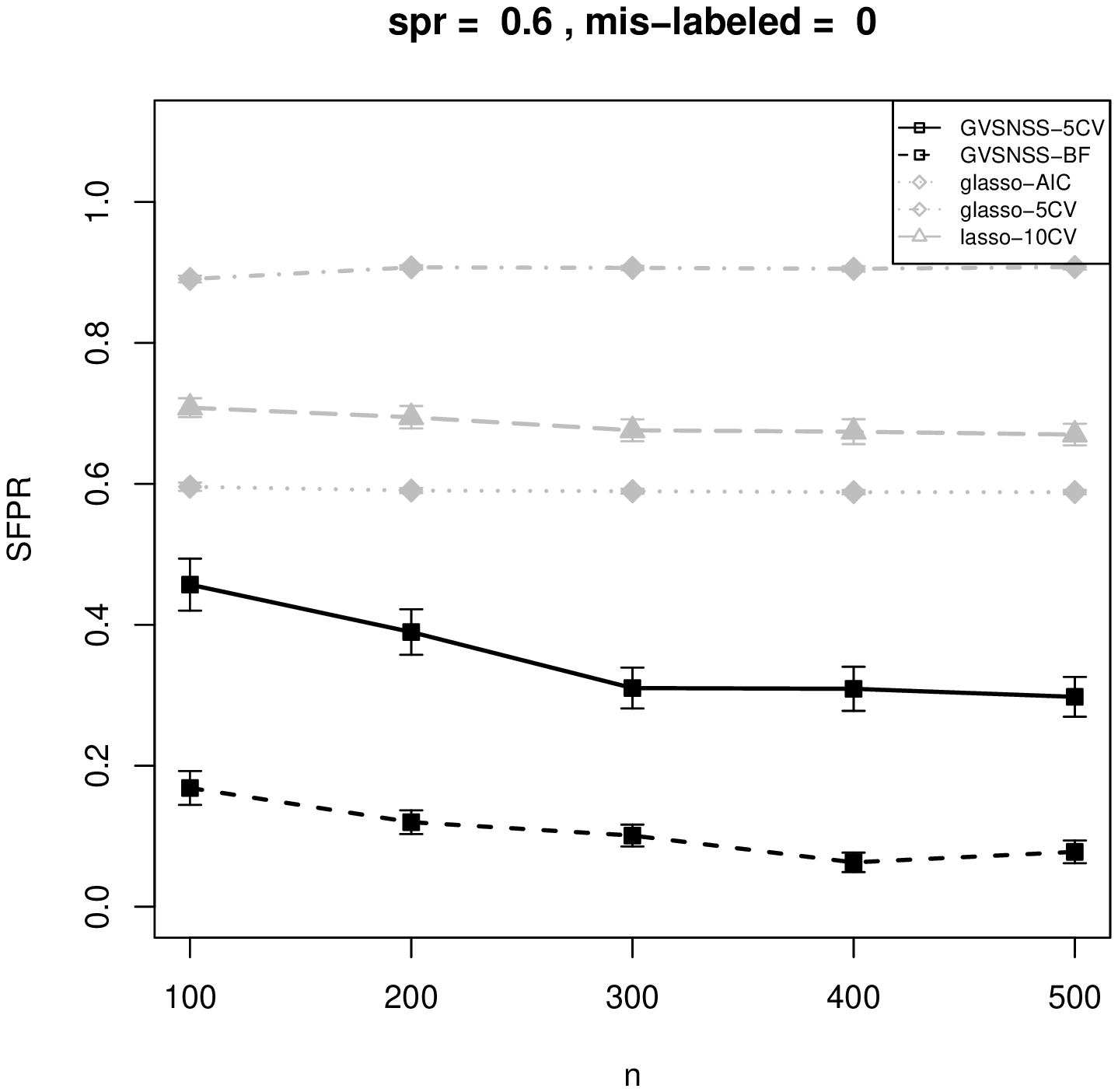}
\includegraphics[width=0.32\textwidth]{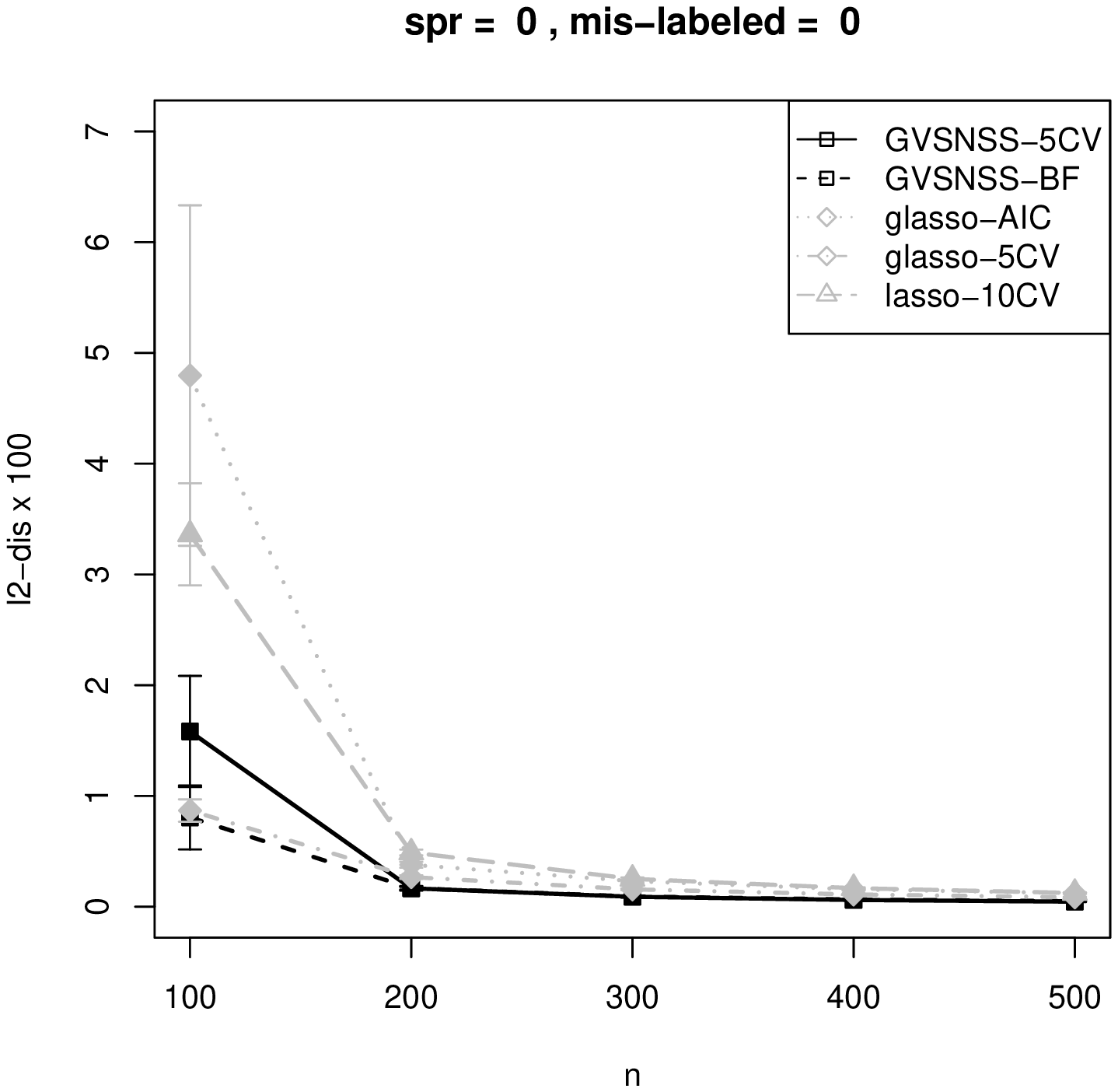}
\includegraphics[width=0.32\textwidth]{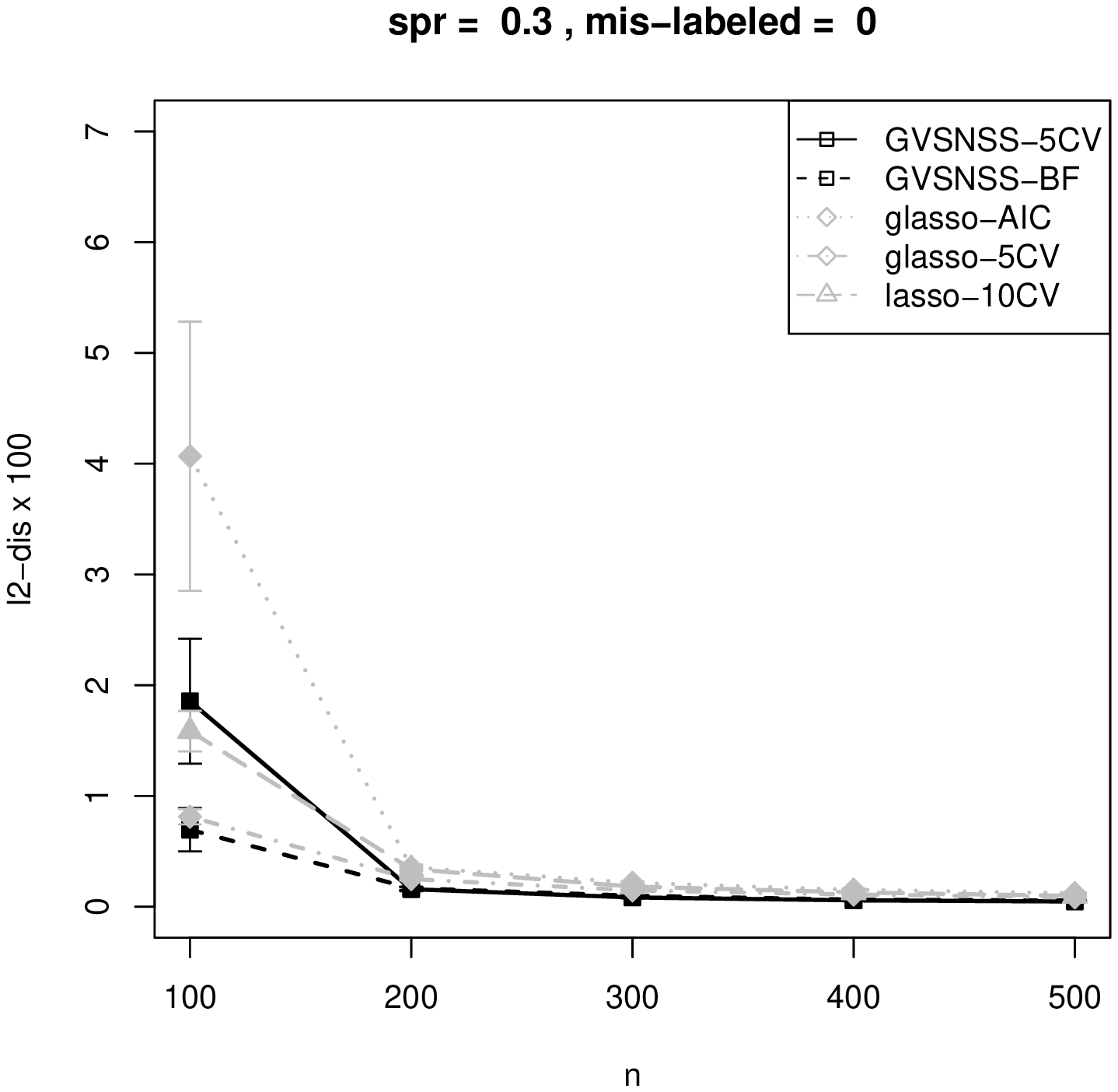}
\includegraphics[width=0.32\textwidth]{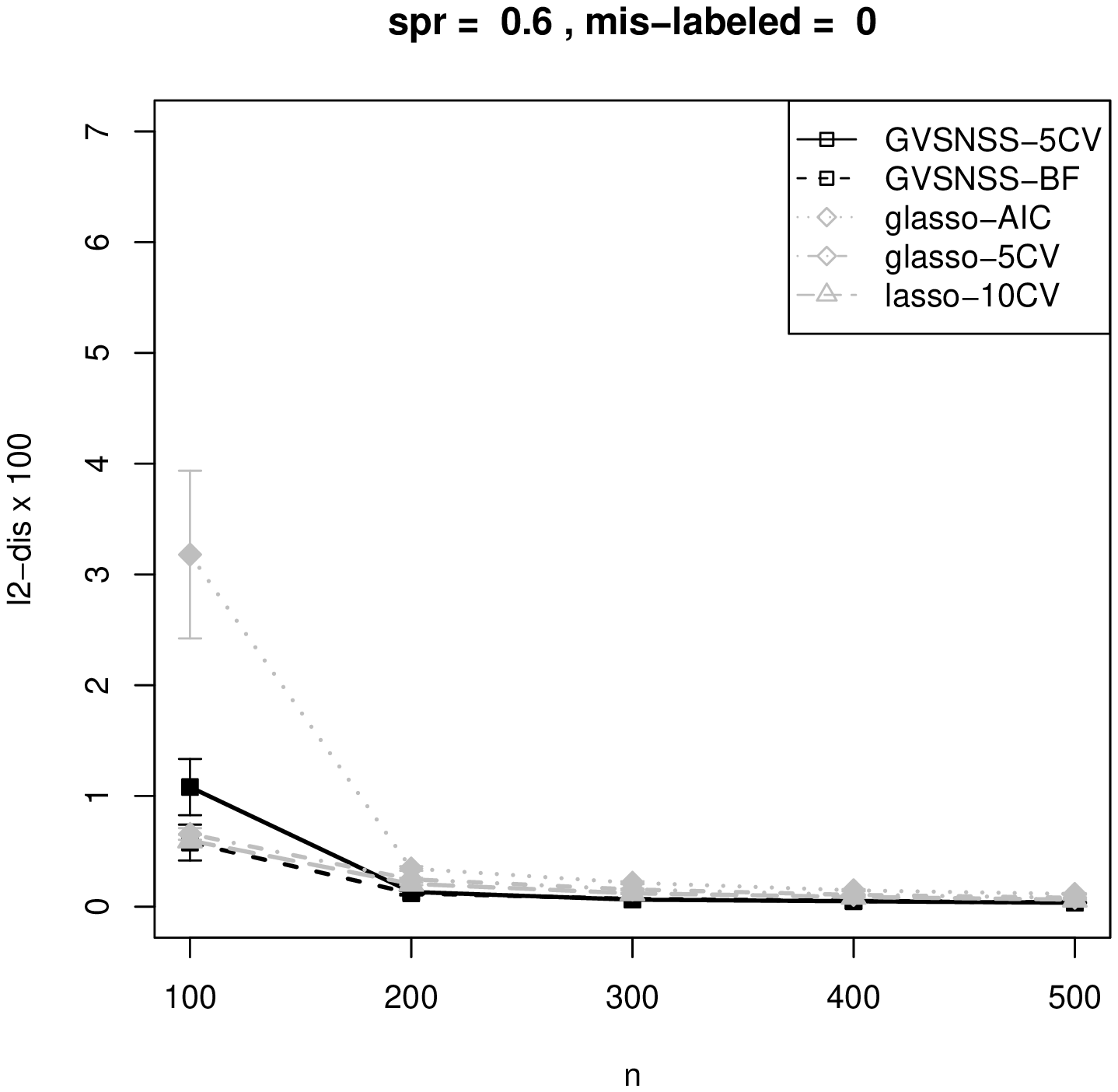}
\includegraphics[width=0.32\textwidth]{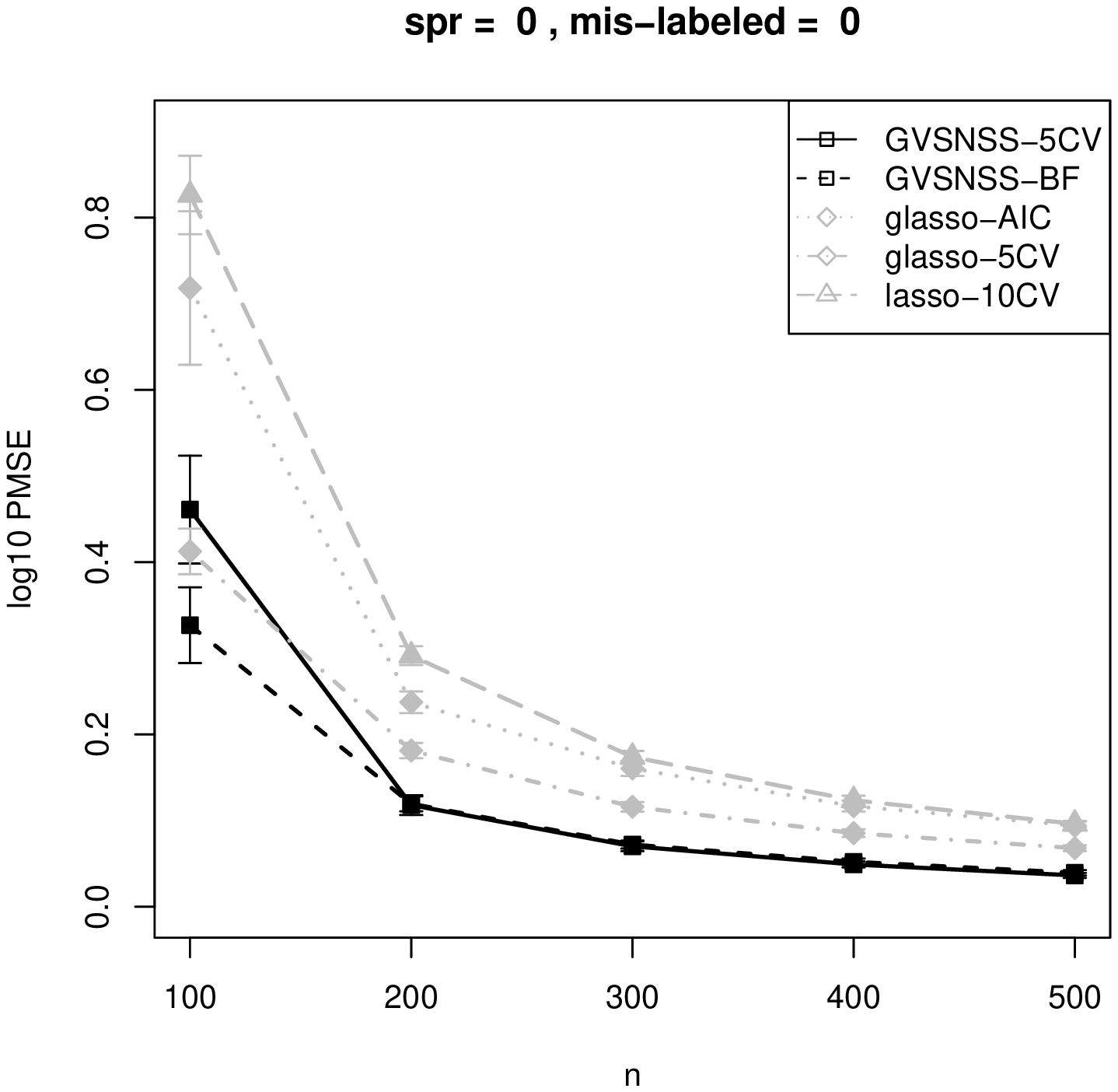}
\includegraphics[width=0.32\textwidth]{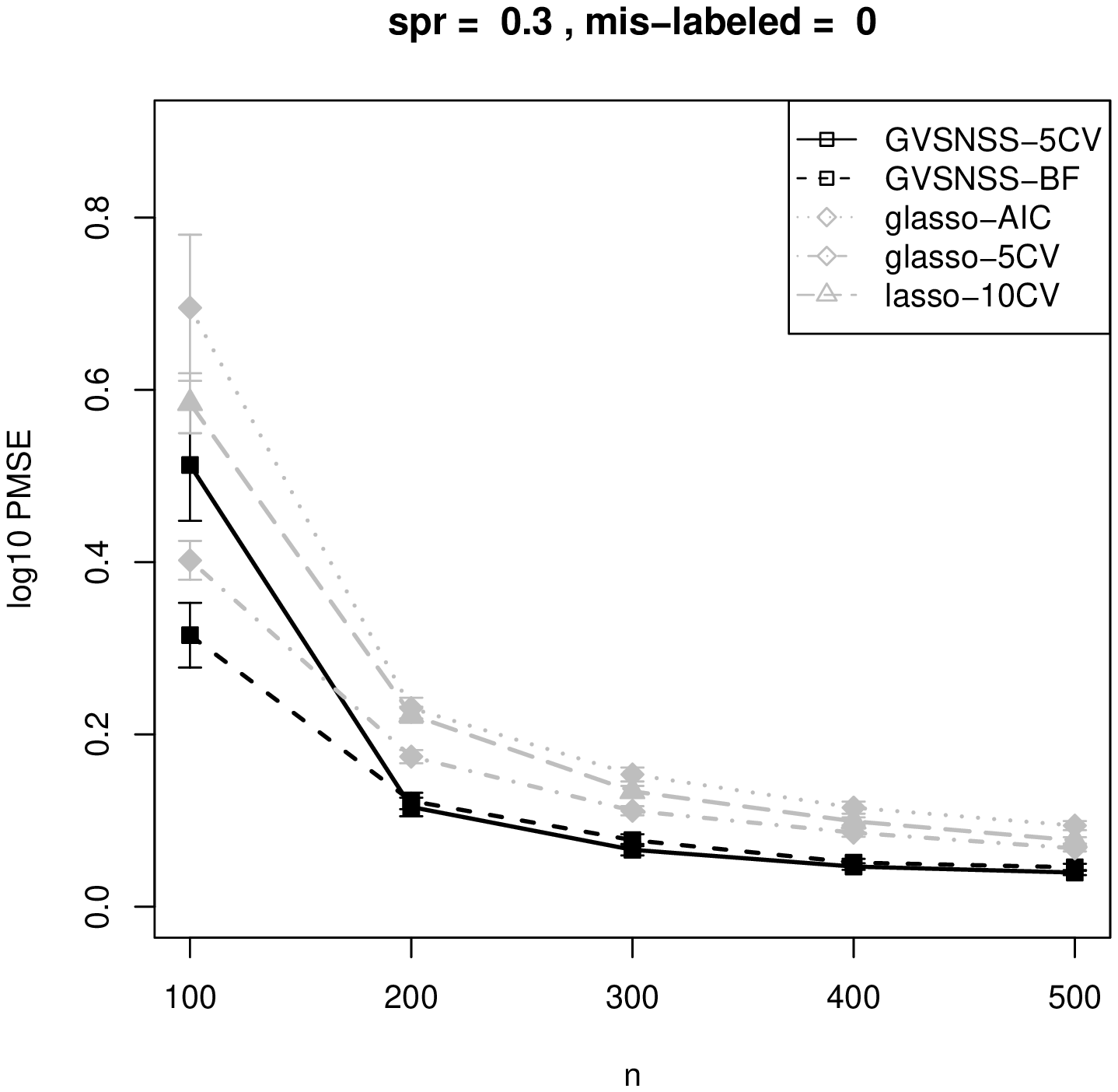}
\includegraphics[width=0.32\textwidth]{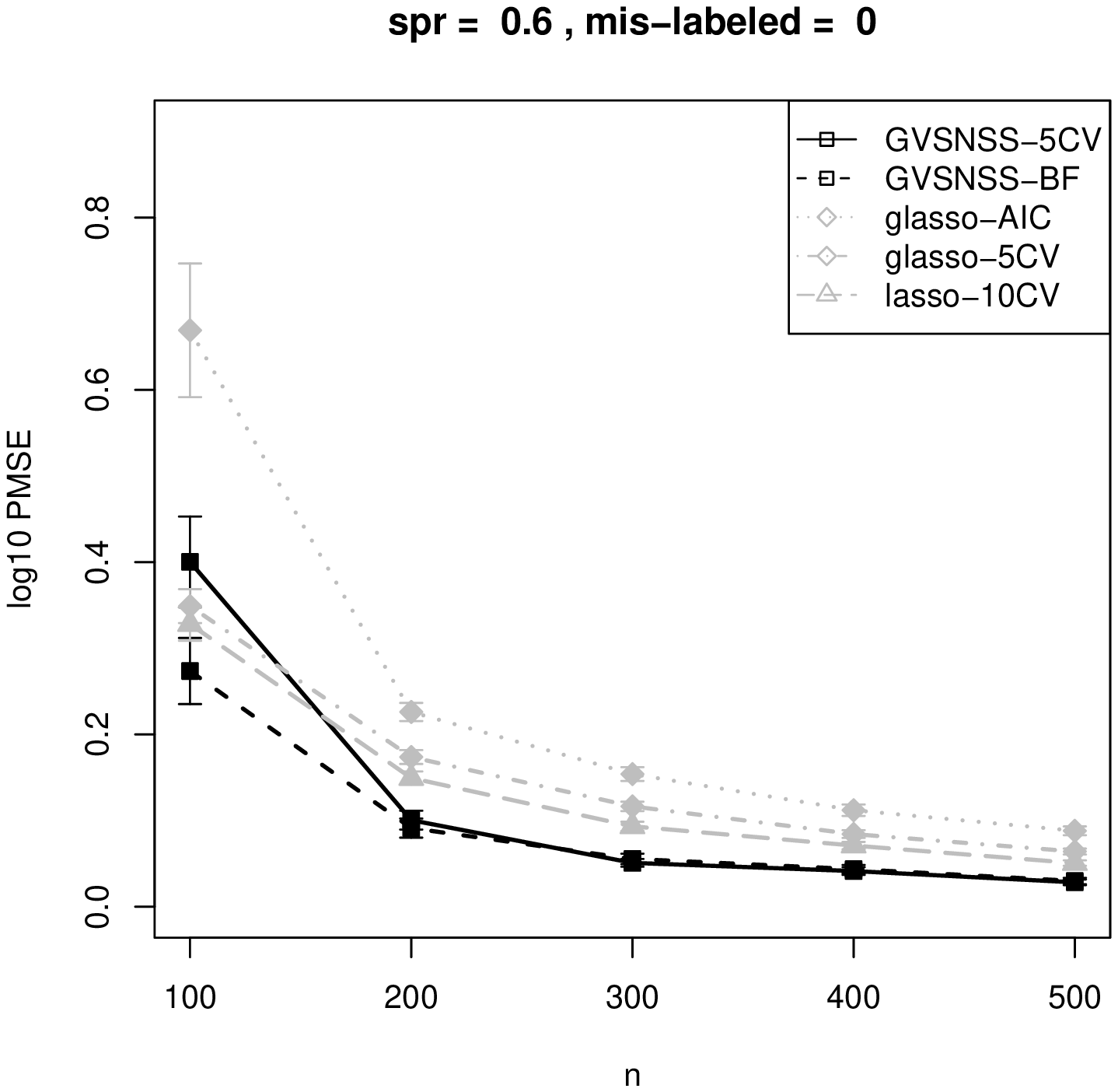}
\end{center}\caption{Estimation results from simulated data. Each point is an average over 100 replicates. For all data sets, we set mis-labeled $=0$, $p = 200$, $m = 10$ and $r = 2$. Left: spr $= 0$; Center: spr $= 0.3$; Right: spr $= 0.6$. Top: SFPR; Middle: $l_{2}$ distance between the estimates and the true values; Bottom: Logarithm of the PMSE with respect to base $10$.} 
\label{figure1}
\end{figure}

We first considered the scenario in which the group switching probability is zero. The results are shown in Figure \ref{figure1}, with the first, second and third rows being the plots of SFPR, $l_{2}$-dis and PMSE, respectively and the first, second and third columns being the plots for cases with spr $=0$, $0.3$, $0.6$, respectively. Each point in the plot is an average over 100 simulation runs. The results show that the gvsnss estimator has relatively good performances over the group lasso in variable selection when the level of within-group-sparsity is increasing. In addition, among the five estimations, the gvsnss estimation using the Bayes factor has relatively small values in squared $l_{2}$ estimation error and PMSE. However, we also noticed that the advantages of using group-based estimations such as the group lasso or gvsnss estimations over the lasso estimation will gradually disappear as the level of within-group-sparsity increases.

\begin{figure}[t]
\begin{center}
\includegraphics[width=0.32\textwidth]{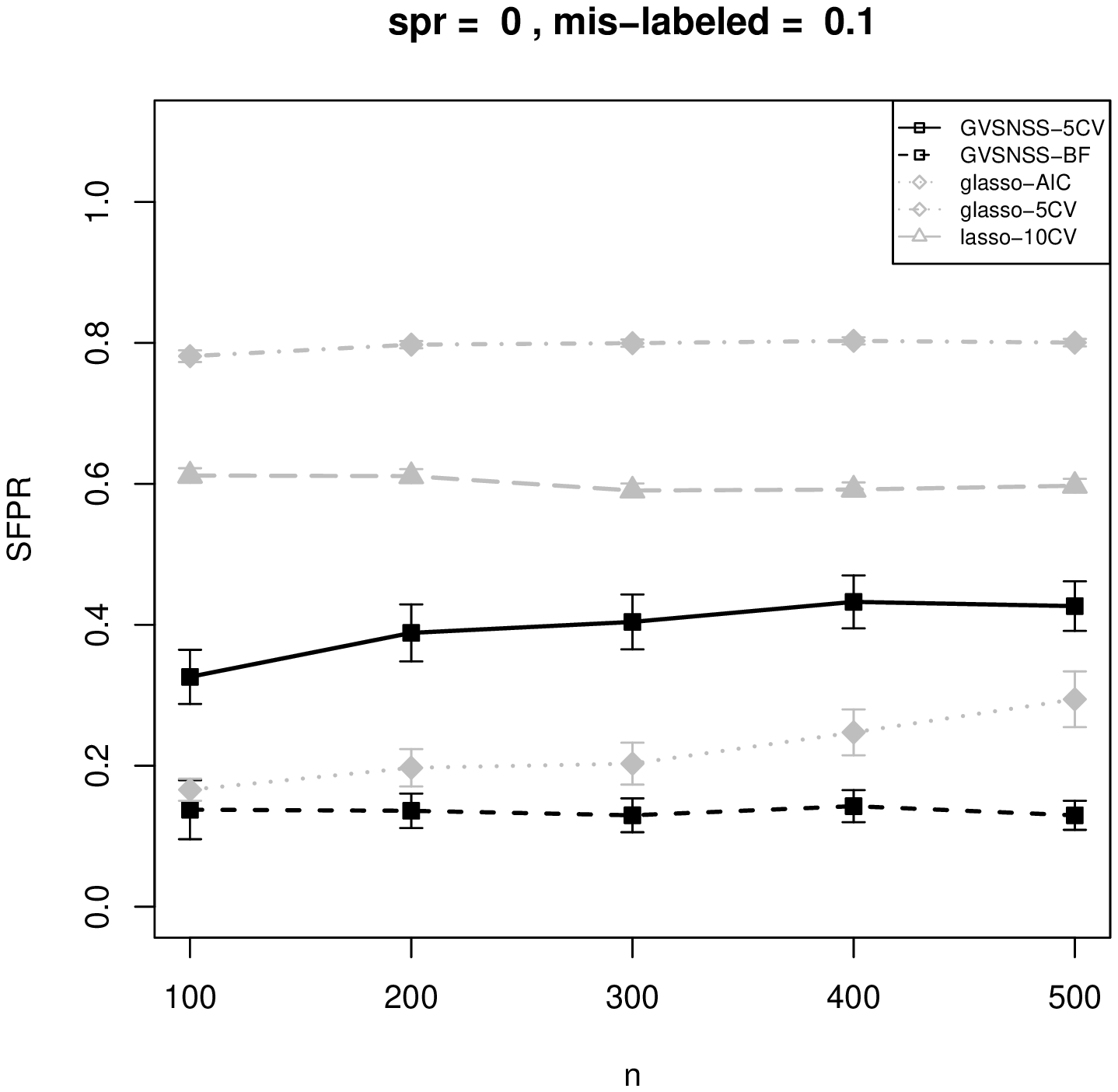}
\includegraphics[width=0.32\textwidth]{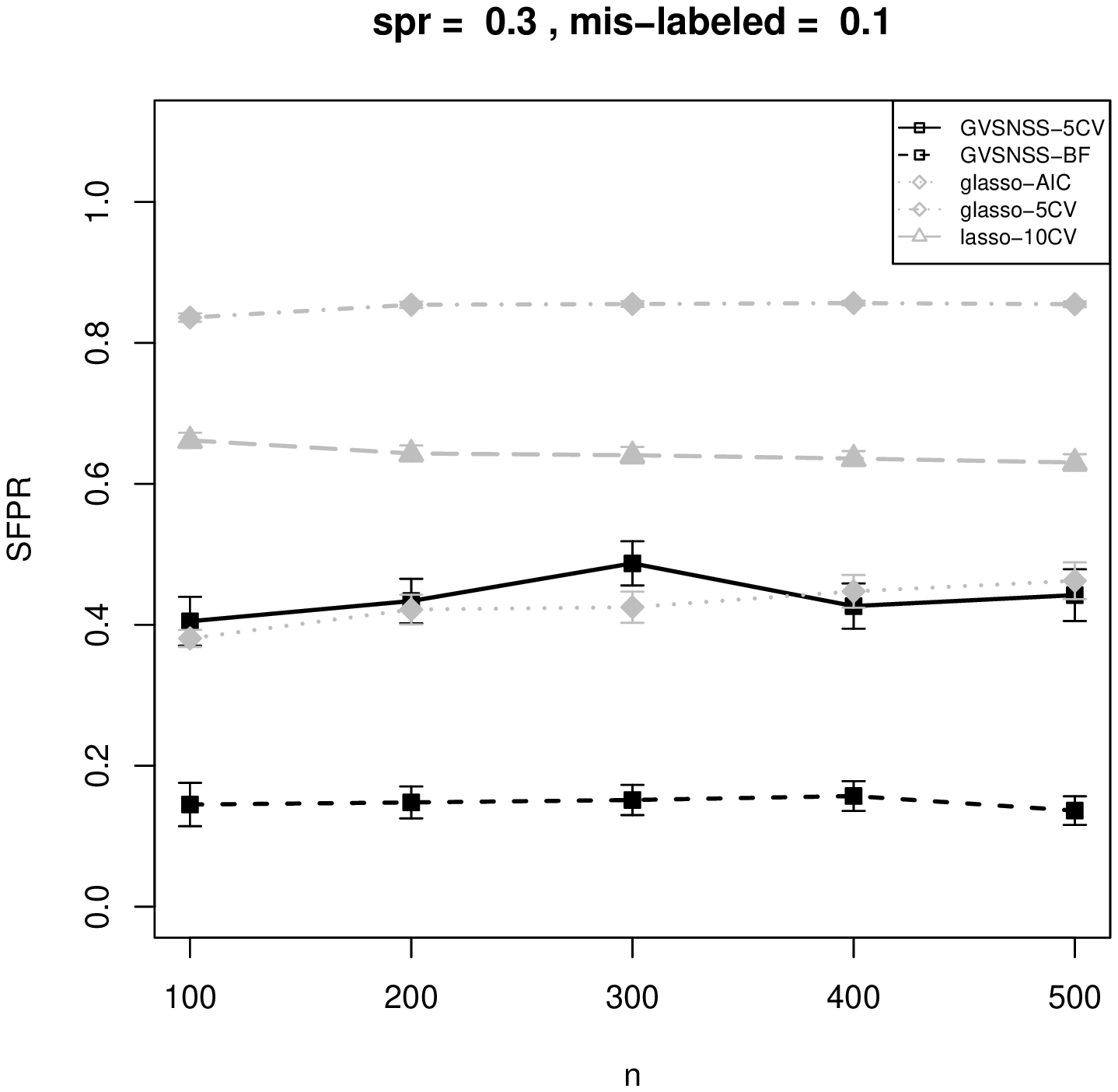}
\includegraphics[width=0.32\textwidth]{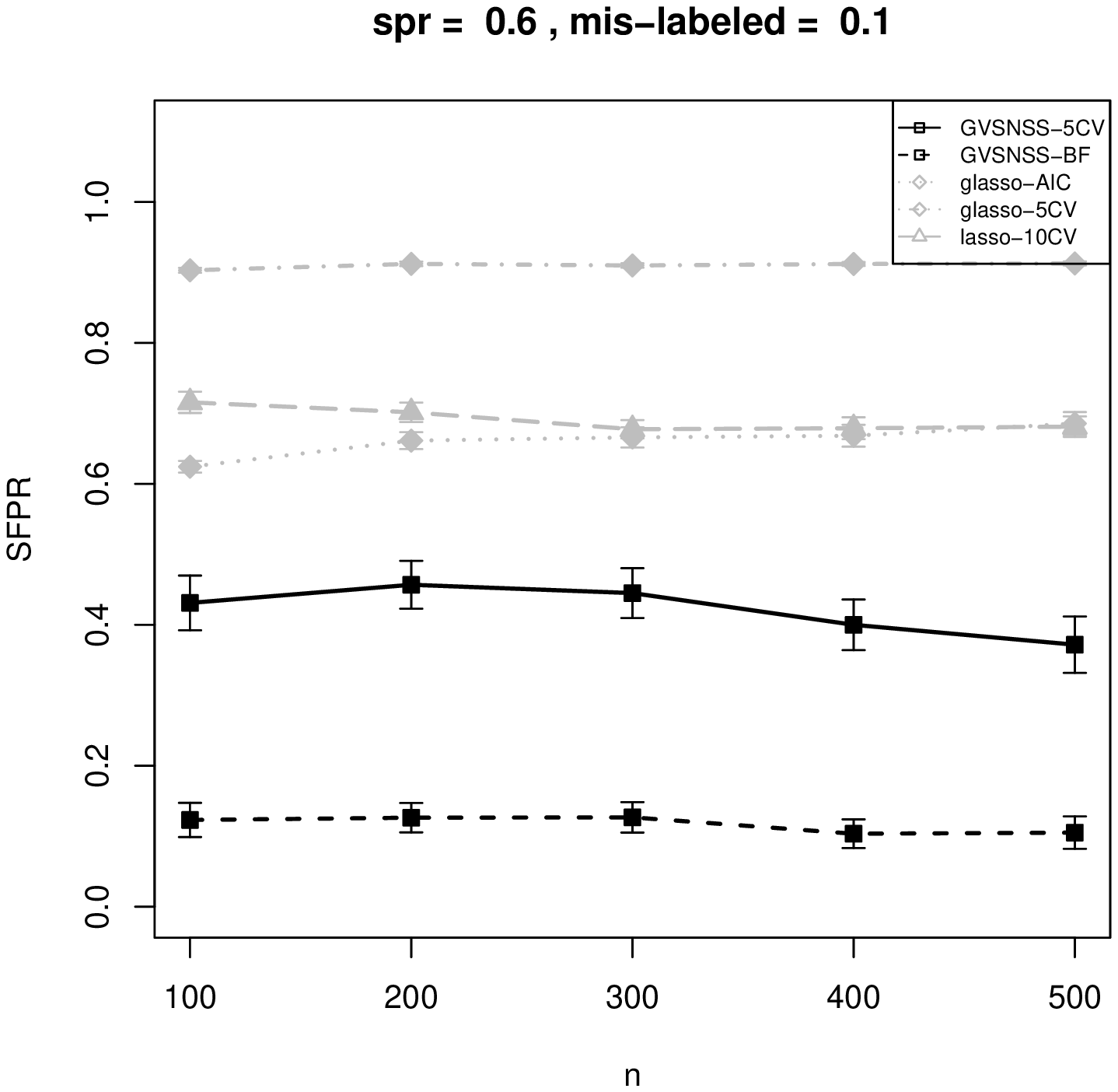}
\includegraphics[width=0.32\textwidth]{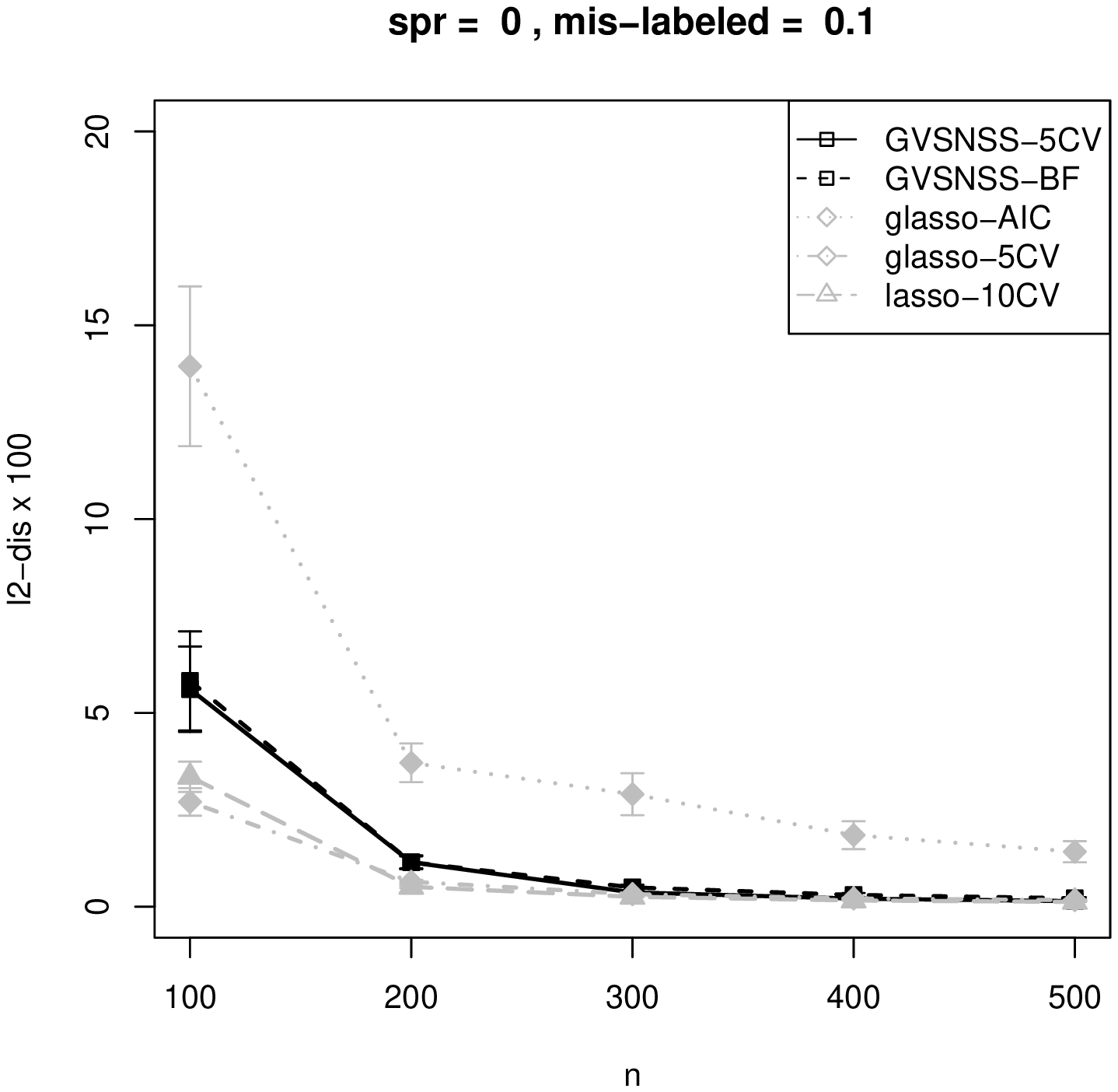}
\includegraphics[width=0.32\textwidth]{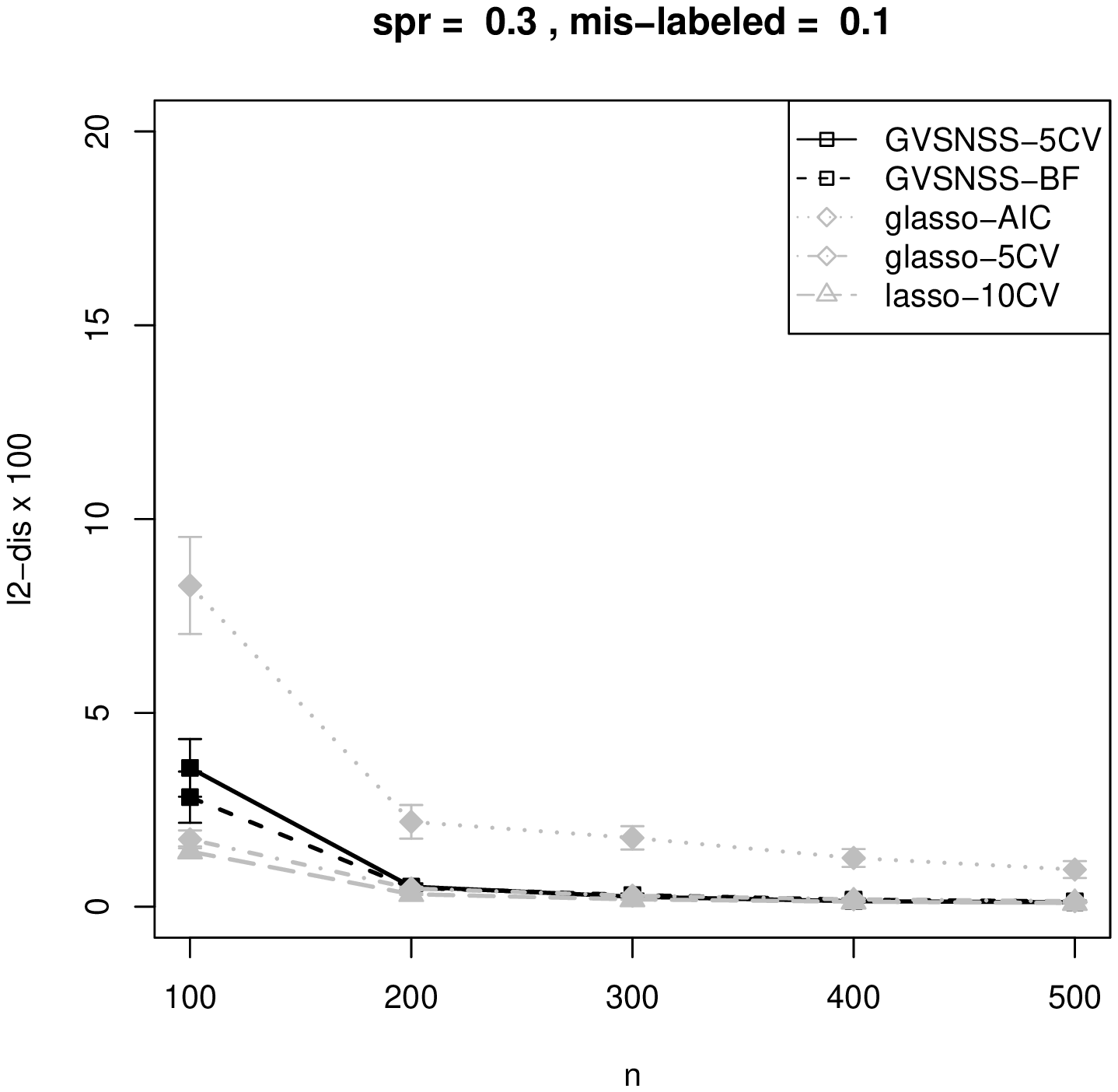}
\includegraphics[width=0.32\textwidth]{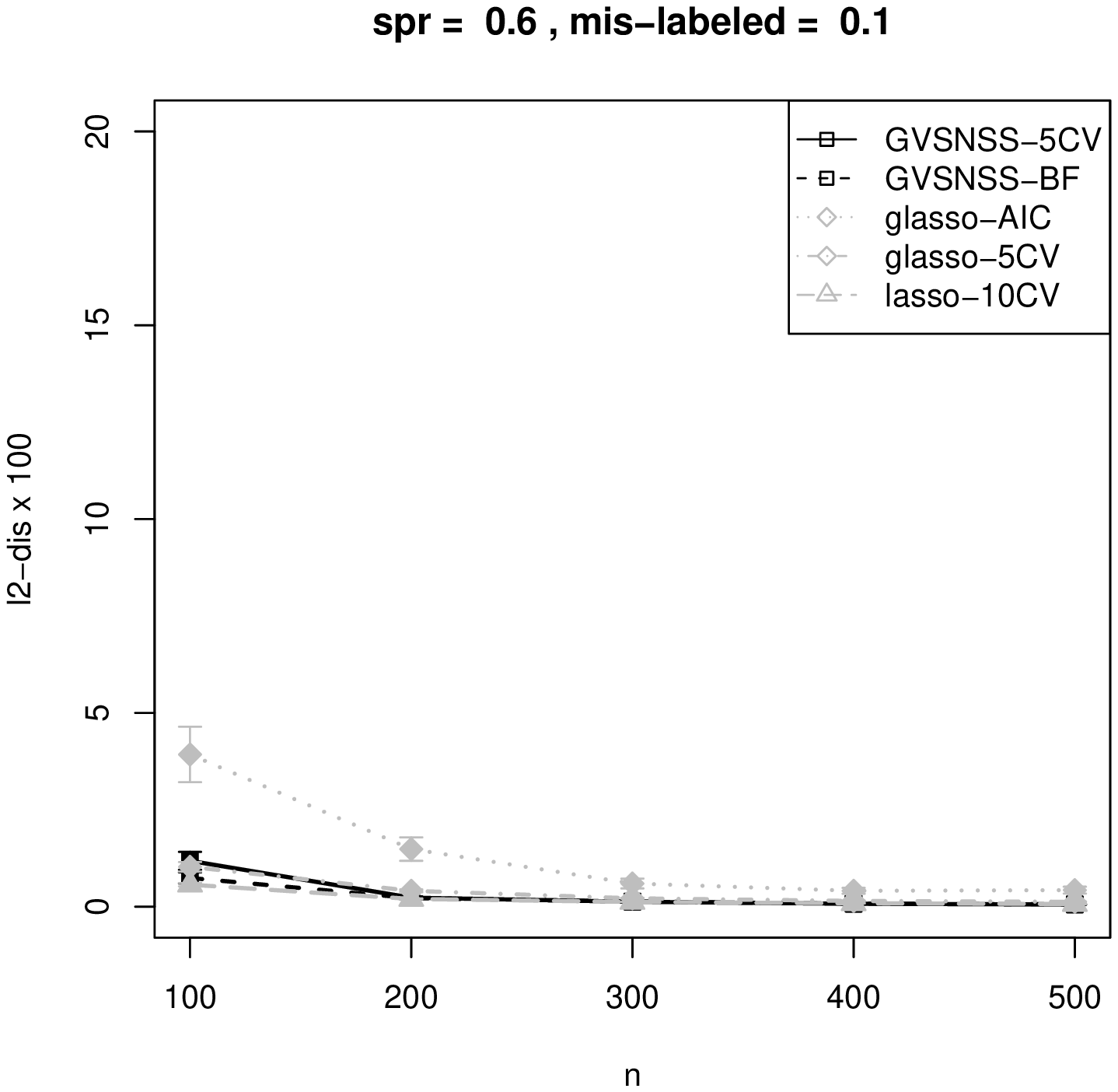}
\includegraphics[width=0.32\textwidth]{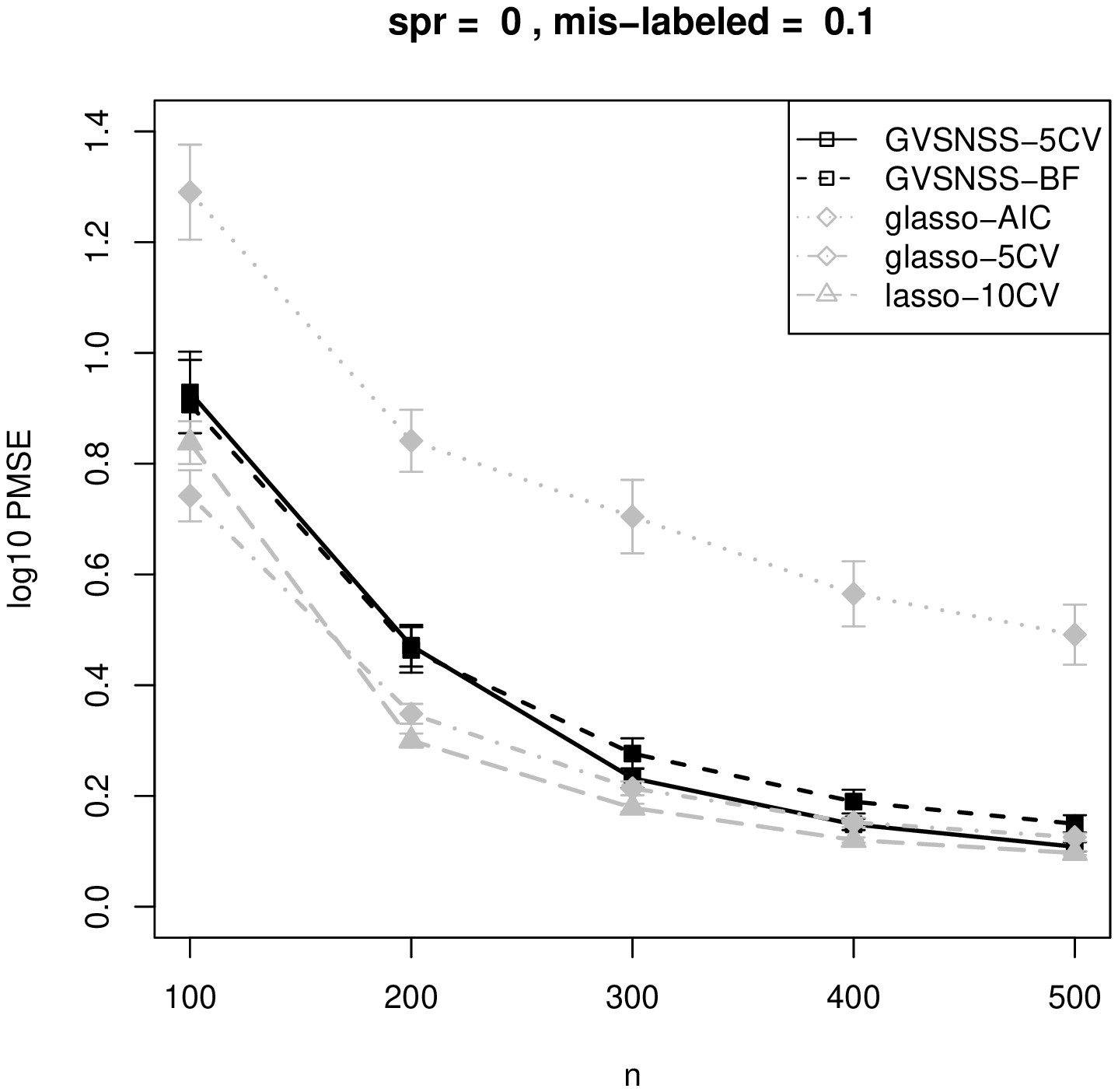}
\includegraphics[width=0.32\textwidth]{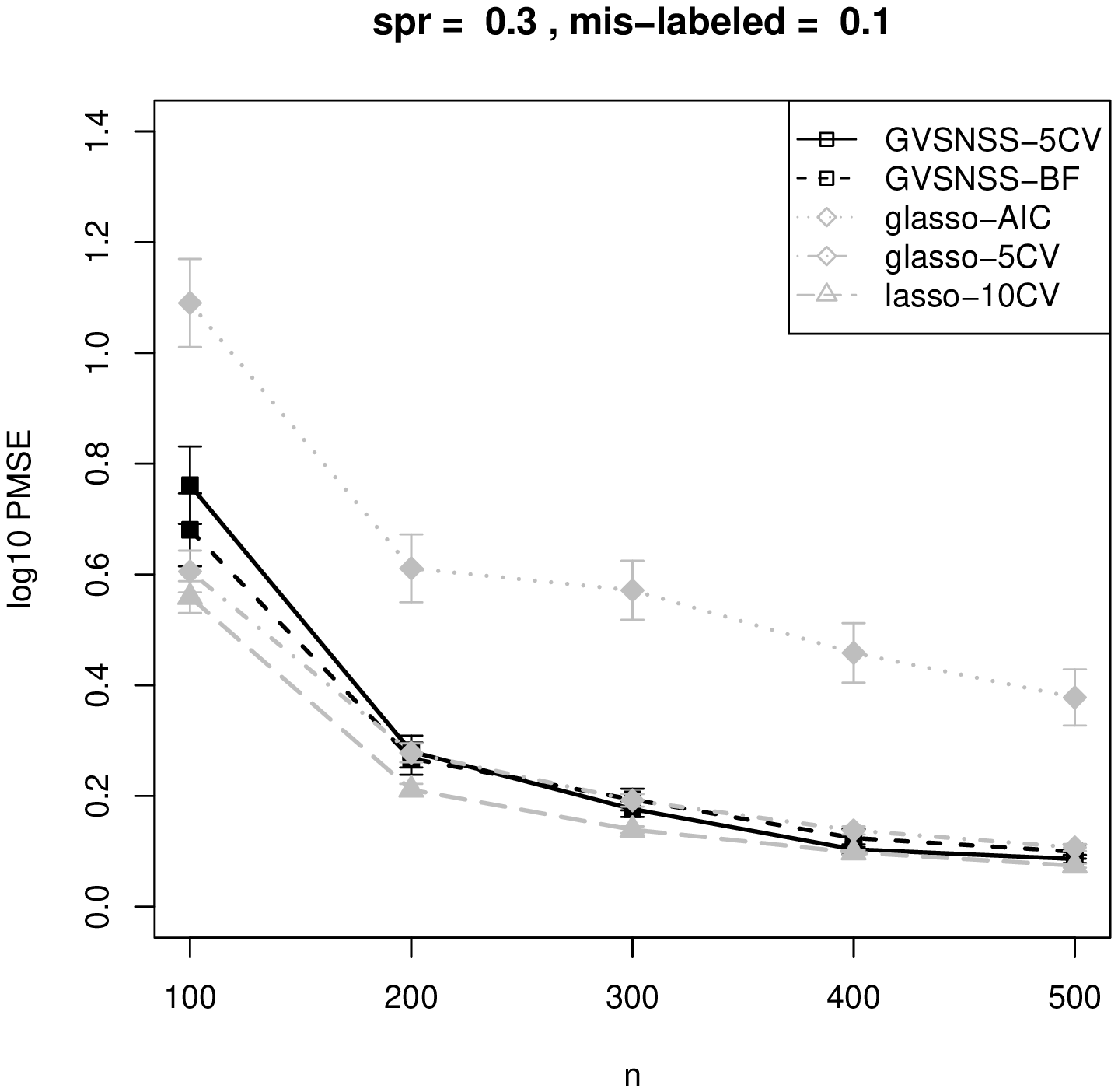}
\includegraphics[width=0.32\textwidth]{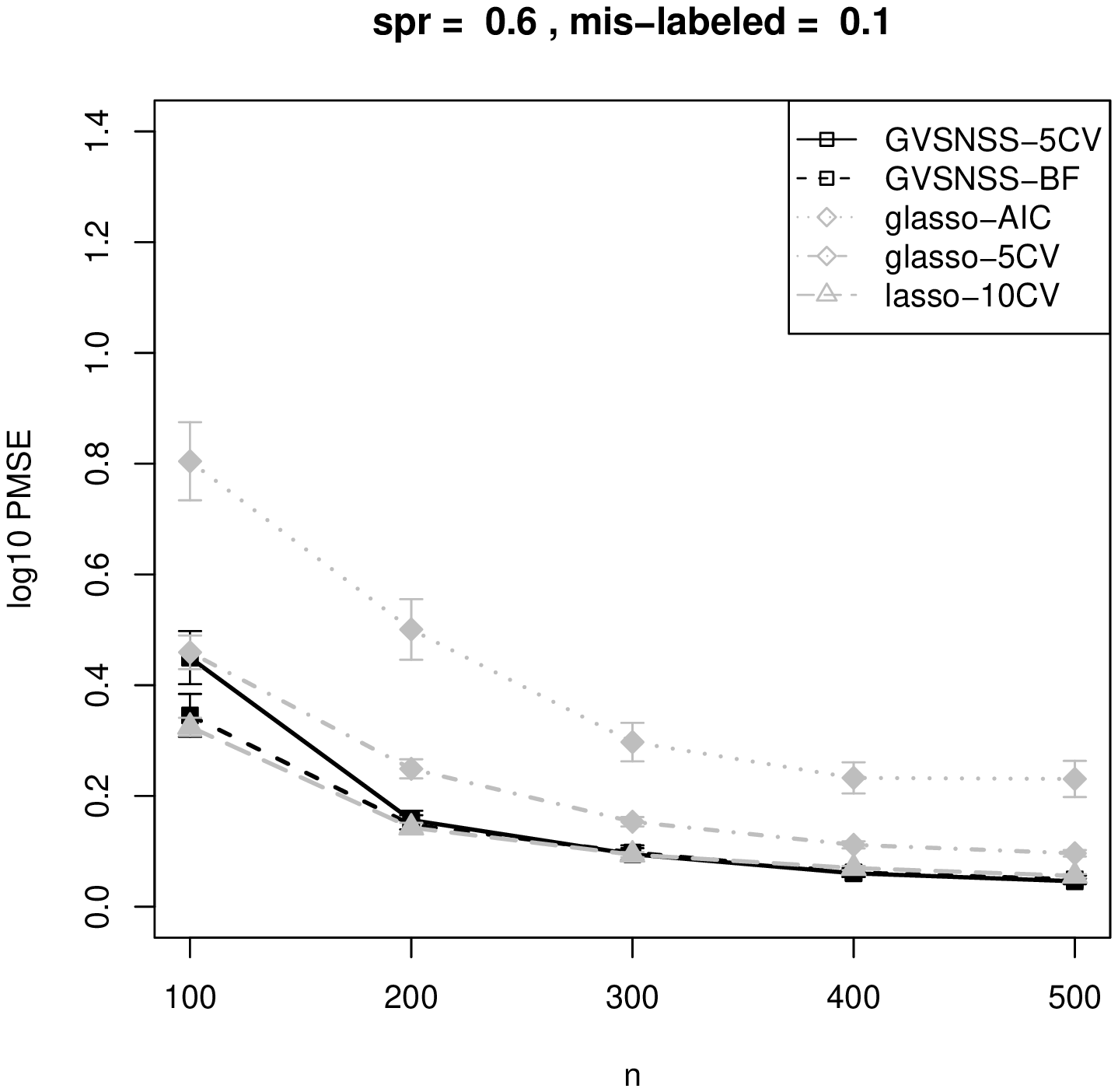}
\end{center}\caption{Estimation results from simulated data. Each point is an average over 100 replicates. For all data sets, we set mis-labeled $=0.1$, $p = 200$, $m = 10$ and $r = 2$. Left: spr $= 0$; Center: spr $= 0.3$; Right: spr $= 0.6$. Top: SFPR; Middle: $l_{2}$ distance between the estimates and the true values; Bottom: Logarithm of the PMSE with respect to base $10$.} 
\label{figure2}
\end{figure}

\begin{figure}[t]
\begin{center}
\includegraphics[width=0.32\textwidth]{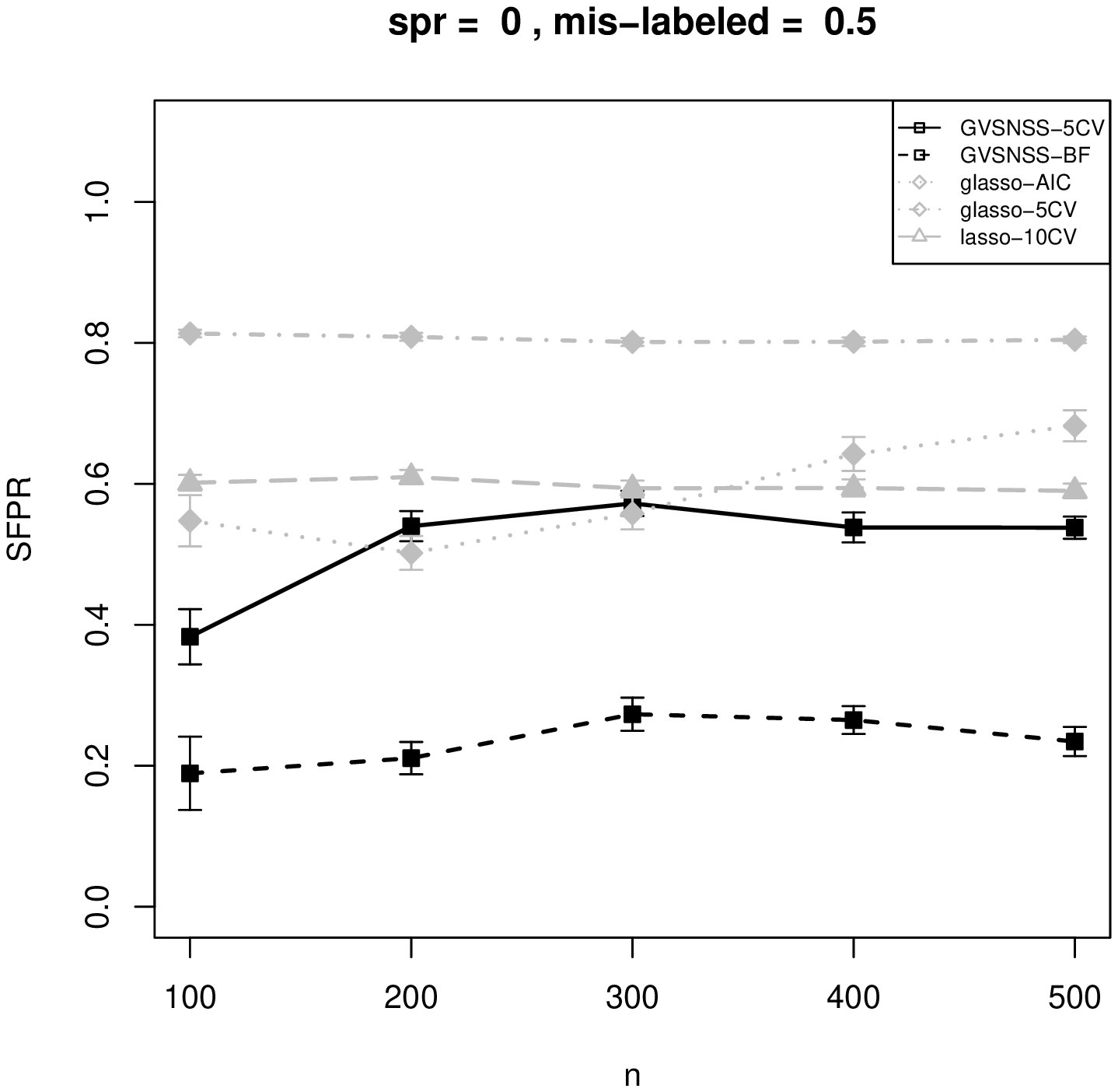}
\includegraphics[width=0.32\textwidth]{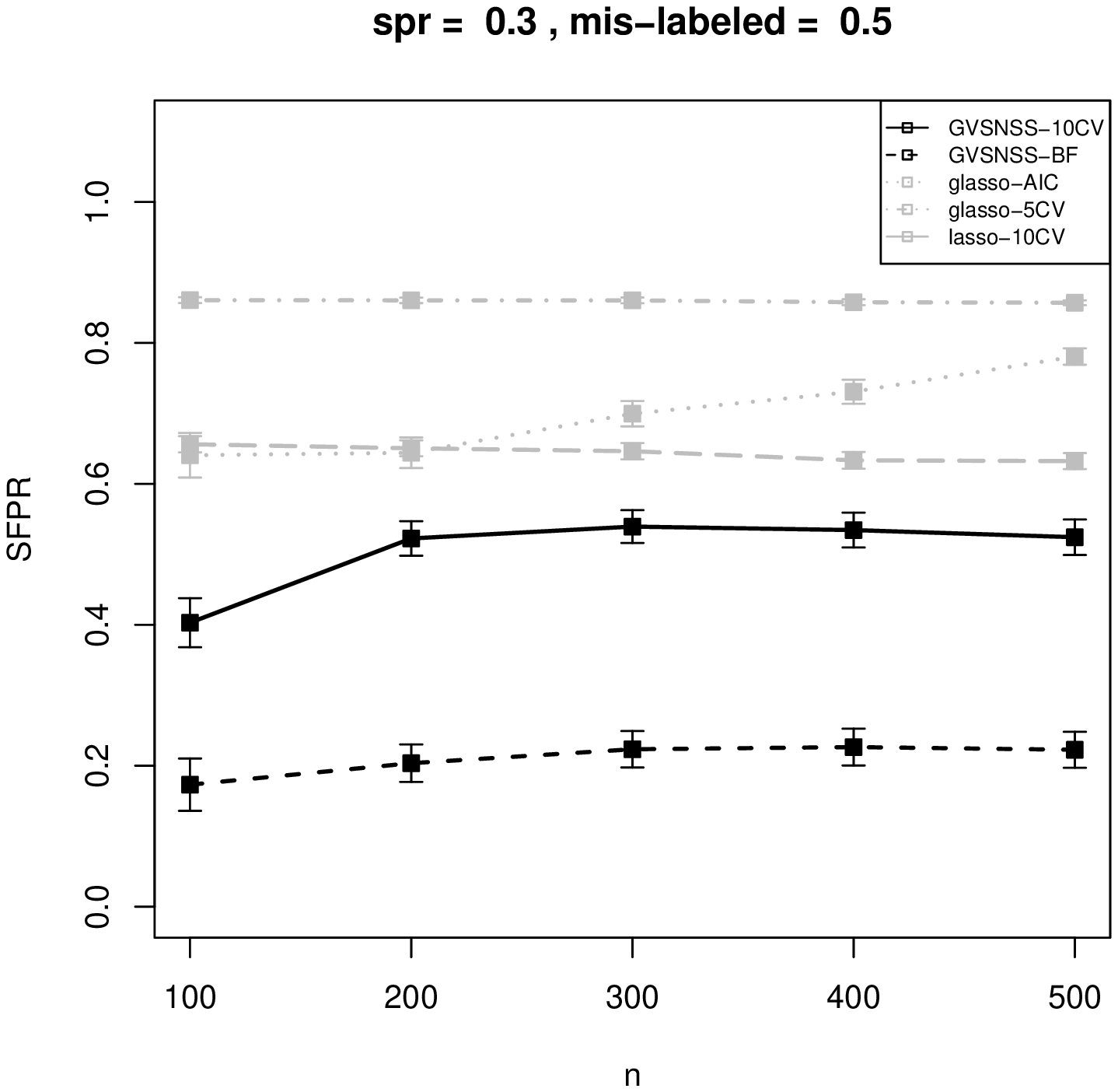}
\includegraphics[width=0.32\textwidth]{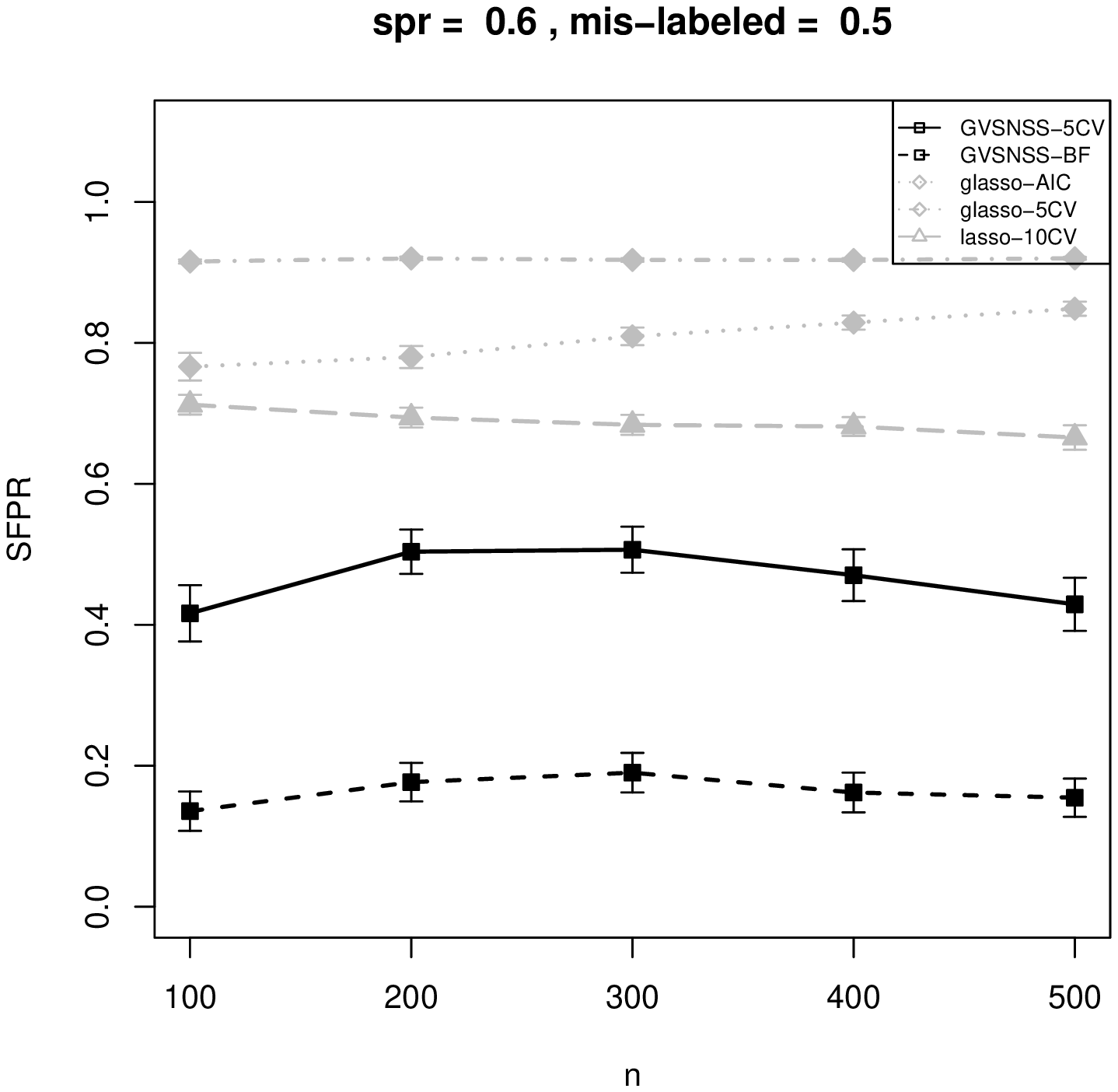}
\includegraphics[width=0.32\textwidth]{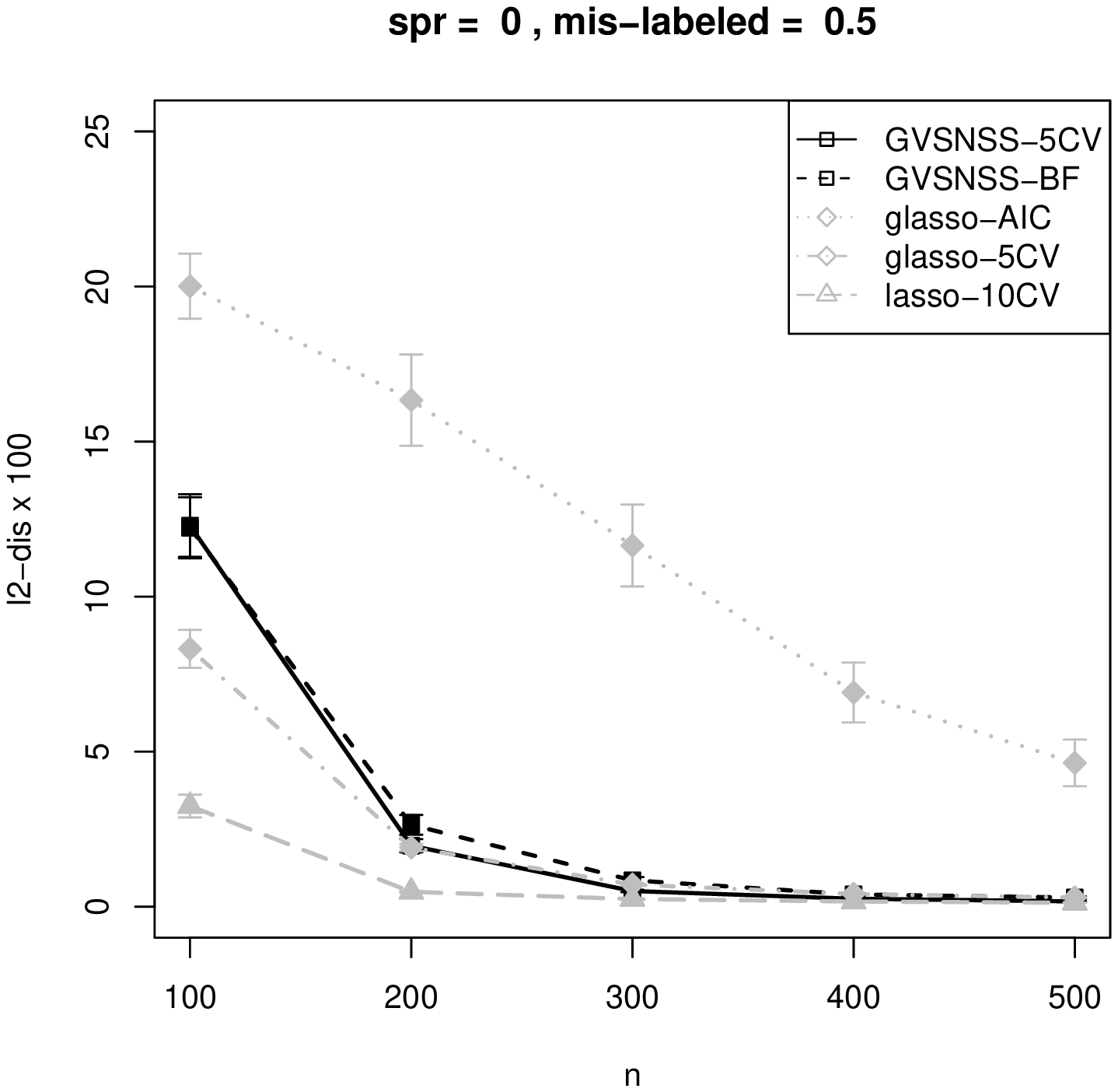}
\includegraphics[width=0.32\textwidth]{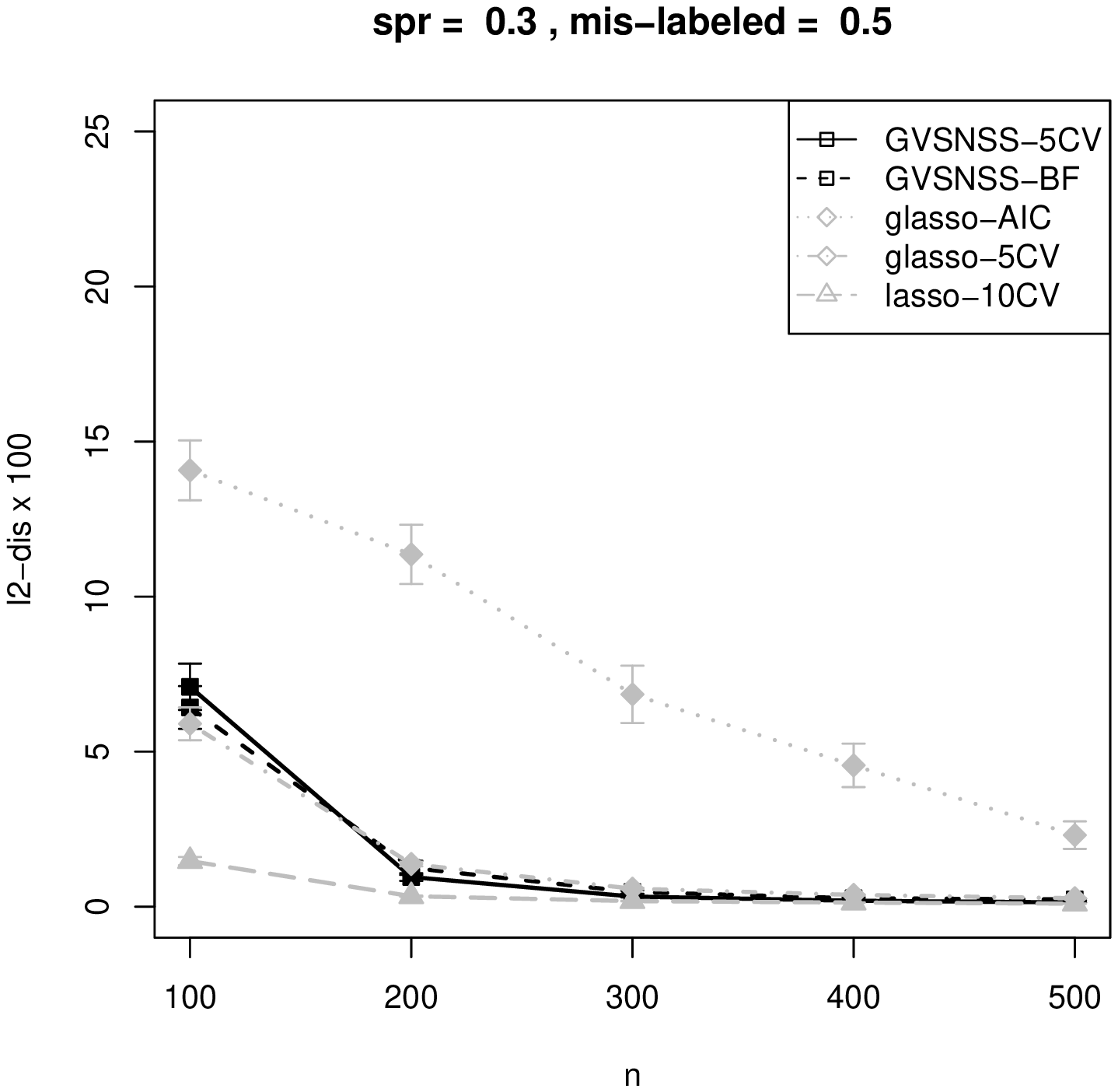}
\includegraphics[width=0.32\textwidth]{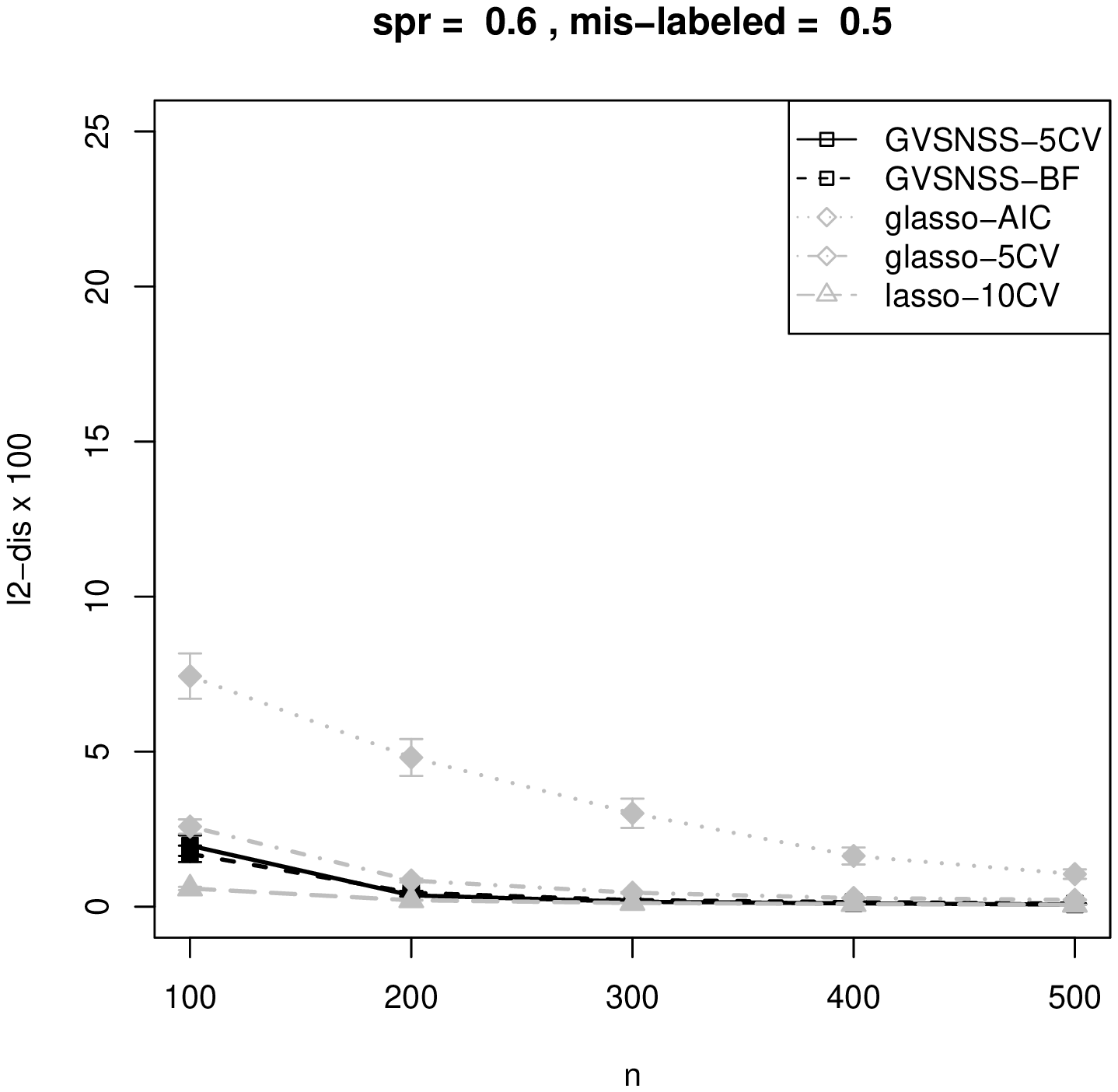}
\includegraphics[width=0.32\textwidth]{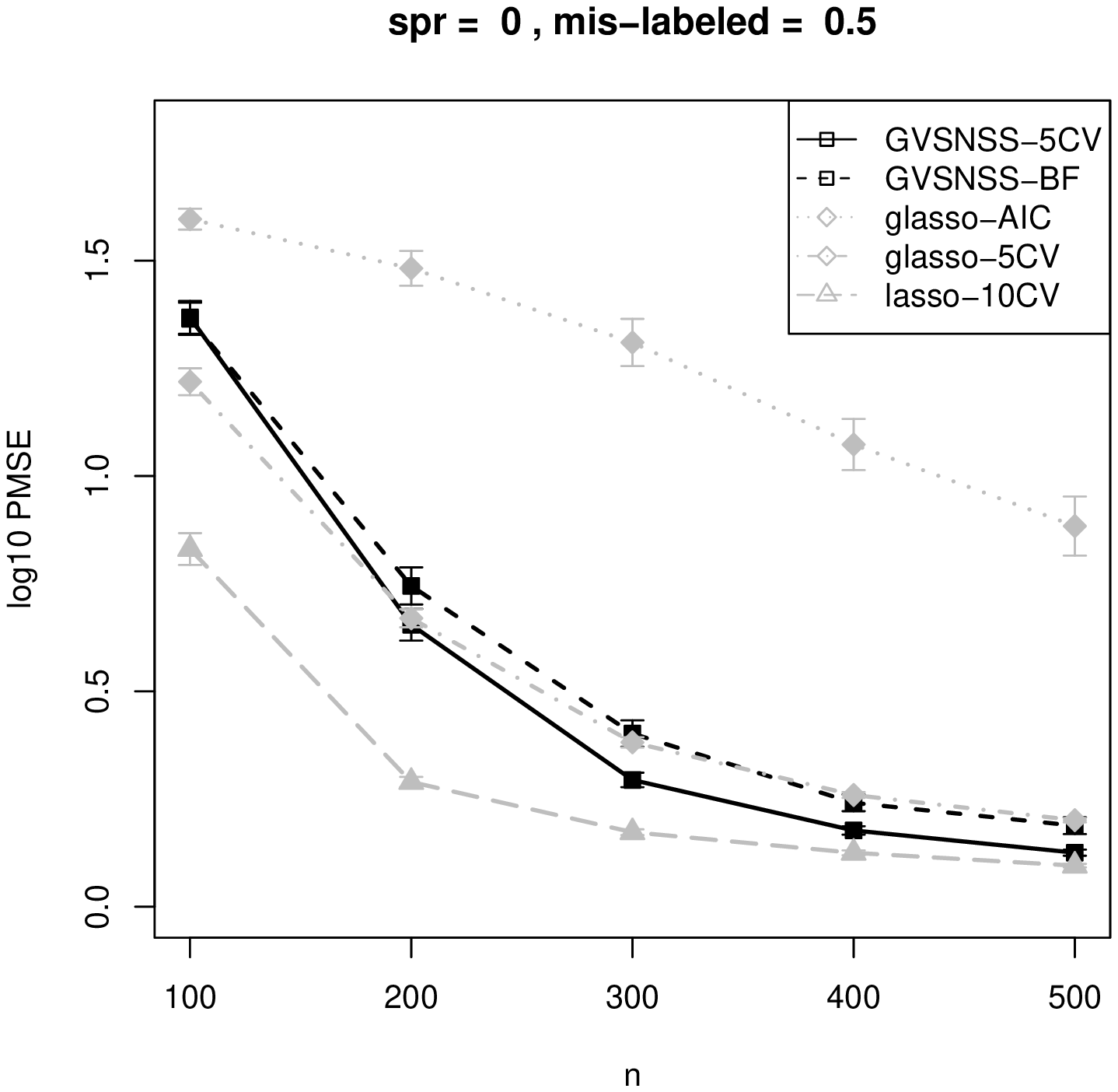}
\includegraphics[width=0.32\textwidth]{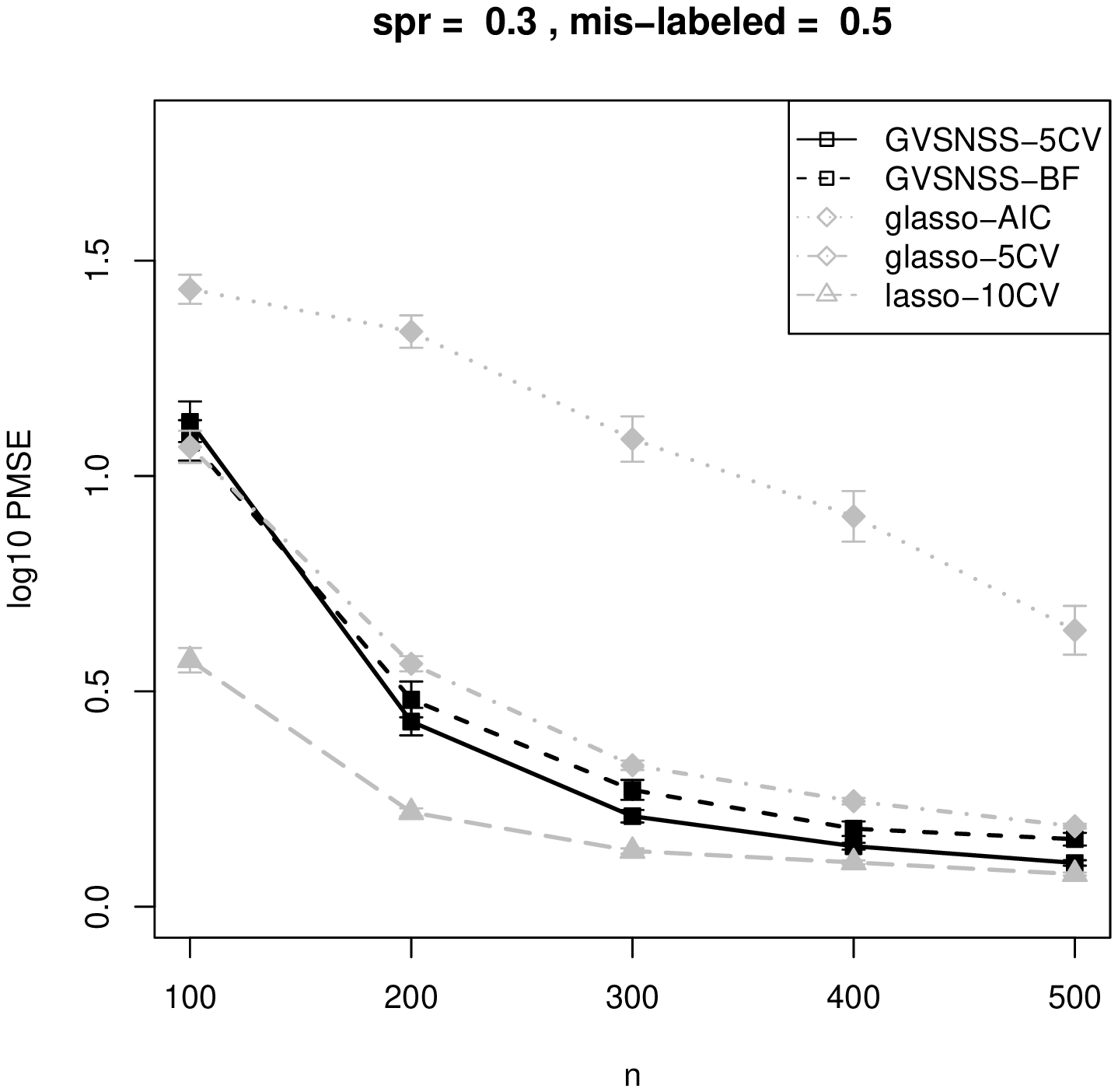}
\includegraphics[width=0.32\textwidth]{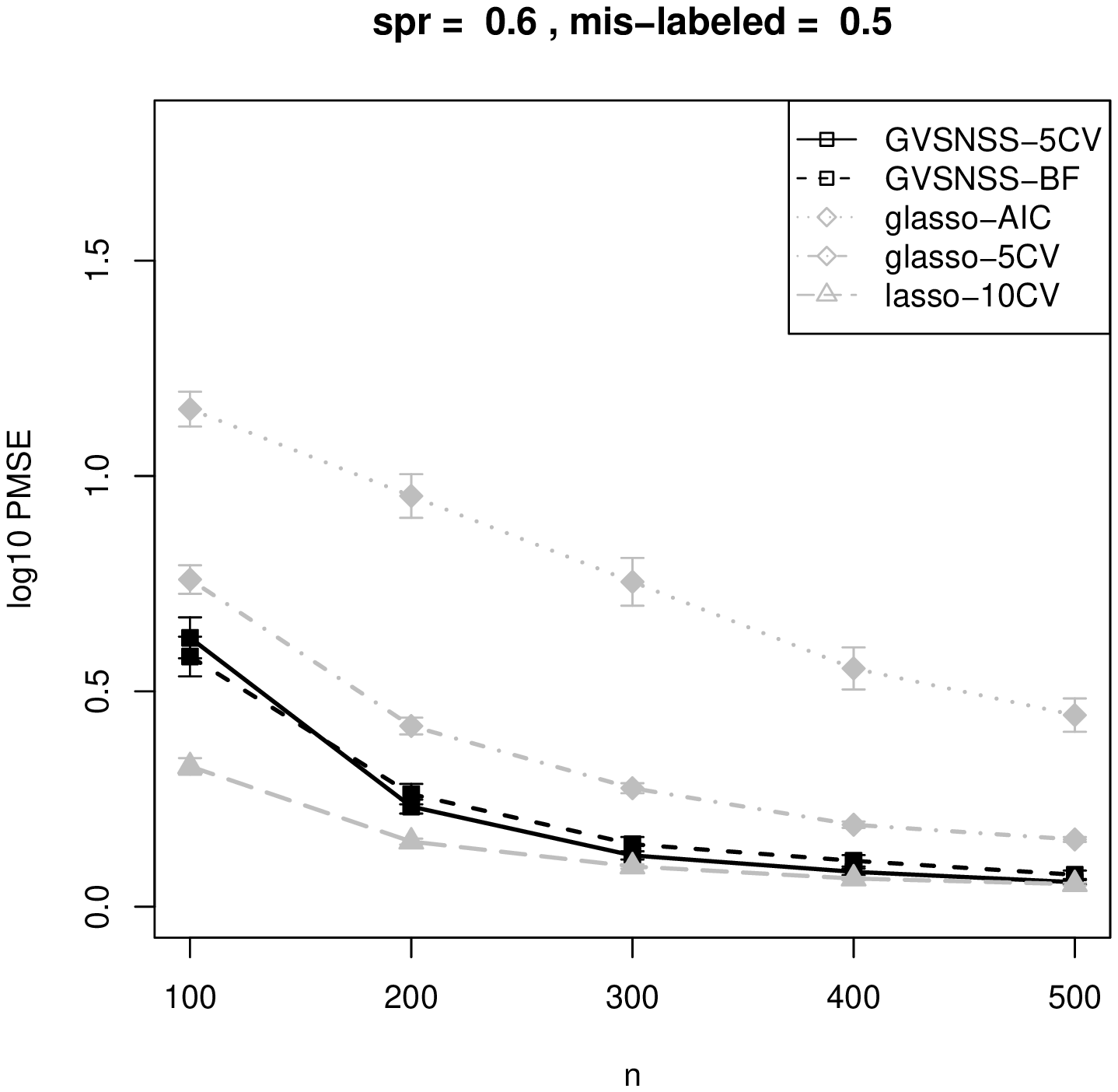}
\end{center}\caption{Estimation results from simulated data. Each point is an average over 100 replicates. For all data sets, we set mis-labeled $=0.5$, $p = 200$, $m = 10$ and $r = 2$. Left: spr $= 0$; Center: spr $= 0.3$; Right: spr $= 0.6$. Top: SFPR; Middle: $l_{2}$ distance between the estimates and the true values; Bottom: Logarithm of the PMSE with respect to base $10$.} 
\label{figure4}
\end{figure}

We then considered scenarios under different group switching probabilities. The results are given in Figures \ref{figure2} and \ref{figure4} for group switching probability equal to $0.1$ and $0.5$, respectively. The results show that the gvsnss estimator can still have relatively good performances over other benchmark estimation methods in variable selection. However, we also noticed the lasso estimation almost dominates performances in $l_{2}$ estimation error and PMSE over group-based estimation methods in these scenarios, especially when the group switching probability is high. A high group switching probability will lead to an increase in $r$, the number of groups that cover the true covariates. In Section \ref{asymptotic_analysis} we will give a theoretical explanation to these simulation results by deriving an upper bound for the $l_{2}$ estimation error.  
\section{Asymptotic analysis}\label{asymptotic_analysis}
In this section we investigate asymptotic behavior of the gvsnss estimator. Before presenting these results, we give some notation definitions. For simplicity, we define $\beta = \beta_{\text{true}}$ throughout this section. Further define $S=\{j:\beta_{j}\neq 0\}$ and $G_{R}=\{G_{k}: k\in R\}$, a collection of disjoint index sets $G_{k}$'s indexed by $R$ that covers $S$, i.e. $S\subseteq G_{R}$. Define $s = |S|$, the number of non-zero coefficients, $q_{R}=|\cup_{k\in R}G_{k}|$, the number of indices covered by $G_{R}$, and $r=|R|$, the number of groups that cover indices for covariates with non-zero coefficients.

Now consider the following function:
\begin{eqnarray}
\label{grand_obj}
V_{\tau}(w^{\prime},\beta^{\prime},G^{\prime})&=&\big|\big|\epsilon^{\prime}-Xw^{\prime}\big|\big|_{2}^{2} + \lambda||w^{\prime}+\beta^{\prime}||_{2}^{2}\nonumber\\
& & + \rho_{1}\sum_{j=1}^{p}\frac{\log(1+\tau^{-1}|w_{j}^{\prime}+\beta_{j}^{\prime}|)}{\log(1+\tau^{-1})}\nonumber\\
& &+\rho_{2}\sum_{k=1}^{m^{\prime}}\sqrt{q_{k}^{\prime}}\frac{\log(1+\tau^{-1}||w_{G_{k}^{\prime}}^{\prime}+\beta_{G_{k}^{\prime}}^{\prime}||_{2})}{\log(1+\tau^{-1})},
\end{eqnarray}
where $\epsilon^{\prime}=y - X\beta^{\prime}$, $G^{\prime} = \{G_{k}^{\prime}:k=1,2,\cdots,m\}$ and $q_{k}^{\prime}= |G_{k}^{\prime}|$. At a fixed $\tau^{\prime}$, we define $\widehat{\beta}^{\tau^{\prime}}$ by 
\begin{eqnarray}
\label{vtau_estimator}
\widehat{\beta}^{\tau^{\prime}} = \arg\min_{\beta^{\prime}}\lim_{\tau\rightarrow\tau^{\prime}}V_{\tau}(0,\beta^{\prime},G).
\end{eqnarray}
Further define $\widehat{S}^{\tau^{\prime}}=\{j:\widehat{\beta}_{j}^{\tau^{\prime}}\neq 0\}$ and $\widehat{s}^{\tau^{\prime}}=|\widehat{S}^{\tau^{\prime}}|$. Note that if we let $\tau\rightarrow 0$, then $V_{\tau}(0,\beta^{\prime},G^{\prime})$ will approach to the objective function (\ref{model07}). Therefore technically we can express the gvsnss estimator as 
\begin{eqnarray}
\label{gvsnss_estimator}
\widehat{\beta}_{\text{gvsnss}} = \arg\min_{\beta^{\prime}}\lim_{\tau\rightarrow 0}V_{\tau}(0,\beta^{\prime},G).
\end{eqnarray} 
We further define $\widehat{S}=\{j:\widehat{\beta}_{\text{gvsnss}}\neq 0\}$ and $\widehat{s}=|\widehat{S}|$. Note that by definition, as $\tau\rightarrow 0$, (\ref{vtau_estimator}) becomes $\widehat{\beta}^{0}=\arg\min_{\beta^{\prime}}\lim_{\tau\rightarrow 0}V_{\tau}(0,\beta^{\prime},G)=\widehat{\beta}_{\text{gvsnss}}$. As a result of that, we have $\widehat{S}^{\tau}\rightarrow\widehat{S}$ and $\widehat{s}^{\tau}\rightarrow\widehat{s}$ as $\tau\rightarrow 0$. 
\subsection{$l_{2}$ estimation error}
One useful concept to justify the advantage of group-based estimation is the strong group sparsity \cite{huangandzhang10}. We say the true coefficient vector $\beta$ is $(s_{0},r_{0})$ strongly group-sparse if there exists a collection of index sets $G_{R}=\{G_{k}:k\in R\}$ such that $S\subseteq G_{R}$ with $q_{R}=|G_{R}|\leq s_{0}$ and $r=|R|\leq r_{0}$. For group lasso $\widehat{\beta}_{\text{GL}}$ defined in (\ref{group_lasso_estimator}), Huang and Zhang \cite{huangandzhang10} showed that if $\beta$ is $(s_{0},r_{0})$ strongly group-sparse, then given some regular conditions hold, with $1-\alpha$ probability, the $l_{2}$ estimation error $||\widehat{\beta}_{\text{GL}}-\beta||_{2}=O(n^{-1/2}\sqrt{s_{0}+r_{0}\log(m/\alpha)})$. The order of magnitude implies that the group lasso estimation can be beneficial if $q_{R}$, the number of indices in $G_{R}$, and $r$, the number of index sets that cover $S$, are small.  

Here we have to note that directly comparing rates of the $l_{2}$ estimation error between the lasso and group lasso is not easy since it requires one to derive the rates under the same assumptions. Lounici et al. \cite{lounicietal11} provided such comparisons for multi-task learning cases and showed that the upper bound for the $l_{2}$ estimation error of the group lasso can have an order of magnitude smaller than the lower bound for the $l_{2}$ estimation error of the lasso.  

Below we start our investigation on the $l_{2}$ estimation error $||\widehat{\beta}_{\text{gvsnss}}-\beta||_{2}$ by deriving a deterministic upper bound for $||\widehat{\beta}^{\tau}-\beta||_{2}$.
\begin{theorem}
\label{theorem_deterministic_bound}
For $\epsilon = y - X\beta$, $\tau\in[0,1)$, and $1\leq\max(q_{R},\widehat{s}^{\tau})\leq p$, we have,
\begin{eqnarray}
\label{theorem1_01}
||\widehat{\beta}^{\tau}-\beta||_{2}&\leq&\frac{q_{R}^{1/2}}{\kappa_{n}+\lambda n^{-1}}\Bigg\{4\bigg[1+\bigg(\frac{\widehat{s}^{\tau}}{4s}\bigg)^{1/2}\bigg]\frac{||X^{T}\epsilon||_{\infty}}{n} + 2\max_{j\in S}|\beta_{j}|\frac{\lambda}{n}\nonumber\\
& &+\bigg[\frac{2c_{2}^{-1}+1}{\log(\tau^{-1})}+c_{3}^{-1}\bigg]\bigg(\frac{\rho_{1}+\rho_{2}}{n}\bigg)\Bigg\},
\end{eqnarray}
where $\kappa_{n}=n^{-1}\min_{w}w^{T}X^{T}Xw$, $c_{2}=\min_{j\in\widehat{S}^{\tau}}|\widehat{\beta}_{j}^{\tau}|$, and $c_{3}=\min_{j\in S}|\beta_{j}|$. \end{theorem}
Theorem \ref{theorem_deterministic_bound} does not rely on any distribution assumption on the error vector $\epsilon$. It is stated in a deterministic way and does not have any probabilistic interpretation. 

Below we will give some conditions that are useful in deriving upper bounds for $||\widehat{\beta}_{\text{gvsnss}}-\beta||_{2}$ in a situation in which some distribution assumption is imposed on $\epsilon$.

\textbf{Assumption 1.} Let $\kappa_{n}$ be the same as the one defined in Theorem \ref{theorem1_01}. We assume $\kappa_{n}+\lambda n^{-1}> 0$ as $n\rightarrow \infty$.

Assumption 1 is similar to Condition A1 in \cite{zouandzhang09}. It mainly serves as a statement to guarantee that the minimum eigenvalue of the matrix $n^{-1}(X^{T}X+\lambda I_{p\times p})$ is positive when $n\rightarrow\infty$. Note that without Assumption 1, $\kappa_{n}$ will be equal to zero when $n<p<\infty$, but the minimum eigenvalue value $\kappa_{n} + n^{-1}\lambda = n^{-1}\lambda$ will remain positive if $\lambda > 0$. Assumption 1 further implies that $\sqrt{n}(\kappa_{n} + \lambda n^{-1})\rightarrow\infty$ when $n\rightarrow\infty$.

\begin{theorem}
\label{theorem_l2_error_bound}
Assume that $\epsilon_{i}$'s are i.i.d. as Normal$(0,\sigma^{2})$. Further assume that $n^{-1}\sum_{i=1}^{n}x_{ij}^{2}=\zeta_{j}$, for $j=1,2,\cdots,p$, $\tau = n^{-1}$, $\lambda = A_{1}\psi_{n}$, $\rho_{1}=A_{2}\psi_{n}$, $\rho_{2}=A_{3}\psi_{n}$ with $A_{1}$, $A_{2}$ and $A_{3}$ being some positive constants, and
\begin{eqnarray}
\label{theorem2_01}
\psi_{n} = 2\sigma\sqrt{2n\max_{j}\zeta_{j}\bigg[\log\bigg(\frac{m}{\alpha}\bigg) + \log\overline{q}\bigg]},
\end{eqnarray}
where $\alpha$ is a non-negative constant and $\overline{q}=m^{-1}\sum_{k=1}^{m}q_{k}$. Then given that Assumption 1 holds, for $1\leq\max(q_{R},\widehat{s})\leq p$, with $1-\alpha$ probability, we have
\begin{eqnarray}
\label{theorem2_02}
||\widehat{\beta}_{\emph{gvsnss}}-\beta||_{2}&\leq&\frac{2\sqrt{2}\sigma\Lambda_{n}\max_{j}\zeta_{j}^{1/2}}{\sqrt{n}(\kappa_{n}+\Omega_{n})}\sqrt{q_{R}\bigg[\log\bigg(\frac{m}{\alpha}\bigg) + \log\overline{q}\bigg]},\nonumber\\
\end{eqnarray}
as $n\rightarrow\infty$, where 
\begin{eqnarray}
\label{theorem2_03}
\Lambda_{n}&=&\bigg\{2\bigg[1+\bigg(\frac{\widehat{s}}{4s}\bigg)^{1/2}\bigg] + 2\max_{j\in S}|\beta_{j}|A_{1}\nonumber\\
& & + (A_{2}+A_{3})\bigg[\frac{2c_{2}^{-1}+1}{\log(n)}+c_{3}^{-1}\bigg]\bigg\},\\
\label{theorem2_04}
\Omega_{n}&=&2A_{1}\sigma\sqrt{\frac{2\max_{j}\zeta_{j}}{n}\bigg[\log\bigg(\frac{m}{\alpha}\bigg) + \log\overline{q}\bigg]},
\end{eqnarray}
where $\widehat{s}=|\widehat{S}|$, $c_{2}$ and $c_{3}$ are defined in Theorem \ref{theorem1_01}. 
\end{theorem}
The deterministic result stated in Theorem \ref{theorem_deterministic_bound} will serve as a bone for deriving upper bound (\ref{theorem2_02}). Note that since we have assumed $\tau = n^{-1}$, therefore effectively we have $\widehat{\beta}^{\tau}\rightarrow\widehat{\beta}_{\text{gvsnss}}$ and $\widehat{S}^{\tau}\rightarrow \widehat{S}$ as $n\rightarrow\infty$. Detailed derivations of Theorem \ref{theorem_deterministic_bound} and Theorem \ref{theorem_l2_error_bound} are given in Appendix \ref{appendix_a}. 

Note that the bound (\ref{theorem2_02}) is proportional to $q_{R}^{1/2}$ and by definition
\begin{eqnarray}
\label{def_qR}
q_{R}=s + \sum_{k\in R}\#\{j\in G_{k}:\beta_{j}=0\}.\nonumber
\end{eqnarray}
Given that $s$ is fixed, the result implies that, if groups that contain the true covariates also contain large numbers of redundant covariates, or if the true covariates are scattered over a large number of groups, like the scenarios with high group switching probabilities we have seen in Section \ref{simulation_study}, then the gvsnss estimator will not perform well. 

Now if we adopt an equal group setting, i.e. $q_{1}=q_{2},\cdots,=q_{m}$, and let $\zeta_{j}=1$ for $j=1,2,\cdots,p$, then $q_{R}=|G_{R}|=|R|\times |G_{k}|=rq_{1}$, and the right hand side of (\ref{theorem2_02}) will have an order of magnitude equal to $n^{-1/2}\sqrt{r\log q_{1} + r\log(m/\alpha)}$. Further note that $\log q_{1}\leq q_{1}$. Therefore with $1-\alpha$ probability, as $n\rightarrow\infty$, we have $||\widehat{\beta}_{\text{gvsnss}}-\beta||_{2}=O(n^{-1/2}\sqrt{s_{0}+r_{0}\log(m/\alpha)})$, where $r_{0}=r$ and $s_{0}=q_{R}$. The result given above implies that the gvsnss estimator can achieve an $l_{2}$ estimation error with an order of magnitude proportional to that of the group lasso established in \cite{huangandzhang10}. 

The following corollary states that if the maximum size of groups is equal to one, then the gvsnss estimator can have an $l_{2}$ estimation error with an order of magnitude similar to that of the lasso established in \cite{meinshausenandyu09,bickeletal09}.  

\begin{corollary}
\label{corollary_l2_error_bound_lasso}
Assume that $\max_{k}q_{k}=1$ and $\zeta_{j}=1$ for $j=1,2,\cdots,p$. Then given that all assumptions stated in Theorem \ref{theorem_l2_error_bound} hold, with $1-\alpha$ probability, we have
\begin{eqnarray}
\label{corollary1_01}
||\widehat{\beta}_{\emph{gvsnss}}-\beta||_{2}&\leq&\frac{2\sqrt{2}\sigma\Lambda_{n}}{\sqrt{n}(\kappa_{n}+\Omega_{n})}\sqrt{s\log\bigg(\frac{p}{\alpha}\bigg)}
\end{eqnarray}
as $n\rightarrow\infty$, where $\Lambda_{n}$ is the same as the one defined in (\ref{theorem2_03}) and
\begin{eqnarray}
\label{corollart1_02}
\Omega_{n}&=&2A_{1}\sigma\sqrt{\frac{2}{n}\log\bigg(\frac{p}{\alpha}\bigg)}.\nonumber
\end{eqnarray}
\end{corollary}
\emph{Proof of Corollary \ref{corollary_l2_error_bound_lasso}.} Obviously given that the maximum group size is one, $q_{R}=s$. In addition, the number of groups is $m=p$. Then by inserting the results given above into the right hand side of (\ref{theorem2_02}), we obtain (\ref{corollary1_01}), which completes the proof. \qed    
\subsection{Label-invariance property}
Here we show that the gvsnss estimator (\ref{gvsnss_estimator}) is asymptotically invariant to group structures. We consider two collections of index sets $G^{*}=\{G_{k}^{*}:k=1,2,\cdots,m^{*}\}$ and $G^{**}=\{G_{l}^{**}:l=1,2,\cdots,m^{**}\}$. In the following discussion as well as in the proof we will see $*$ and $**$ attached to various vector-valued quantities and the presence of $*$ (or $**$) in a given vector means that the entries of the vector are indexed by $G_{k}^{*}$ (or $G_{k}^{**}$) in the original vector. 

Our result relies on the fact that the third term in $V_{\tau}(0,\beta^{\prime},G^{\prime})$ allows the gvsnss estimation to produce zero estimates for coefficients whose covariates are in active groups. Without this setting, we would be unable to establish the label-invariance property for some cases, and $\widehat{\beta}_{\text{gvsnss}}^{*}=\arg\min_{\beta^{\prime}}\lim_{\tau\rightarrow 0}V_{\tau}(0,\beta^{\prime},G^{*})$ might never be a solution to the subgradient equations of $\lim_{\tau\rightarrow 0}V_{\tau}(0,\beta^{\prime},G^{**})$, where $G^{**}$ is an arbitrary collection of index sets. Therefore we assume $\rho_{1} > 0$. In addition, our result relies on evaluating the difference between the log-sum penalties involving $l_{2}$-norms in $V_{\tau}(0,\beta^{\prime},G^{*})$ and $V_{\tau}(0,\beta^{\prime},G^{**})$. Since $\rho_{2}$ and the size of a group play a crucial role in the evaluation process, we will also impose an assumption on their orders of magnitude.
\begin{theorem}
\label{theorem_label_inv_property}
Assume that
\begin{eqnarray}
\label{linv01}
\widehat{\beta}^{\tau*}=\arg\min_{\beta^{\prime}} V_{\tau}(0,\beta^{\prime}, G^{*})\nonumber
\end{eqnarray}
is the unique solution to the subgradient equations of $V_{\tau}(0,\beta^{\prime},G^{*})$ for all $\tau\in[0,1)$. Further assume that $\rho_{1}>0$, $\rho_{2}\max_{k}\sqrt{q_{k}}=o(\log n)$, and $\tau= n^{-1}$. Then as $n\rightarrow\infty$, $\widehat{\beta}_{\emph{gvsnss}}^{*}=\arg\min_{\beta^{\prime}}\lim_{\tau\rightarrow 0}V_{\tau}(0,\beta^{\prime},G^{*})$ is the minimizer of $\lim_{\tau\rightarrow 0}V_{\tau}(0,\beta^{\prime},G^{**})$, where $G^{**}$ is an arbitrary collection of index sets.
\end{theorem}
\subsection{Variable selection and sign consistency}
Here we study asymptotic behavior of the gvsnss estimator in variable selection. In particular, we focus on sign consistency of the estimated coefficients. We explain the idea of sign consistency first. An estimator $\widehat{\beta}(n)$ is said to be sign consistent in estimating $\beta$ if probability $\mathbb{P}\{\text{sign}(\widehat{\beta}(n))=\text{ sign}(\beta)\}$ approaches to one as $n\rightarrow\infty$. Given the sign consistency holds, the estimated index set $\widehat{S}(n)=\{j:\widehat{\beta}_{j}(n)\neq 0\}$ will be the same as the true index set $S$, therefore the sign consistency implies variable selection consistency, that is, asymptotically with probability one, non-zero valued coefficients will have non-zero estimated values, and zero-valued coefficients will be estimated with zero values. 

Below we derive a lower bound for $\mathbb{P}\{\text{sign}(\widehat{\beta}^{\tau})=\text{ sign}(\beta)\}$. Then with $\tau=n^{-1}$, we have $\widehat{\beta}^{\tau}\rightarrow\widehat{\beta}_{\text{gvsnss}}$ as $n\rightarrow\infty$, and in turn, the lower bound for $\mathbb{P}\{\text{sign}(\widehat{\beta}_{\text{gvsnss}})=\text{ sign}(\beta)\}$ can be established asymptotically. The following assumptions on eigenvalues of matrices are useful in deriving the lower bound.

\textbf{Assumption 2.} Define $C_{SS}=n^{-1}(X_{S}^{T}X_{S}+\lambda I_{s\times s})$. Define $\kappa_{\min}=\min_{w} wC_{SS}w$. We assume $0<\kappa_{\min}<\infty$ as $n\rightarrow \infty$.

\textbf{Assumption 3.} Define $\varsigma_{\max}=\max_{w} n^{-1}wX_{S}X_{S}^{T}w$. We assume $0< \varsigma_{\max} <\infty$ as $n\rightarrow\infty$.

\textbf{Assumption 4.} Define $\nu_{\max,k}=\max_{w}n^{-1}wX_{G_{k}}X_{G_{k}}^{T}w$ and $\nu_{\max}= \max_{k}\nu_{\max,k}$. For $k=1,2,\cdots,m$, we assume $0< \nu_{\max,k}<\infty$ as $n\rightarrow\infty$. 

\begin{theorem}
\label{theorem_sign_consistency}
Assume that $\epsilon_{i}$'s are i.i.d. as Normal$(0,\sigma^{2})$. Further assume that $n^{-1}\sum_{i=1}^{n}x_{ij}^{2}=1$ for $j=1,2,\cdots,p$, $\tau = n^{-1}$, $\lambda=O(n^{1/2})$, $\rho_{1}= O(n^{1/2})$, $\rho_{2}=O(n^{1/2})$, and $p=o(n(\log(n+1))^{-2})$. Then given that Assumptions 2, 3 and 4 hold, the probability $\mathbb{P}\big\{\emph{sign}(\widehat{\beta}^{\tau})=\emph{sign}(\beta)\big\}$ can be bounded from below in a way such that
\begin{eqnarray}
\label{sign01}
& &\mathbb{P}\big\{\emph{sign}(\widehat{\beta}^{\tau})=\emph{sign}(\beta)\big\}\nonumber\\
& &\geq 1-\exp\bigg\{-n\bigg(\frac{\psi_{1,n}^{2}\kappa_{\min}^{2}}{2\sigma^{2}}-\frac{\log s}{n}\bigg)\bigg\}\nonumber\\
& &-\exp\bigg\{-n\bigg[\frac{\psi_{2,n}^{2}\kappa_{\min}^{2}}{8n(\varsigma_{\max} + \kappa_{\min})^{2}\sigma^{2}}-\frac{\log s_{1}^{c}}{n}\bigg]\bigg\}\nonumber\\
& &-\exp\bigg\{-n\bigg[\frac{\kappa_{\min}^{2}\psi_{3,n}^{2}}{16n^{2}\nu_{\max}(\varsigma_{\max}+\kappa_{\min})^{2}\sigma^{2}}-0.35 - \frac{\log r^{c}}{n}\bigg]\bigg\},
\end{eqnarray}
where $s_{1}^{c}=|S_{1}^{c}|$ with $S_{1}^{c}= S^{c}\cap G_{R}$, $r^{c} = |R^{c}|$, $\psi_{1,n}$, $\psi_{2,n}$ and $\psi_{3,n}$ are non-negative constants and as $n\rightarrow\infty$, $\psi_{1,n}=O(1)$, $\psi_{2,n}=O(n^{3/2}(\log n)^{-1})$ and $\psi_{3,n}=O(n^{3/2}(\log n)^{-1})$.
\end{theorem}
The proof can be found in Appendix \ref{appendix_c}. The proof will start by exploring the KKT conditions associated to the minimization problem involving objective function (\ref{grand_obj}). Note that in Theorem \ref{theorem_sign_consistency} we do not assume that the irrepresentable-type conditions \cite{zhaoandyu06} should hold.

\begin{corollary}
\label{sign_consistency_for_gvsnss}
Assume that all assumptions and results stated in Theorem \ref{theorem_sign_consistency} hold. Then  
\begin{eqnarray}
\label{sign02}
\mathbb{P}\big\{\emph{sign}(\widehat{\beta}_{\emph{gvsnss}})=\emph{sign}(\beta)\big\}\rightarrow 1\nonumber
\end{eqnarray}
as $n\rightarrow\infty$.
\end{corollary}

\emph{Proof of Corollary \ref{sign_consistency_for_gvsnss}.} Note that $s\leq p = o(n(\log(1+n))^{-2})$, therefore $n^{-1}\log s\rightarrow 0$ as $n\rightarrow\infty$. In addition, $\psi_{1,n}=O(1)$, therefore $(2\sigma^{2})^{-1}\psi_{1,n}^{2}\kappa_{\min}^{2} > 0$. Then as $n\rightarrow\infty$, the first exponential term in (\ref{sign01}) will approach to zero. For the second exponential term in (\ref{sign01}), since $\psi_{2,n}=O(n^{3/2}(\log(n))^{-1})$, therefore we have $n^{-1}\psi_{2,n}^{2} =O(n^{2}(\log(n))^{-2})\rightarrow \infty$ as $n\rightarrow\infty$. In addition, $s_{1}^{c}\leq p = o(n(\log(1+n))^{-2})$, therefore $n^{-1}\log s_{1}^{c}\rightarrow 0$ as $n\rightarrow\infty$. Then as $n\rightarrow\infty$, the second exponential term in (\ref{sign01}) will approach to zero. Furthermore, since $n^{-2}\psi_{3,n}^{2}=O(n(\log n)^{-2})\rightarrow\infty$ and $n^{-1}\log r^{c}\rightarrow 0$ as $n\rightarrow\infty$, therefore the third exponential term in (\ref{sign01}) will approach to zero as $n\rightarrow\infty$. Finally note that since $\tau=n^{-1}$, therefore $\widehat{\beta}^{\tau}\rightarrow\widehat{\beta}_{\text{gvsnss}}$ as $n\rightarrow\infty$. The results given above imply that $\mathbb{P}\big\{\text{sign}(\widehat{\beta}_{\text{gvsnss}})=\text{sign}(\beta)\big\}\rightarrow 1$ as $n\rightarrow\infty$, which completes the proof.

\section{Real data examples}\label{real_data_examples}
\subsection{The U.S. industrial product index}
The data set we consider here contains the monthly-based U.S. industrial production index and 125 macroeconomic variables, spanning from July 1964 to December 2010. The industrial production index is an important indicator for economic policy-making. Our aim here is to predict the growth rate of the industrial production index from the 125 macroeconomic variables. Similar data set was used in \cite{stockandwatson2002,baietal2008,ludvigsonandng2009}. The 125 macroeconomic variables are essentially a subset of the 132 variables used by Bai and Ng \cite{baietal2008}. For the 125 macroeconomic variables, we follow a benchmark categorization to divide them into 8 groups: 1) output and income (OI), 2) labor market (LM), 3) housing (H), 4) consumption, orders and inventories (COI), 5) money and credits (MC), 6) bond and exchange rates (BE), 7) prices (P), 8) stock market (SM).

\begin{figure}[h]
\begin{center}
\includegraphics[scale=0.4]{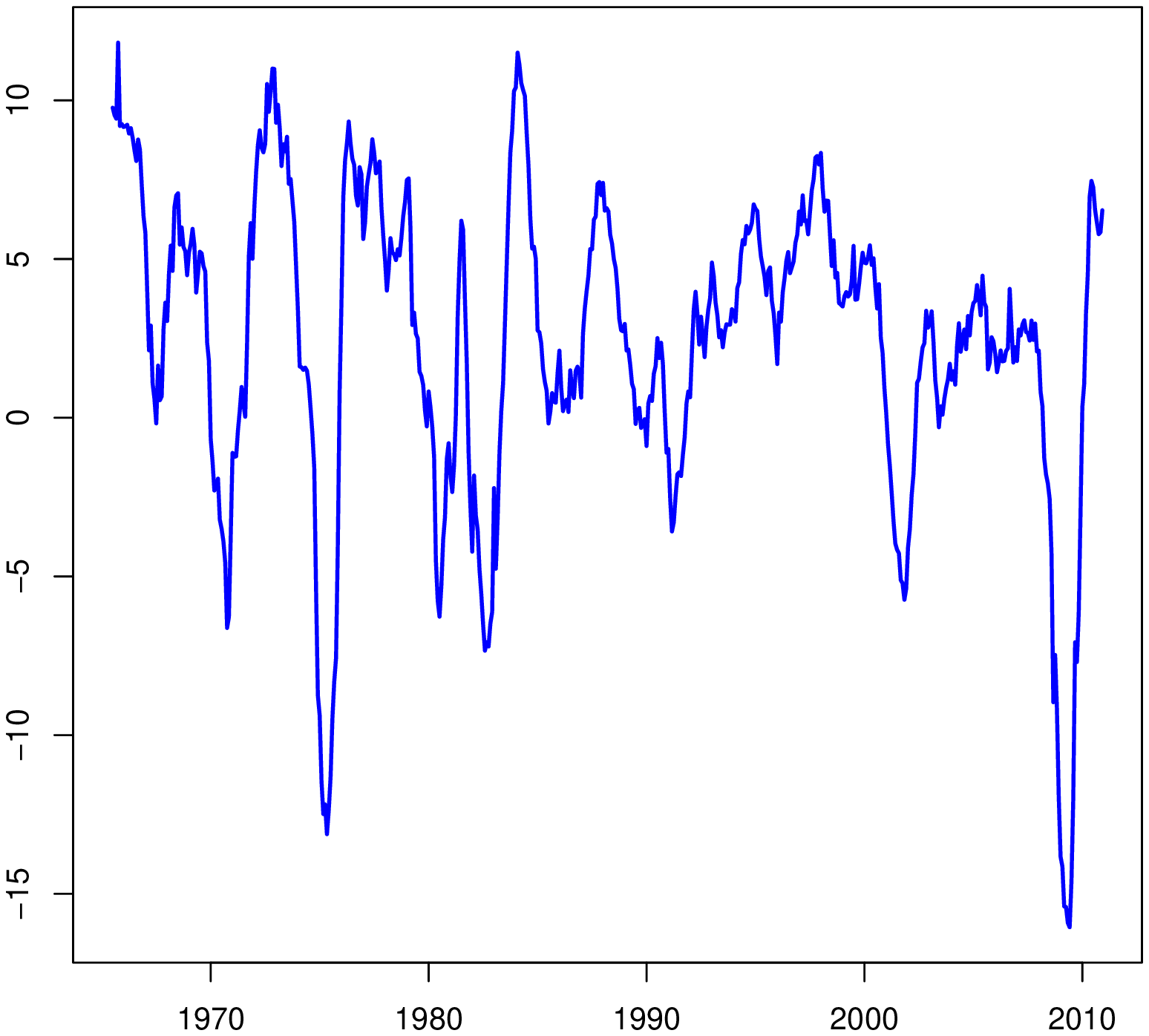}
\includegraphics[scale=0.4]{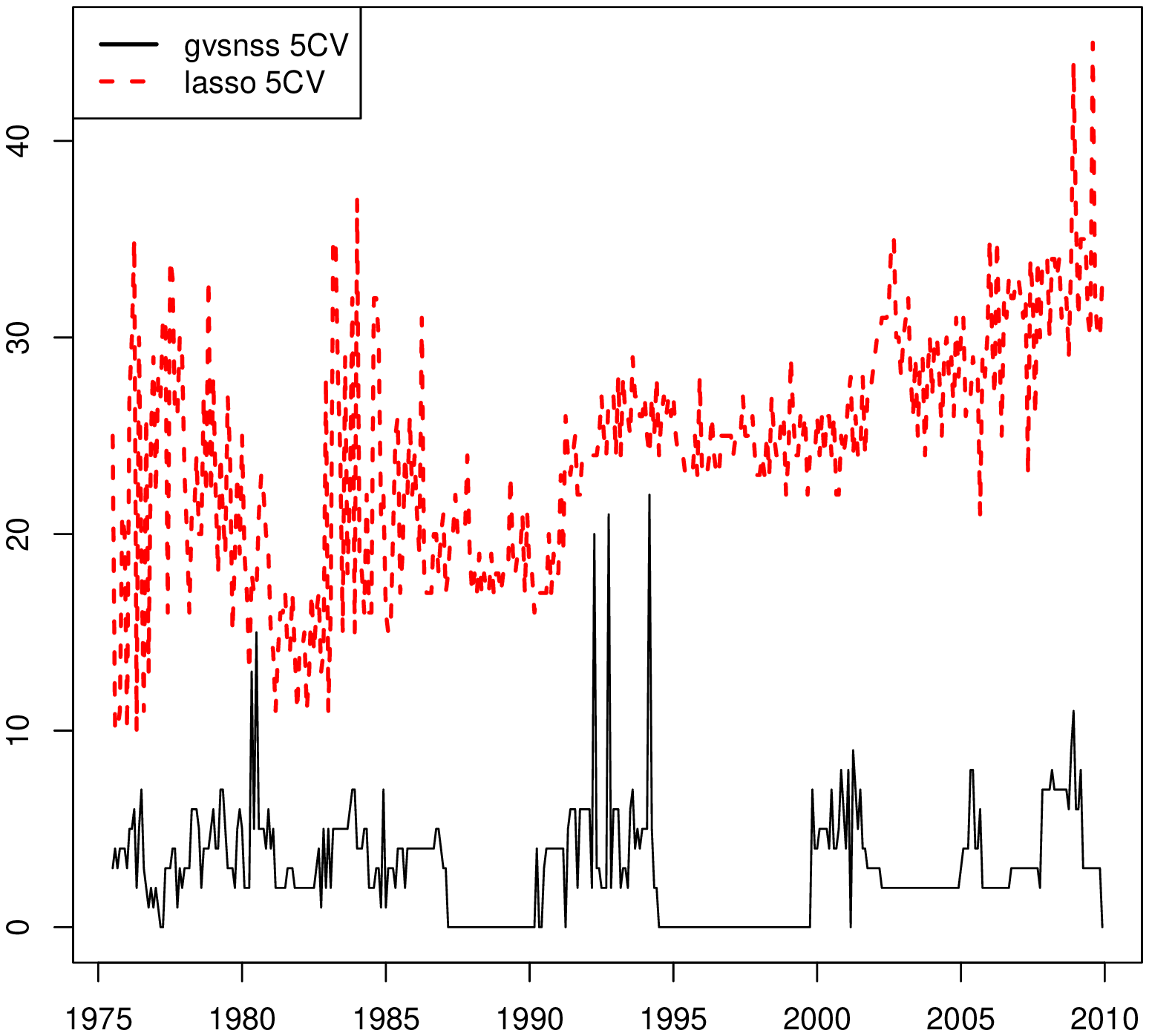}
\includegraphics[scale=0.4]{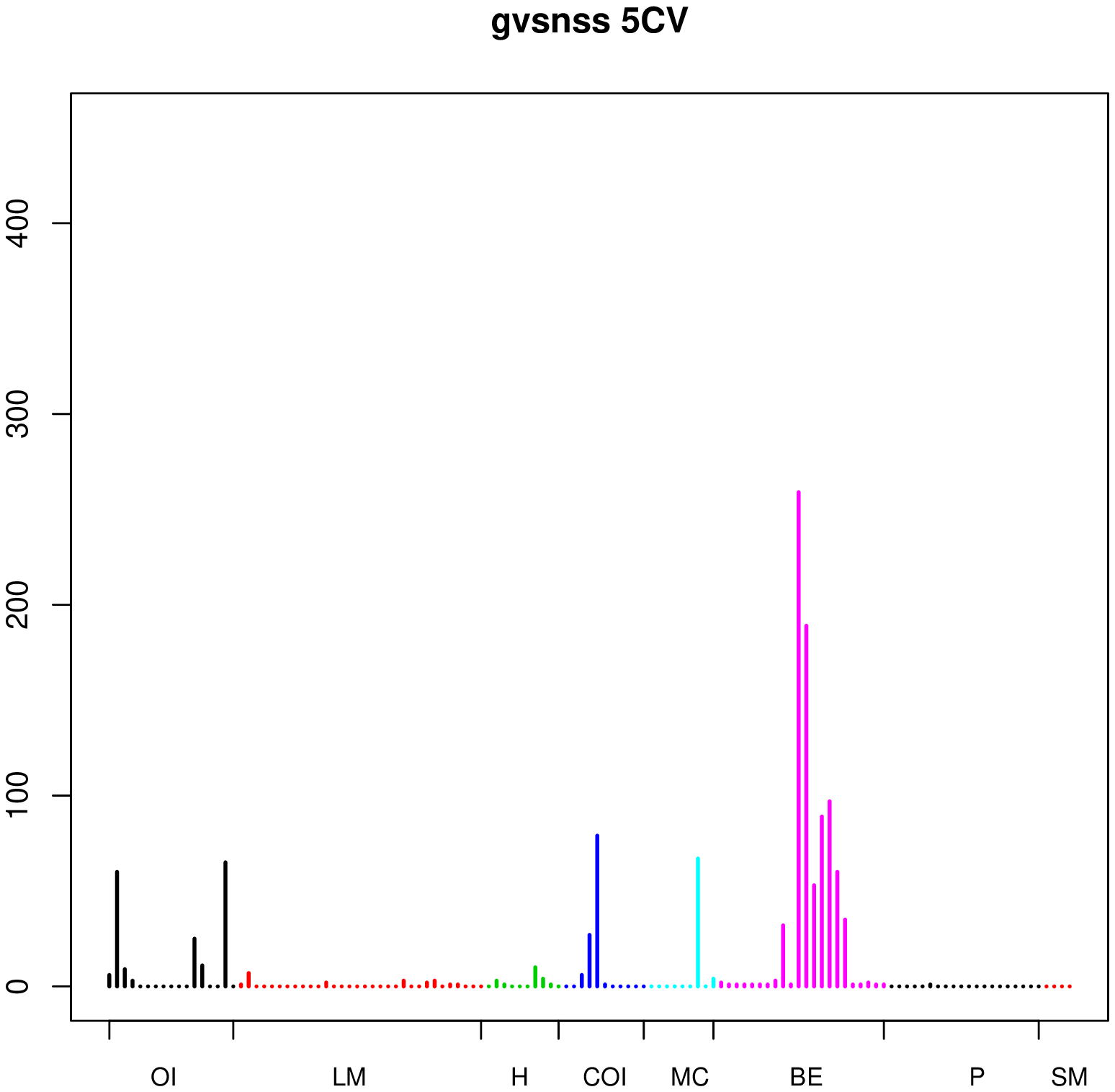}
\includegraphics[scale=0.4]{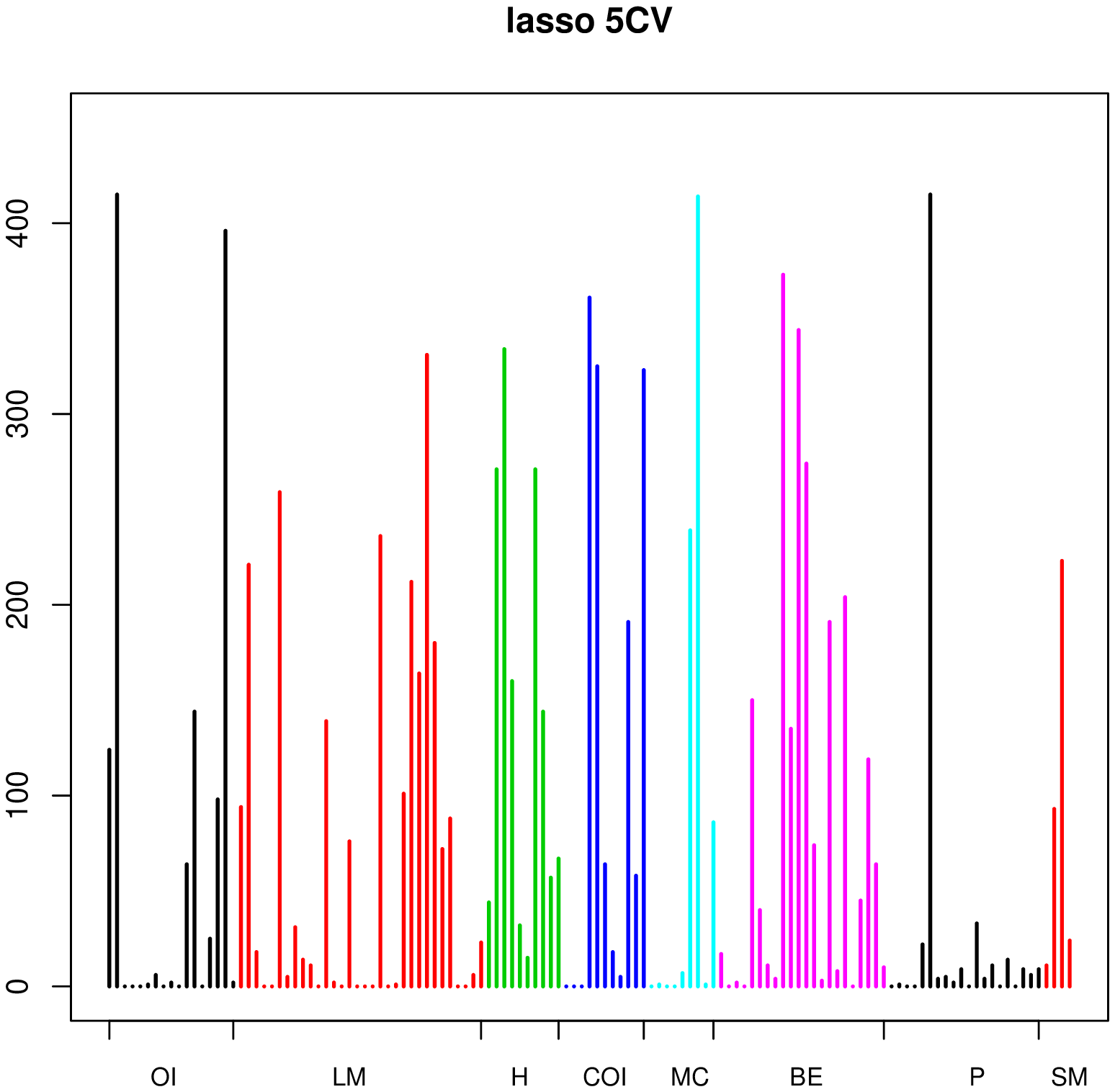} 
\end{center}
\caption{Top Left: Percentage change of the U.S. industrial production index. The change is defined as $100[\log(IP_{t})-\log(IP_{t-12})]$. Top Left: The number of selected variables for the 415 time blocks. Bottom Left: Frequencies of variables being selected under the gvsnss. Bottom Right: Frequencies of variables being selected under the lasso. OI: output and income; LM: labor market; H: housing; COI: consumption, orders and inventories; MC: money and credits; BE: bond and exchange rates; P: prices; SM: stock market.}
\label{figure_real_01}
\end{figure}

Now let $IP_{t}$ denote the level of the industrial production index at time $t$. We define the growth rate at time $t+t^{\prime}$ by $y_{t+t^{\prime}}=(t^{\prime})^{-1}1200[\log(IP_{t+t^{\prime}}) -\log(IP_{t})]$. The plot in the top left panel of Figure \ref{figure_real_01} shows the corresponding time series trend. We further model the growth rate $y_{t+t^{\prime}}$ by 
\begin{eqnarray}
\label{ts_model}
y_{t+t^{\prime}}=\eta_{0} +\sum_{l=0}^{3}z_{t-l}\eta_{l+1} +\sum_{k=1}^{8}\sum_{j\in G_{k}}x_{tj}\beta_{j} +\varepsilon_{t+t^{\prime}},  
\end{eqnarray}
where $z_{t-l}= 1200[\log( IP_{t-l}) -\log(IP_{t-l-1})]$ is the $l$th lag term, $x_{tj}$ is the $j$th macroeconomic variable at time $t$, $G_{k}$ is the index set corresponding to the $k$th macroeconomic group, and $\varepsilon_{t+t^{\prime}}$ is the error term. 

We adopt an expanding window scheme to carry out real time estimation for model (\ref{ts_model}). That is, we estimate parameters $\eta_{l}$'s and $\beta_{j}$'s with information from time 1 to time $t$. Note that in such setting, at time $t$, dependent variable $y_{t^{\prime\prime}+t^{\prime}}$ is only available for $t^{\prime\prime}=1,\ldots,t-t^{\prime}$. Let $\widehat{\eta}_{l}^{1,t-t^{\prime}}$'s and $\widehat{\beta}_{j}^{1,t-t^{\prime}}$'s denote the corresponding estimates. With model (\ref{ts_model}) and the estimates, at time $t$, we predict $y_{t+t^{\prime}}$ by
\begin{eqnarray}
\widehat{y}_{t+t^{\prime}}=\widehat{\eta}_{0}^{1,t-t^{\prime}} +\sum_{l=0}^{3}z_{t-l}\widehat{\eta}_{l+1}^{1,t-t^{\prime}} +\sum_{k=1}^{8}\sum_{j\in G_{k}}x_{tj}\widehat{\beta}_{j}^{1,t-t^{\prime}}.  
\end{eqnarray}       

In practice, we let $t^{\prime}=12$, which corresponds to one year change. The prediction is started from $t=132$ (June 1975) and ended at $t=546$ (December 2009). Under this setting, there are 415 time blocks. For each time block, we applied two methods to estimate parameters in model (\ref{ts_model}). The first method used the gvsnss to select the 125 macroeconomic variables and then re-estimate regression coefficients of the selected variables with the ordinary least squares method. For the gvsnss estimation, we used five fold cross validation to select the tuning parameter. The second method is similar to the first one but using the lasso for variable selection. For the lasso estimation, we also used five fold cross validation to select the tuning parameter.

In addition, we also used principal components (PCs) of the selected variables to construct models for prediction. For simplicity, we use the first four PCs for the prediction. If the number of selected variables is less than four, we use the selected variables as the predictors. 

The plot in the top right panel of Figure \ref{figure_real_01} shows the number of selected variables for the 415 time blocks while plots in the bottom panel of Figure \ref{figure_real_01} show frequencies of selected variables for each macroeconomic group under the gvsnss and the lasso, respectively. The results show that the gvsnss estimation selected less variables and produced stronger between-group-sparsity and within-group-sparsity than the lasso.

\begin{figure}[h]
\begin{center}
\includegraphics[scale=0.4]{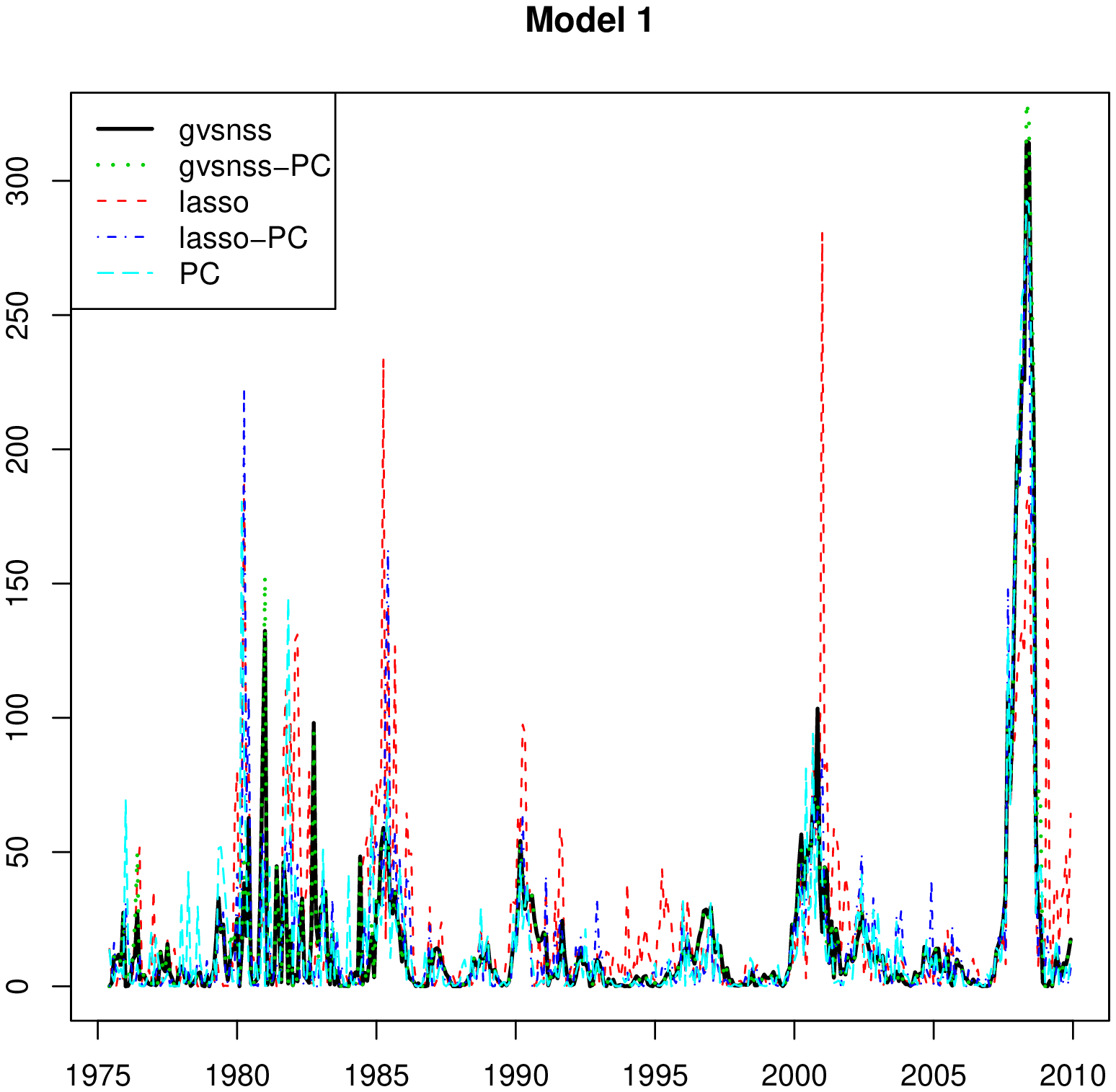}
\includegraphics[scale=0.4]{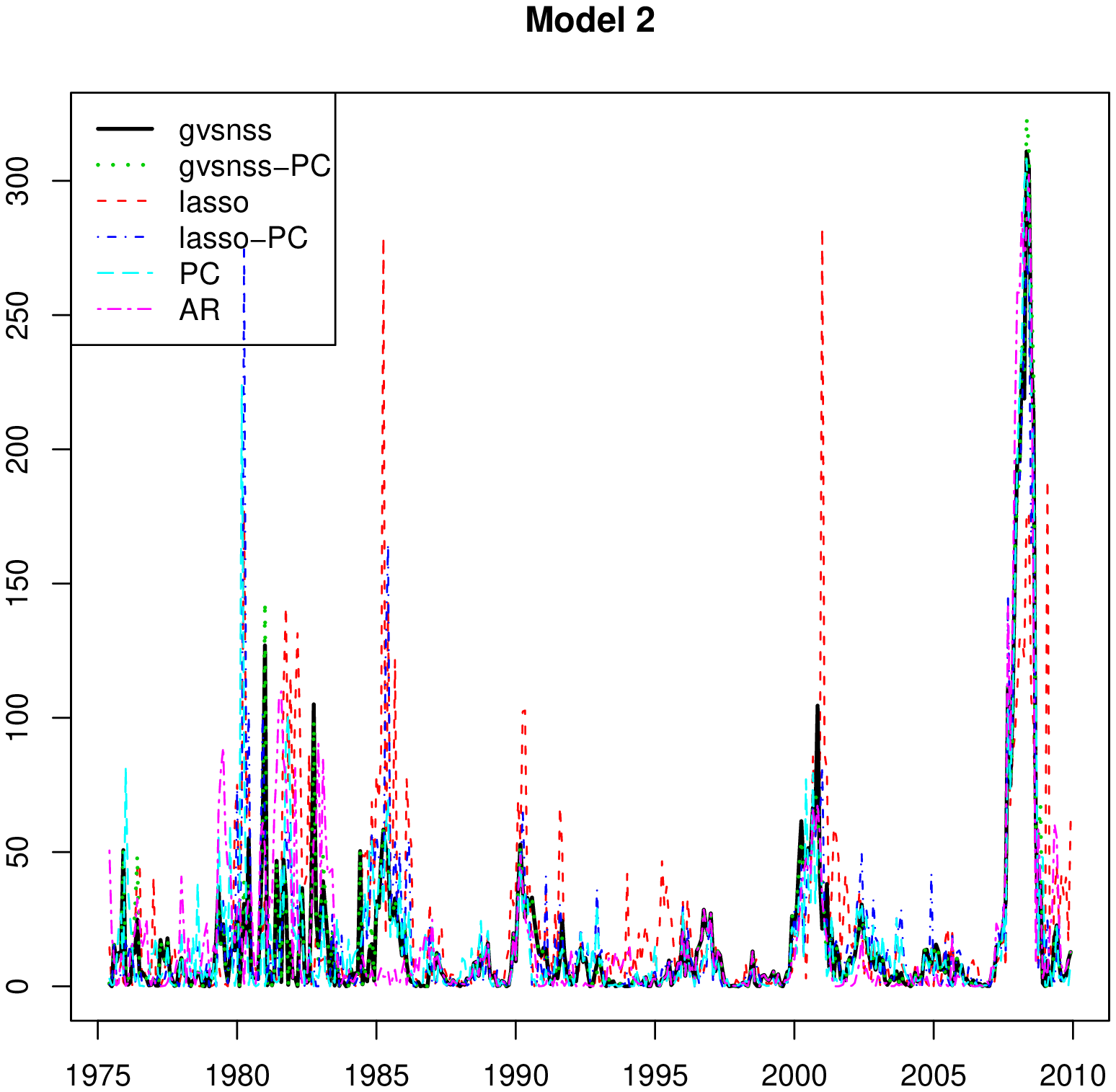} 
\end{center}
\caption{Left: The out-of-sample squared error of Model 1 for the 415 time blocks. Right: The out-of-sample squared error of Model 2 for the 415 time blocks.}
\label{figure_real_02}
\end{figure}

In addition, we also reported the out-of-sample mean squared error under the two estimation methods. The out-of-sample mean squared error is defined as
\begin{eqnarray}
MSE_{OS}^{t^{\prime}}=\frac{1}{T-t^{\prime}}\sum_{t=1}^{T-t^{\prime}}(y_{t+t^{\prime}}-\widehat{y}_{t+t^{\prime}})^{2}.
\end{eqnarray}
The results are shown in Table \ref{table2} and Figure \ref{figure_real_02}, where Model 1 is the model without the lag terms, Model 2 is the model with the lag terms, PC is the model using the first four PCs of all macroeconomic variables, and AR is the model with the lag terms but without the grouped variable terms. The results suggest that including the macroeconomic variables can slightly improve the prediction results.
\subsection{Retirement plan data}
The data set, adopted from \cite{bryantandsmith1995,ruppertetal2003}, contains information about employee retirement plans of 92 firms. The retirement plans are managed by a company called Best Retirement Inc. (BRI). The response variable is the contribution to retirement plan at the end of the first year. It is measured at the logarithm scale. Let $y_{i}$ denote the response variable corresponding to the $i$th retirement plan. Our aim here is to help the company to assess whether the presence of a specially trained sales, named Susan Shepard, has a positive effect on $y_{i}$. For the $i$th retirement plan, we define $x_{i9}=1$ if Susan Shepard is present and $x_{i9}=0$ otherwise. The data set also contains eight other variables. To fully assess the presence of Susan Shepard on $y_{i}$, we will consider interactions between $x_{i9}$ and the eight variables in the regression model. We call the collection of $x_{i9}$ and the interaction terms the "Susan Shepard Effect" group. Let $G_{\text{SSE}}$ denote the set that contains indices of covariates in the Susan Shepard Effect group. We will jointly estimate regression coefficients of the covariates with indices in $G_{\text{SSE}}$. After some calculations, we excluded one interaction variable that has the same value for all retirement plans. The set $G_{\text{SSE}}$ therefore only contains indices of eight variables. 

\begin{figure}[h]
\begin{center}
\includegraphics[width=0.32\textwidth]{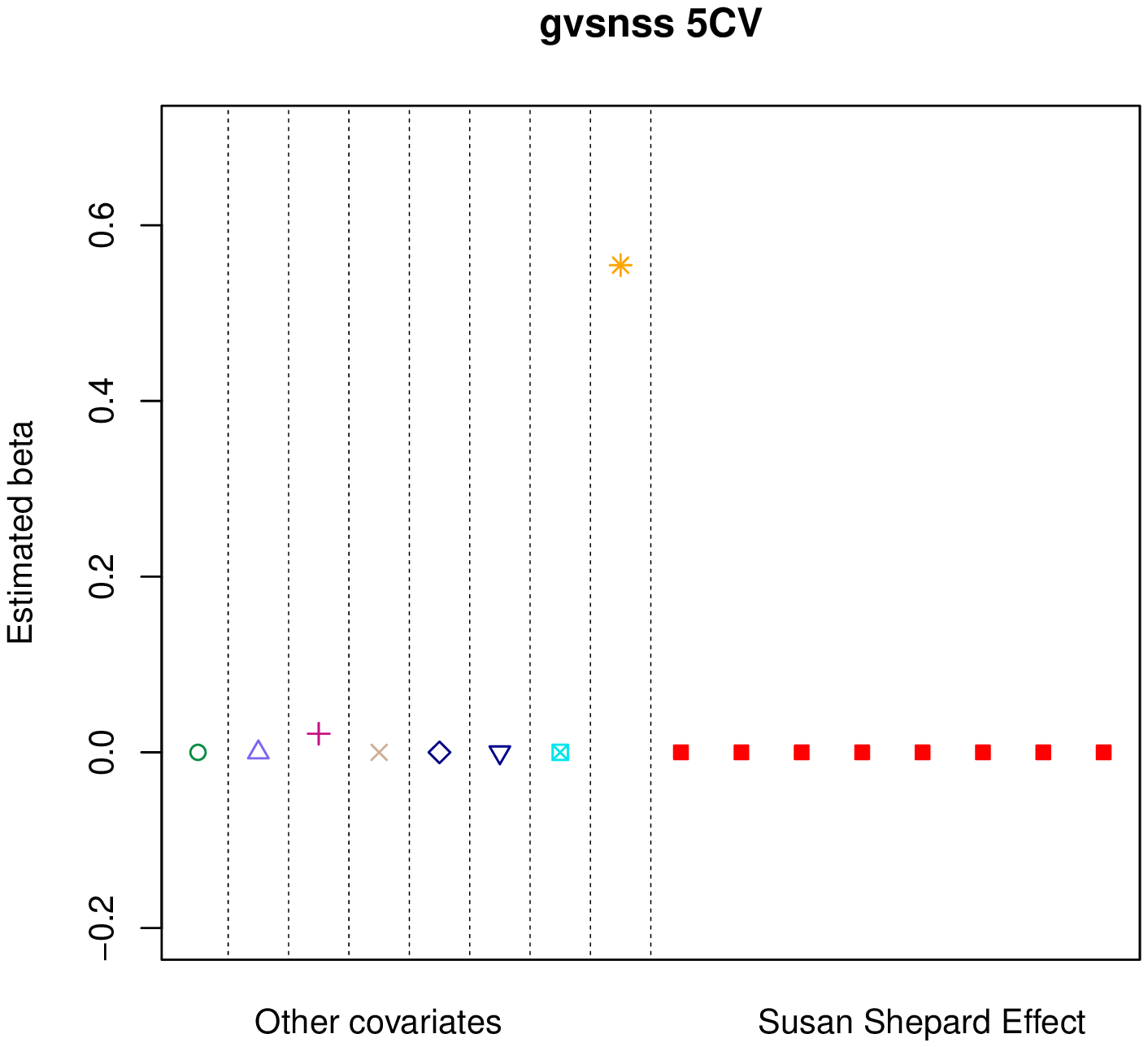}
\includegraphics[width=0.32\textwidth]{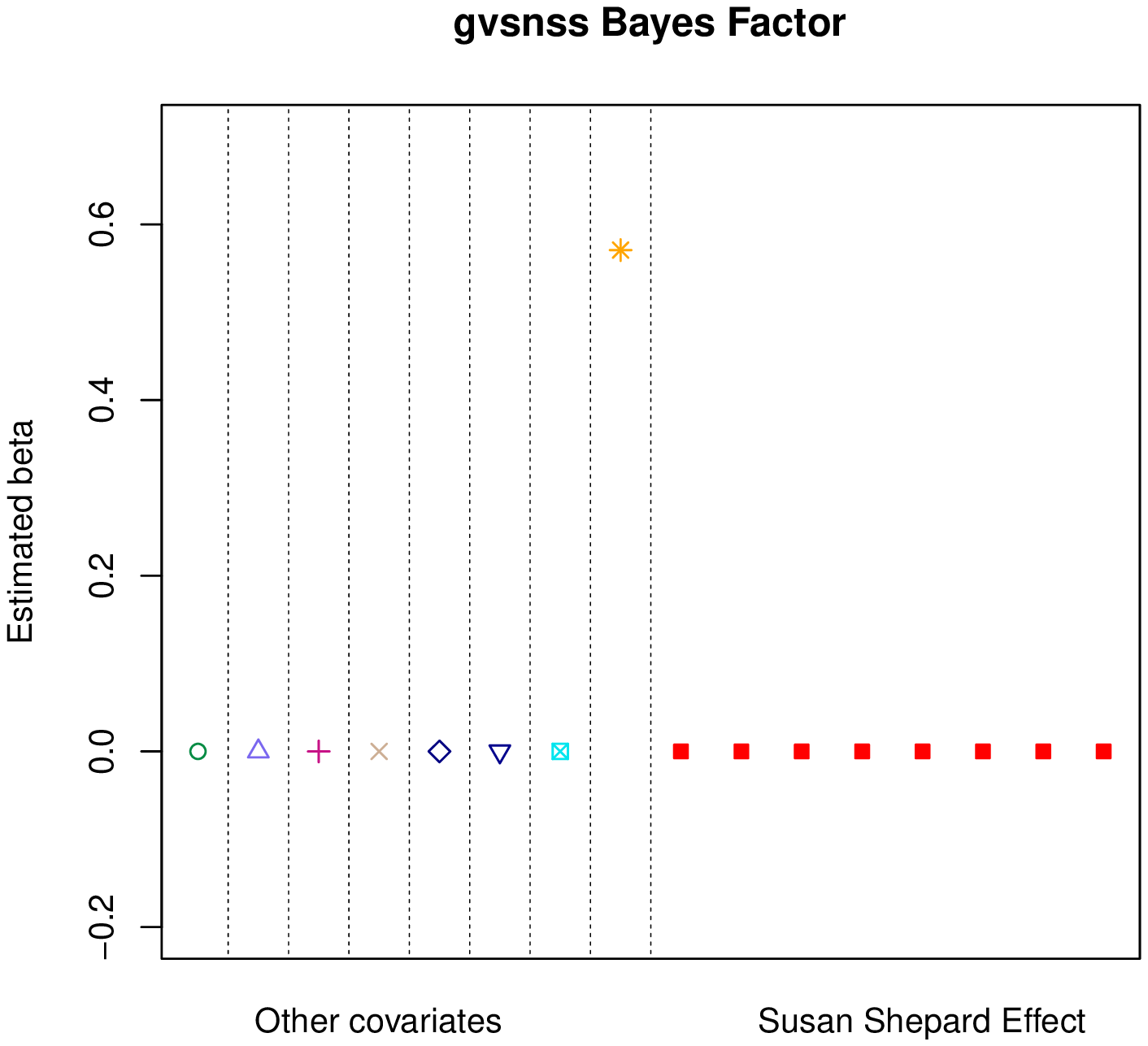}
\includegraphics[width=0.32\textwidth]{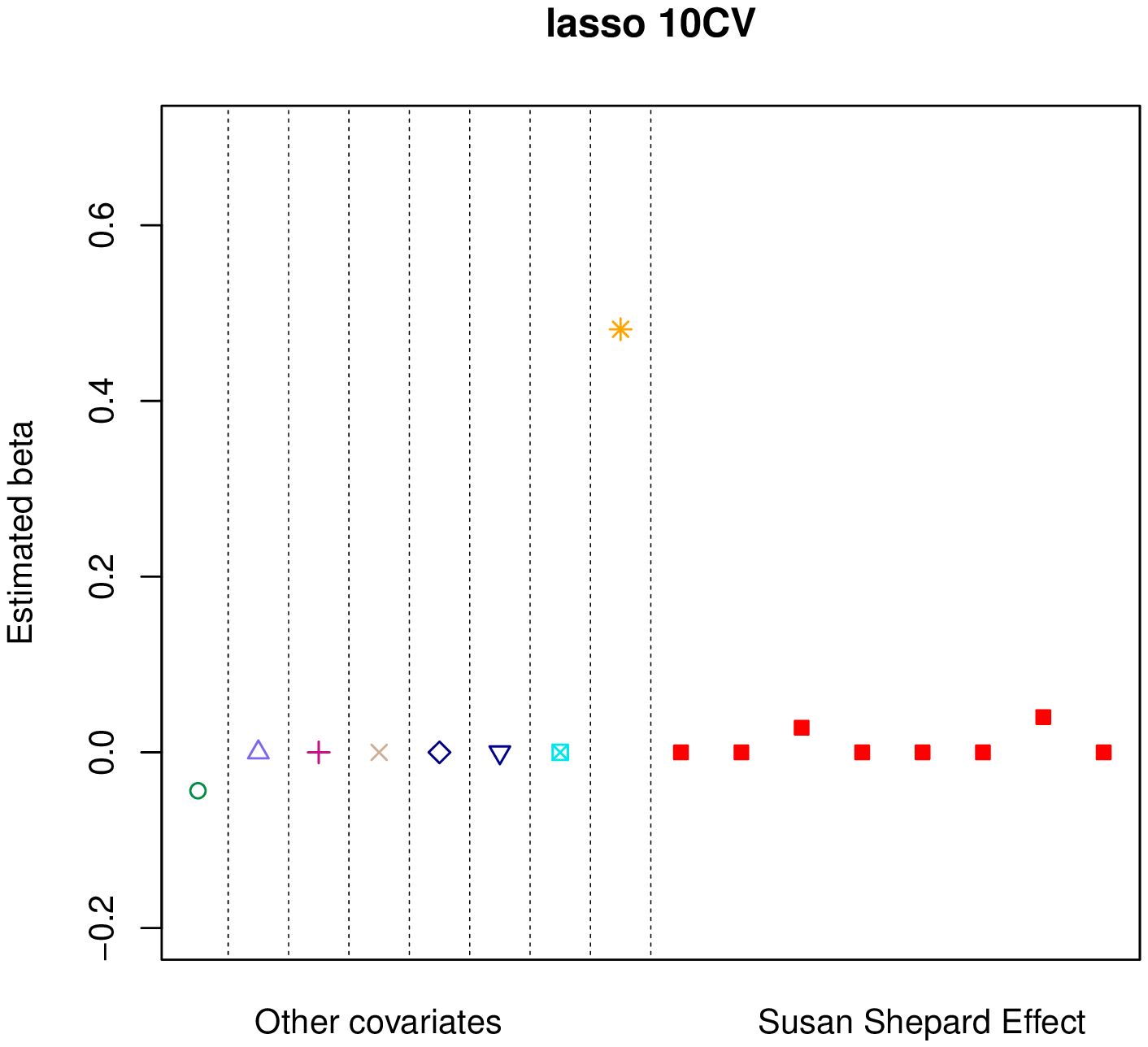} 
\end{center}
\caption{Estimation results from the retirement plan data. Left: The gvsnss estimation with five fold cross validation. Middle: The gvsnss estimation with the Bayes factor. Right: The lasso estimation with ten fold cross validation.}
\label{figure_real_03}
\end{figure}

We model the expectation of the response variable $\mu_{i}=\mathbb{E}(y_{i}|\ \beta, x_{i})$ by
\begin{eqnarray}
\label{real_example02_01}
\mu_{i} = \sum_{j=1}^{8}x_{ij}\beta_{j} + \sum_{j\in G_{\text{SSE}}}x_{ij}\beta_{j}.
\end{eqnarray}
We applied three methods, the gvsnss with five fold cross validation, the gvsnss with the Bayes factor, and the lasso with ten fold cross validation to estimate parameters in model (\ref{real_example02_01}). To carry out the parameter estimations, each column of design matrix $X$ was standardized to have mean zero and variance one. The results are shown in Figure \ref{figure_real_03}. The estimation results under the lasso suggest that covariates in the Susan Shepard Effect group do have positive effects on the response variable while the results under the two gvsnss estimations imply that covariates in the Susan Shepard Effect group do not have such effects. 

We also carried out 100 sub-sampling estimations for the model. At each sub-sampling instance, we randomly split two thirds of the data into the training set and one third of the data into the test set. We used data from the training set to estimate parameters in model (\ref{real_example02_01}) and data from the test set to compute the predictive mean squared error. We also computed the number of covariates with non-zero estimated coefficients and the number of covariates with positive estimated coefficients in the Susan Shepard Effect group. The results are shown in Table \ref{table_real_02}.
\section{Discussion}\label{discussion}
We have proposed a specified prior, called the nested spike and slab prior, to model collective behavior of regression coefficients in grouped variable selection. We have developed numerical procedures for solving the optimization problem related to maximum a posteriori estimation for the model. Simulation studies showed that the proposed estimator performs relatively well in variable selection when within-group-sparsity is present. However, we have found the proposed estimator will loss its advantage in parameter estimation if groups that contain the true covariates also contain too many redundant covariates. Subsequent asymptotic analysis also confirmed our findings. 

With suitable modifications, the nested spike and slab prior can be extended to tackle grouped variable selection problems in the generalized linear models, time series models such as autoregressive and moving average models, or graphical models in covariance matrix estimation. 
\section*{Acknowledgments}
Tso-Jung Yen is supported by grants NSC 97-3112-B-001-020 and NSC 98-3112-B-001-027 in the National Research Program
for Genomic Medicine and Academia Sinica grant AS-100-TP2-C01. Yu-Min Yen would like to thank Professor Oliver Linton for his encouragement and helpful suggestions.

\clearpage
\appendix
\section{Proof of Theorems \ref{theorem_deterministic_bound} and \ref{theorem_l2_error_bound}}\label{appendix_a}
\emph{Proof of Theorem \ref{theorem_deterministic_bound}.} Now define $w = \widehat{\beta}^{\tau}-\beta$. It can be shown that $w$ is the minimizer of the objective function $V_{\tau}(w^{*},\beta, G)$ defined in (\ref{grand_obj}) with respect to $w^{*}$. Therefore $V_{\tau}(w,\beta,G)\leq V_{\tau}(0,\beta,G)$. Here $V_{\tau}(0,\beta,G)$ can be explicitly expressed as 
\begin{eqnarray}
V_{\tau}(0,\beta,G)&=&||\epsilon||_{2}^{2} + \lambda||\beta||_{2}^{2}\nonumber\\
& & + \rho_{1}\sum_{j=1}^{p}\frac{\log(1+\tau^{-1}|\beta_{j}|)}{\log(1+\tau^{-1})}+\rho_{2}\sum_{k=1}^{m}\sqrt{q_{k}}\frac{\log(1+\tau^{-1}||\beta_{G_{k}}||_{2})}{\log(1+\tau^{-1})}.\nonumber
\end{eqnarray}
where $\epsilon = y- X\beta$. Further note that
\begin{eqnarray}
& &\big|\big|\epsilon-Xw\big|\big|_{2}^{2} +\lambda ||w+\beta||_{2}^{2}\nonumber\\
&=&\epsilon^{T}\epsilon + w^{T}X^{T}Xw - 2w^{T}X^{T}\epsilon+\lambda (w^{T}w + 2w^{T}\beta + \beta^{T}\beta)\nonumber\\
&=&||\epsilon||_{2}^{2} + w^{T}(X^{T}X+\lambda)w - 2w^{T}(X^{T}\epsilon - \lambda\beta)+\lambda||\beta||_{2}^{2}.\nonumber
\end{eqnarray}
With the results given above, we can compute $V_{\tau}(w,\beta,G)-V_{\tau}(0,\beta,G)$. In addition, since $V_{\tau}(w,\beta,G)-V_{\tau}(0,\beta,G)\leq 0$, therefore by rearranging the terms in $V_{\tau}(w,\beta,G)-V_{\tau}(0,\beta,G)$, we obtain 
\begin{eqnarray}
\label{basic_inq01}
& &w^{T}(X^{T}X+\lambda)w\\
\label{basic_inq02}
&\leq&2w^{T}(X^{T}\epsilon - \lambda\beta)\\
\label{basic_inq03}
& &+\rho_{1}\sum_{j=1}^{p}\bigg[\frac{\log(1+\tau^{-1}|\beta_{j}|)}{\log(1+\tau^{-1})}-\frac{\log(1+\tau^{-1}|w_{j}+\beta_{j}|)}{\log(1+\tau^{-1})}\bigg]\\
\label{basic_inq04}
& &+\rho_{2}\sum_{k=1}^{m}\sqrt{q_{k}}\bigg[\frac{\log(1+\tau^{-1}||\beta_{G_{k}}||_{2})}{\log(1+\tau^{-1})}-\frac{\log(1+\tau^{-1}||w_{G_{k}}+\beta_{G_{k}}||_{2})}{\log(1+\tau^{-1})}\bigg].
\end{eqnarray}
Note that by Assumption 1, (\ref{basic_inq01}) can be bounded from below in a way such that
\begin{eqnarray}
\label{assumption3}
w^{T}(X^{T}X+\lambda I)w\geq n(\kappa_{n} +\lambda n^{-1})||w||_{2}^{2}. 
\end{eqnarray}
In the following discussion we derive inequalities to bound (\ref{basic_inq02}), (\ref{basic_inq03}) and (\ref{basic_inq04}). 
\\
\\
\textbf{Deriving an upper bound for (\ref{basic_inq03}).} We first derive an inequality to bound the difference $\sum_{j=1}^{p}[\log(1+\tau^{-1}|\beta_{j}|)-\log(1+\tau^{-1}|w_{j}+\beta_{j}|)$. For $j\in \widehat{S}^{\tau}=\{j:\widehat{\beta}_{j}^{\tau} \neq 0\}$, $|w_{j}+\beta_{j}|=|\widehat{\beta}_{j}^{\tau}| > 0$. Then given that $\tau\in[0,1)$, for $j\in\widehat{S}^{\tau}$, we have  
\begin{eqnarray}
\label{ub_inq03_01}
\log\bigg(\frac{1+\tau^{-1}|\beta_{j}|}{1+\tau^{-1}|w_{j}+\beta_{j}|}\bigg)&=&\log\bigg(1 + \frac{|\beta_{j}|-|w_{j}+\beta_{j}|}{\tau + |w_{j}+\beta_{j}|}\bigg)\nonumber\\
&\leq&\frac{|\beta_{j}|-|w_{j}+\beta_{j}|}{\tau + |w_{j}+\beta_{j}|}\nonumber\nonumber\\
&\leq&\frac{|\beta_{j}|-|w_{j}+\beta_{j}| + |w_{j}|}{\tau + |w_{j}+\beta_{j}|}.
\end{eqnarray}
Now for $j\in \widehat{S}^{\tau}\cap S^{c}$, we have $\beta_{j}=0$, therefore for $j\in \widehat{S}^{\tau}\cap S^{c}$, the right hand side of (\ref{ub_inq03_01}) is zero. For $j\in\widehat{S}^{\tau}\cap S$, note that $|\beta_{j}|-|w_{j}+\beta_{j}| \leq |\beta_{j}-w_{j}-\beta_{j}|=|w_{j}|$. Then with the result given above, we have
\begin{eqnarray}
\label{ub_inq03_02}
\sum_{j\in \widehat{S}^{\tau}}\log\bigg(\frac{1+\tau^{-1}|\beta_{j}|}{1+\tau^{-1}|w_{j}+\beta_{j}|}\bigg)&\leq&\sum_{j\in \widehat{S}^{\tau}\cap S}\frac{|\beta_{j}-w_{j}-\beta_{j}| + |w_{j}|}{\tau + |w_{j}+\beta_{j}|}\nonumber\\
&\leq& 2c_{2}^{-1}\sum_{j\in \widehat{S}^{\tau}\cap S}|w_{j}|\nonumber\\
&\leq& 2c_{2}^{-1}\sum_{j\in S}|w_{j}|\nonumber\\
&\leq&2c_{2}^{-1}s^{1/2}||w||_{2},
\end{eqnarray}
where $c_{2}=\min_{j\in \widehat{S}^{\tau}}|\widehat{\beta}_{j}|$. 

Now consider the summation over indices $j\in(\widehat{S}^{\tau})^{c}$. Note that for $j\in(\widehat{S}^{\tau})^{c}\cap S^{c}$, we have $\widehat{\beta}_{j}^{\tau}=\beta_{j}=0$, therefore the difference $\log(1+\tau^{-1}|\beta_{j}|)-\log(1+\tau^{-1}|w_{j}+\beta_{j}|) = 0$. On the other hand, for $j\in(\widehat{S}^{\tau})^{c}\cap S$, we have $|w_{j}+\beta_{j}|=|\widehat{\beta}_{j}^{\tau}-\beta_{j}+\beta_{j}|=0$ and $|\beta_{j}|=|\widehat{\beta}_{j}^{\tau}-\beta_{j}|=|w_{j}|$. Therefore for $j\in (\widehat{S}^{\tau})^{c}\cap S$, we have
\begin{eqnarray}
\log(1+\tau^{-1}|\beta_{j}|)-\log(1+\tau^{-1}|w_{j}+\beta_{j}|)=\log(\tau + |w_{j}|)+\log(\tau^{-1}),\nonumber
\end{eqnarray}
In addition, for $\tau\in[0,1)$, $\log(\tau + |w_{j}|)\leq \log(1+|w_{j}|)\leq |w_{j}|$. Now with $c_{3}=\min_{j\in S}|\beta_{j}|$, we have $c_{3}\leq\min_{j\in (\widehat{S}^{\tau})^{c}\cap S}|\beta_{j}|=\min_{j\in (\widehat{S}^{\tau})^{c}\cap S}|w_{j}|\leq |w_{j}|$ for any $j\in (\widehat{S}^{\tau})^{c}\cap S$. Therefore with the results given above, we have
\begin{eqnarray}
\label{ub_inq03_03}
& &\sum_{j\in (\widehat{S}^{\tau})^{c}}\log(1 + \tau^{-1}|\beta_{j}|)-\log(1+\tau^{-1}|w_{j}+\beta_{j}|)\nonumber\\
&\leq&\sum_{j\in \widehat{S}^{c}\cap S}|w_{j}|\big[1 + c_{3}^{-1}\log(\tau^{-1})\big]\nonumber\\
&\leq&\big[1+c_{3}^{-1}\log(\tau^{-1})\big]\sum_{j\in S}|w_{j}|\nonumber\\
&\leq&\big[1+c_{3}^{-1}\log(\tau^{-1})\big]s^{1/2}||w||_{2}.
\end{eqnarray}
For $\tau\in [0,1)$, we have $[\log(1+\tau^{-1})]^{-1}\leq[\log(\tau^{-1})]^{-1}$. Now combining results in (\ref{ub_inq03_02}) and (\ref{ub_inq03_03}), we can bound (\ref{basic_inq03}) in a way such that
\begin{eqnarray}
\label{bound_inq03}
& &\rho_{1}\sum_{j=1}^{p}\Bigg[\frac{\log(1+\tau^{-1}|\beta_{j}|)}{\log(1+\tau^{-1})}-\frac{\log(1+\tau^{-1}|w_{j}+\beta_{j}|)}{\log(1+\tau^{-1})}\Bigg]\nonumber\\
&\leq&\frac{\rho_{1}}{\log(\tau^{-1})}\Bigg[\sum_{j\in \widehat{S}}\log\bigg(\frac{1+\tau^{-1}|\beta_{j}|}{1+\tau^{-1}|w_{j}+\beta_{j}|}\bigg) + \sum_{j\in\widehat{S}^{c}}\log\bigg(\frac{1+\tau^{-1}|\beta_{j}|}{1+\tau^{-1}|w_{j}+\beta_{j}|}\bigg)\Bigg]\nonumber\\
&\leq&\frac{\rho_{1}}{\log(\tau^{-1})}\Big\{2c_{2}^{-1}s^{1/2}||w||_{2}+\big[1+c_{3}^{-1}\log(\tau^{-1})\big]s^{1/2}||w||_{2}\Big\}\nonumber\\
&=&\rho_{1}\bigg[\frac{2c_{2}^{-1}+1}{\log(\tau^{-1})}+c_{3}^{-1}\bigg]s^{1/2}||w||_{2}.
\end{eqnarray}
\textbf{Deriving an upper bound for (\ref{basic_inq04}).} Similarly, for $k\in \widehat{R}^{\tau}=\{k:||\widehat{\beta}_{G_{k}}^{\tau}||_{2}>0\}$, we have $||w_{G_{k}}+\beta_{G_{k}}||_{2}=||\widehat{\beta}_{G_{k}}^{\tau}||_{2}>0$. In turn, we have
\begin{eqnarray}
\label{ub_inq04_01}
\log\bigg(\frac{1+\tau^{-1}||\beta_{G_{k}}||_{2}}{1+\tau^{-1}||w_{G_{k}}+\beta_{G_{k}}||_{2}}\bigg)&\leq&\frac{||\beta_{G_{k}}||_{2}-||w_{G_{k}}+\beta_{G_{k}}||_{2}+||w_{G_{k}}||_{2}}{\tau + ||w_{G_{k}}+\beta_{G_{k}}||_{2}}.
\end{eqnarray}
for $k\in\widehat{R}^{\tau}$. 

Now if $k\in \widehat{R}^{\tau}\cap R^{c}$, where $R^{c}=\{k:||\beta_{G_{k}}||_{2}=0\}$, then the right hand side of (\ref{ub_inq04_01}) is zero. On the other hand, for $j\in G_{\widehat{R}^{\tau}}\cap S$, we have $c_{2}=\min_{j\in \widehat{S}^{\tau}}|\widehat{\beta}_{j}|\leq\min_{k\in \widehat{R}^{\tau}}||\widehat{\beta}_{G_{k}}||_{2}\leq ||\widehat{\beta}_{G_{k}}||_{2}$. In addition, $||\beta_{G_{k}}||_{2}-||w_{G_{k}}+\beta_{G_{k}}||_{2}\leq ||\beta_{G_{k}}-w_{G_{k}}-\beta_{G_{k}}||_{2}=||w_{G_{k}}||_{2}$. Then with the results given above, we can further obtain
\begin{eqnarray}
\label{ub_inq04_02}
\sum_{k\in \widehat{R}^{\tau}}\sqrt{q_{k}}\log\bigg(\frac{1+\tau^{-1}||\beta_{G_{k}}||_{2}}{1+\tau^{-1}||w_{G_{k}}+\beta_{G_{k}}||_{2}}\bigg)
&\leq&\sum_{k\in \widehat{R}^{\tau}\cap R}\sqrt{q_{k}}\frac{2||w_{G_{k}}||_{2}}{\tau + ||w_{G_{k}}+\beta_{G_{k}}||_{2}}\nonumber\\
&\leq&\sum_{k\in \widehat{R}^{\tau}\cap R}\sqrt{q_{k}}\frac{2||w_{G_{k}}||_{2}}{c_{2}}\nonumber\\
&\leq&2c_{2}^{-1}\bigg(\sum_{k\in R}\sqrt{q_{k}}^{2}\bigg)^{1/2}\bigg(\sum_{k\in R}||w_{G_{k}}||_{2}^{2}\bigg)^{1/2}\nonumber\\
&\leq&2c_{2}^{-1}q_{R}^{1/2}||w||_{2},
\end{eqnarray}
where $q_{R}=|G_{R}|=\sum_{k\in R}q_{k}$ is the number of indices covered by $G_{R}$. We now consider the summation over indices $k\in (\widehat{R}^{\tau})^{c}$. If $k\in (\widehat{R}^{\tau})^{c}$, $||\widehat{\beta}_{G_{k}}^{\tau}||_{2}=0$. Therefore, we have $||w_{G_{k}}+\beta_{G_{k}}||_{2}=||\widehat{\beta}_{G_{k}}^{\tau}-\beta_{G_{k}}+\beta_{G_{k}}||_{2}=0$ and $||\beta_{G_{k}}||_{2}=||\widehat{\beta}_{G_{k}}^{\tau}-\beta_{G_{k}}||_{2}=||w_{G_{k}}||_{2}$. In turn,
\begin{eqnarray}
\log(1+\tau^{-1}||\beta_{G_{k}}||_{2})-\log(1+\tau^{-1}||w_{G_{k}}+\beta_{G_{k}}||_{2})=\log(\tau + ||w_{G_{k}}||_{2})+\log(\tau^{-1})\nonumber
\end{eqnarray}
for $k\in (\widehat{R}^{\tau})^{c}$. In addition, for $\tau\in[0,1)$, $\log(\tau + ||\beta_{G_{k}}||_{2})\leq \log(1+||\beta_{G_{k}}||_{2})\leq ||\beta_{G_{k}}||_{2}$. Further note that 
\begin{eqnarray}
c_{3}=\min_{j\in S}|\beta_{j}|\leq\min_{k\in (\widehat{R}^{\tau})^{c}, ||\beta_{G_{k}}||_{2}\neq 0}||\beta_{G_{k}}||_{2}=\min_{k\in (\widehat{R}^{\tau})^{c}, ||w_{G_{k}}||_{2}\neq 0}||w_{G_{k}}||_{2}.\nonumber
\end{eqnarray}
Moreover, for an arbitrary index $k\in (\widehat{R}^{\tau})^{c}$, $||w_{G_{k}}||_{2}= ||\beta_{G_{k}}||_{2}$, therefore $||w_{G_{k}}||_{2}\neq 0$ implies $||\beta_{G_{k}}||_{2}\neq 0$ and the index $k\in R$. Now by applying the results given above, we have
\begin{eqnarray}
\label{ub_inq04_03}
& &\sum_{k\in (\widehat{R}^{\tau})^{c}}\sqrt{q_{k}}\big[\log(1 + \tau^{-1}||\beta_{G_{k}}||_{2})-\log(1+\tau^{-1}||w_{G_{k}}+\beta_{G_{k}}||_{2})\big]\nonumber\\
&=&\sum_{k\in (\widehat{R}^{\tau})^{c}, ||w_{G_{k}}||_{2}\neq 0}\sqrt{q_{k}}\big[\log(\tau + ||w_{G_{k}}||_{2})+\log(\tau^{-1})\big]\nonumber\\
& &+\sum_{k\in (\widehat{R}^{\tau})^{c}, ||w_{G_{k}}||_{2}=0}\sqrt{q_{k}}\big[\log(\tau + ||w_{G_{k}}||_{2})+\log(\tau^{-1})\big]\nonumber\\
&\leq&\big[1 + c_{3}^{-1}\log(\tau^{-1})\big]\sum_{k\in (\widehat{R}^{\tau})^{c}, k\in R}\sqrt{q_{k}}||w_{G_{k}}||_{2}\nonumber\\
&\leq&\big[1+c_{3}^{-1}\log(\tau^{-1})\big]q_{R}^{1/2}||w||_{2}.
\end{eqnarray}
Combining the results in (\ref{ub_inq04_02}) and (\ref{ub_inq04_03}), we can bound (\ref{basic_inq04}) in a way such that
\begin{eqnarray}
\label{bound_inq04}
& &\rho_{2}\sum_{k=1}^{m}\sqrt{q_{k}}\bigg[\frac{\log(1+\tau^{-1}||\beta_{G_{k}}||_{2})}{\log(1+\tau^{-1})}-\frac{\log(1+\tau^{-1}||w_{G_{k}}+\beta_{G_{k}}||_{2})}{\log(1+\tau^{-1})}\bigg]\nonumber\\
&\leq&\frac{\rho_{2}}{\log(\tau^{-1})}\Bigg[\sum_{k\in \widehat{R^{\tau}}}\sqrt{q_{k}}\log\bigg(\frac{1+\tau^{-1}||\beta_{G_{k}}||_{2}}{1+\tau^{-1}||w_{G_{k}}+\beta_{G_{k}}||_{2}}\bigg)\nonumber\\
& &+ \sum_{k\in(\widehat{R}^{\tau})^{c}}\sqrt{q_{k}}\log\bigg(\frac{1+\tau^{-1}||\beta_{G_{k}}||_{2}}{1+\tau^{-1}||w_{G_{k}}+\beta_{G_{k}}||_{2}}\bigg)\Bigg]\nonumber\\
&\leq&\frac{\rho_{2}}{\log(\tau^{-1})}\Big\{2c_{2}^{-1}q_{R}^{1/2}||w||_{2}+\big[1+c_{3}^{-1}\log(\tau^{-1})\big]q_{R}^{1/2}||w||_{2}\Big\}\nonumber\\
&=&\rho_{2}\bigg[\frac{2c_{2}^{-1}+1}{\log(\tau^{-1})}+c_{3}^{-1}\bigg]q_{R}^{1/2}||w||_{2}.
\end{eqnarray} 
\textbf{Deriving an upper bound for (\ref{basic_inq02}).} First note that
\begin{eqnarray}
\label{bound02_01}
w^{T}X^{T}\epsilon \leq ||w||_{1}||X^{T}\epsilon||_{\infty}.\nonumber
\end{eqnarray}
Now for $||w||_{1}$, we can decompose it as 
\begin{eqnarray}
\label{bound02_02}
||w||_{1}&=&||w_{\widehat{S}^{\tau}\cap S}||_{1} + ||w_{\widehat{S}^{\tau}\cap S^{c}}||_{1} + ||w_{(\widehat{S}^{\tau})^{c}\cap S}||_{1} + 
||w_{(\widehat{S}^{\tau})^{c}\cap S^{c}}||_{1}.
\end{eqnarray}
Note that for the first and third terms on the right hand side of (\ref{bound02_02}), we have $||w_{\widehat{S}^{\tau}\cap S}||_{1} \leq ||w_{S}||_{1}$ and $||w_{(\widehat{S}^{\tau})^{c}\cap S}||_{1}\leq ||w_{S}||_{1}$.  For the second term on the right hand side of (\ref{bound02_02}), we have $||w_{\widehat{S}^{\tau}\cap S^{c}}||_{1}\leq ||w_{\widehat{S}^{\tau}}||_{1}$. The fourth term on the right hand side of (\ref{bound02_02}) is zero since $(\widehat{S}^{\tau})^{c}\cap S^{c}$ is an intersection of indices for entries with zero values in $\beta$ and entries with zero values in $\widehat{\beta}^{\tau}$. With the results given above, we can further bound $||w||_{1}$ in a way such that
\begin{eqnarray}
\label{bound02_03}
||w||_{1}&\leq&2||w_{S}||_{1} + ||w_{\widehat{S}^{\tau}}||_{1}\nonumber\\
&\leq&s^{1/2}\bigg[2+\bigg(\frac{\widehat{s}^{\tau}}{s}\bigg)^{1/2}\bigg]||w||_{2}.
\end{eqnarray}
With the result in (\ref{bound02_03}), we can bound  (\ref{basic_inq02}) in a way such that
\begin{eqnarray}
\label{bound_inq02}
2w^{T}(X^{T}\epsilon-\lambda\beta)&\leq&2||w||_{1}||X^{T}\epsilon||_{\infty} + 2\lambda|w^{T}\beta|\nonumber\\
&\leq&2s^{1/2}\bigg[2+\bigg(\frac{\widehat{s}^{\tau}}{s}\bigg)^{1/2}\bigg]||w||_{2}||X^{T}\epsilon||_{\infty}\nonumber\\ 
& &+2\lambda||w||_{2}s^{1/2}\max_{j\in S}|\beta_{j}|.
\end{eqnarray}
Combining the results (\ref{bound_inq03}), (\ref{bound_inq04}) and (\ref{bound_inq02}), we obtain
\begin{eqnarray}
\label{combined_result01_1}
n(\kappa_{n}+\lambda n^{-1})||w||_{2}^{2}&\leq&2s^{1/2}\bigg[2 + \bigg(\frac{\widehat{s}^{\tau}}{s}\bigg)^{1/2}\bigg]||w||_{2}||X^{T}\epsilon||_{\infty}+2\lambda s^{1/2}\max_{j\in S}|\beta_{j}|||w||_{2}\nonumber\\
& &+\rho_{1}\bigg[\frac{2c_{2}^{-1}+1}{\log(\tau^{-1})}+c_{3}^{-1}\bigg]s^{1/2}||w||_{2}\nonumber\\
& &+\rho_{2}\bigg[\frac{2c_{2}^{-1}+1}{\log(\tau^{-1})}+c_{3}^{-1}\bigg]q_{R}^{1/2}||w||_{2}.
\end{eqnarray}
Then by using the fact that $s=|S|\leq |G_{R}|=q_{R}$ and doing some rearrangement in (\ref{combined_result01_1}), we obtain the inequality (\ref{theorem1_01}), which completes the proof.\qed

\emph{Proof of Theorem \ref{theorem_l2_error_bound}.} We start our proof by showing that with at least $1-\alpha$ probability, the inequality $2||X^{T}\epsilon||_{\infty}< \psi_{n}$ will hold, where $\psi_{n}$ is defined in (\ref{theorem2_01}). Note that $\{2||X^{T}\epsilon||_{\infty}< \psi_{n}\}$ is equivalent to the following event:
\begin{eqnarray}
\label{group_sp01}
\mathcal{A} = \bigcap_{k=1}^{m}\bigg\{2||X_{G_{k}}^{T}\epsilon||_{\infty}< \psi_{n}\bigg\}.\nonumber
\end{eqnarray}
We will establish the inequality $\mathbb{P}(\mathcal{A})=1-\mathbb{P}(\mathcal{A}^{c})\geq 1-\alpha$ by showing that given $\psi_{n}$ is defined in (\ref{theorem2_01}), $\mathbb{P}(\mathcal{A}^{c})\leq \alpha$. The technique we use to derive the inequality $\mathbb{P}(\mathcal{A}^{c})\leq \alpha$ is borrowed from Lemma B.1 of \cite{bickeletal09}. Note that the tail probability $\mathbb{P}(\mathcal{A}^{c})$ can be bounded in a way such that
\begin{eqnarray}
\label{group_sp02}
\mathbb{P}(\mathcal{A}^{c})&=&\mathbb{P}\Bigg(\bigcup_{k=1}^{m}\bigg\{2||X_{G_{k}}^{T}\epsilon||_{\infty}\geq \psi_{n}\bigg\}\Bigg)\nonumber\\
&\leq&\sum_{k=1}^{m}\mathbb{P}\bigg(||X_{G_{k}}^{T}\epsilon||_{\infty}\geq\frac{\psi_{n}}{2}\bigg)\leq \sum_{k=1}^{m}\sum_{j\in G_{k}}\mathbb{P}\bigg(\bigg|\sum_{i=1}^{n}x_{ij}\epsilon_{i}\bigg|\geq \frac{\psi_{n}}{2}\bigg).
\end{eqnarray}
Under assumptions given in Theorem \ref{theorem_l2_error_bound}, $\epsilon_{i}$'s are i.i.d. normal variables with mean zero and variance $\sigma^{2}$, therefore $\sum_{i=1}^{n}x_{ij}\epsilon_{i}$ is a normal variable with mean zero and variance $\sigma^{2}\sum_{i=1}^{n}x_{ij}^{2}=n\zeta_{j}\sigma^{2}$. In turn, we can express $|\sum_{i=1}^{n}x_{ij}\epsilon_{i}| = \sqrt{n\zeta_{j}}\sigma |Z|$, where $Z$ is a standard normal variable. By using the Chernoff bound argument on the tail probability of a standard normal variable, we can bound the right hand side of (\ref{group_sp02}) in a way such that
\begin{eqnarray}
\label{group_sp05}
\sum_{k=1}^{m}\sum_{j\in G_{k}}\mathbb{P}\bigg(\bigg|\sum_{i=1}^{n}x_{ij}\epsilon_{i}\bigg|\geq \frac{\psi_{n}}{2}\bigg)&\leq&
\sum_{k=1}^{m}q_{k}\mathbb{P}\bigg(|Z|\geq\frac{\psi_{n}}{2\sqrt{n\max_{j}\zeta_{j}}\sigma}\bigg)\nonumber\\ &\leq&m\sum_{k=1}^{m}\frac{q_{k}}{m}\exp\bigg(-\frac{\psi_{n}^{2}}{8n\max_{j}\zeta_{j}\sigma^{2}}\bigg)\nonumber\\
&\leq&m\exp\bigg(-\frac{\psi_{n}^{2}}{8n\sigma^{2}\max_{j}\zeta_{j}}+\log\overline{q}\bigg),\nonumber\\
\end{eqnarray}
where $\overline{q}=m^{-1}\sum_{k=1}^{k}q_{k}$. With $\psi_{n}$ defined in (\ref{theorem2_01}), the right hand side of (\ref{group_sp05}) is equal to $\alpha$, and further with (\ref{group_sp02}), we obtain $\mathbb{P}(\mathcal{A}^{c})\leq \alpha$, which implies that with $\psi_{n}$ defined in (\ref{theorem2_01}), $\mathbb{P}(\mathcal{A})=1-\mathbb{P}(\mathcal{A}^{c})\geq 1-\alpha$.

To complete the proof, note that since we have assumed $\tau = n^{-1}$, therefore effectively we have $\widehat{\beta}^{\tau}\rightarrow\widehat{\beta}_{\text{gvsnss}}$ and $\widehat{s}^{\tau}\rightarrow \widehat{s}$ as $n\rightarrow\infty$. Therefore with the result from Theorem \ref{theorem_deterministic_bound} and the assumptions on $\lambda$, $\rho_{1}$, $\rho_{2}$ and $\tau$, as $n\rightarrow\infty$, the inequality
\begin{eqnarray}
\label{group_sp06}
||\widehat{\beta}_{\text{gvsnss}}-\beta||_{2}\leq\frac{q_{R}^{1/2}}{(\kappa_{n}+\Omega_{n})}\frac{\Lambda_{n}\psi_{n}}{n}
\end{eqnarray}
will hold with $1-\alpha$ probability, where $\Lambda_{n}$ is defined in (\ref{theorem2_03}) and $\Omega_{n}=n^{-1}\lambda$ is defined in (\ref{theorem2_04}) and $\psi_{n}$ defined in (\ref{theorem2_01}), which completes the proof. \qed
\section{Proof of Theorem \ref{theorem_label_inv_property}}\label{appendix_b}
\emph{Proof of Theorem \ref{theorem_label_inv_property}.} Now define
\begin{eqnarray}
\label{as_results02}
U_{\tau}(\beta^{\prime}, G^{*}, G^{**})&=& \rho_{2}\sum_{k=1}^{m^{*}}\sqrt{q_{k}^{*}}\frac{\log(1+\tau^{-1}||\beta_{G_{k}^{*}}^{\prime}||_{2})}{\log(1+\tau^{-1})}\nonumber\\
& & - \rho_{2}\sum_{l=1}^{m^{**}}\sqrt{q_{l}^{**}}\frac{\log(1+\tau^{-1}||\beta_{G_{l}^{**}}^{\prime}||_{2})}{\log(1+\tau^{-1})}.
\end{eqnarray}
where $\beta_{G_{k}^{*}}^{\prime}$ is the coefficient vector in which the elements are those indexed by $G_{k}^{*}$ in the vector $\beta^{\prime}$. The vector $\beta_{G_{l}^{**}}^{\prime}$ follows a similar definition. The function (\ref{as_results02}) is the difference between the log-sum penalties involving $l_{2}$-norms indexed by $G^{*}$ and $G^{**}$. Note that, with (\ref{as_results02}), the objective function $V_{\tau}(0, \beta^{\prime},G^{*})$ in (\ref{linv01}) can be re-expressed as
\begin{eqnarray}
\label{as_results03}
V_{\tau}(0,\beta^{\prime}, G^{*}) = V_{\tau}(0,\beta^{\prime}, G^{**}) + U_{\tau}(\beta^{\prime}, G^{*}, G^{**}).
\end{eqnarray} 
Since $\widehat{\beta}^{\tau*}$ is the minimizer of $V_{\tau}(0,\beta^{\prime},G^{*})$, therefore it must be the solution to the following subgradient equations:
\begin{eqnarray}
\label{as_results04}
2X^{T}(y - X\beta^{\prime}) - 2\lambda\beta^{\prime}- \rho_{1}g^{\prime}- \rho_{2}\sqrt{q^{**}}u^{**}- \rho_{2}(\sqrt{q^{*}}u^{*} -\sqrt{q^{**}}u^{**})= 0,
\end{eqnarray}
where
\begin{eqnarray}
(g^{\prime})_{j} = \frac{h_{j}^{\prime}}{[\tau\log(1+\tau^{-1})](1+\tau^{-1}|\beta_{j}^{\prime}|)},\nonumber
\end{eqnarray} 
with $h_{j}^{\prime} =$ sign$(\beta_{j}^{\prime})$ if $\beta_{j}^{\prime}\neq 0$ and $-1\leq h_{j}^{\prime}\leq 1$ if $\beta_{j}^{\prime} = 0$, and
\begin{eqnarray}
(u^{*})_{j} = \frac{v^{*}_{j}}{[\tau\log(1+\tau^{-1})](1+\tau^{-1}||\beta_{G_{k_{j}}^{*}}^{\prime}||_{2})}\nonumber
\end{eqnarray}
with $v^{*}_{j} = \beta_{j}^{\prime}/||\beta_{G_{k_{j}}^{*}}^{\prime}||_{2}$ if $||\beta_{G_{k_{j}}^{*}}^{\prime}||_{2}> 0$ and $\sum_{j\in G_{k_{j}}^{*}}(v_{j}^{*})^{2}\leq 1$ if $||\beta_{G_{k_{j}}^{*}}^{\prime}||_{2}=0$, where $k_{j}$ is the index for the group that $j$ belongs to, i.e. if $j\in G_{k^{\prime}}^{*}$, then $k_{j}=k^{\prime}$. The quantity $(u^{**})_{j}$ follows a similar definition. In addition, $(q^{*})_{j}= q_{k_{j}}^{*}$ and $(q^{**})_{j}= q_{l_{j}}^{**}$. 

Note that the derivation of the subgradient equations (\ref{as_results04}) has explicitly used representation (\ref{as_results03}), and after some simple arrangement, (\ref{as_results04}) becomes
\begin{eqnarray}
\label{as_results05}
2X^{T}(y - X\beta) - 2\lambda\beta- \rho_{1}g^{\prime}- \rho_{2}\sqrt{q^{**}}u^{**}=\rho_{2}(\sqrt{q^{*}}u^{*} -\sqrt{q^{**}}u^{**}),
\end{eqnarray}
where
\begin{eqnarray}
\label{as_results06}
(\sqrt{q^{*}}u^{*} - \sqrt{q^{**}}u^{**})_{j}&=&\frac{1}{\log(1+\tau^{-1})}\Bigg[\frac{(\tau + ||\beta_{G_{l_{j}}^{**}}^{\prime}||_{2})\sqrt{q_{k_{j}}^{*}}v_{j}^{*}}{(\tau+||\beta_{G_{k_{j}}^{*}}^{\prime}||_{2})(\tau+||\beta_{G_{l_{j}}^{**}}^{\prime}||_{2})}\nonumber\\
& &-\frac{(\tau + ||\beta_{G_{k_{j}}^{*}}^{\prime}||_{2})\sqrt{q_{l_{j}}^{**}}v_{j}^{**}}{(\tau+||\beta_{G_{k_{j}}^{*}}^{\prime}||_{2})(\tau+||\beta_{G_{l_{j}}^{**}}^{\prime}||_{2})}\Bigg].
\end{eqnarray}
For each $j$, one of the following four cases will occur: (i) $||\beta_{G_{k_{j}}^{*}}^{\prime}||_{2} = 0$ and $||\beta_{G_{l_{j}}^{**}}^{\prime}||_{2} = 0$; (ii) $||\beta_{G_{k_{j}}^{*}}^{\prime}||_{2} > 0$ and $||\beta_{G_{l_{j}}^{**}}^{\prime}||_{2} = 0$; (iii) $||\beta_{G_{k_{j}}^{*}}^{\prime}||_{2} = 0$ and $||\beta_{G_{l_{j}}^{**}}^{\prime}||_{2} > 0$; and (iv) $||\beta_{G_{k_{j}}^{*}}^{\prime}||_{2} > 0$ and $||\beta_{G_{l_{j}}^{**}}^{\prime}||_{2} > 0$. In the following discussion, we will evaluate (\ref{as_results06}) under the four cases. 

We consider case (i) first. If (i) occurs, then all regression coefficients with indices in $G_{k_{j}}^{*}$ or $G_{l_{j}}^{**}$ will be zero. It implies that $\beta_{j}^{\prime} = 0$ and by definitions, $v_{j}^{*}$ is an arbitrary quantity such that $0\leq (v_{j}^{*})^{2}\leq \sum_{j\in G_{k_{j}}^{*}}(v_{j}^{*})^{2}\leq 1$. The same property applies to $v_{j}^{**}$. For practical purposes, we choose $v_{j}^{*}=\tau$ and $v_{j}^{**}=\tau$. Then under case (i), 
\begin{eqnarray}
\label{as_results071}
(\sqrt{q^{*}}u^{*} - \sqrt{q^{**}}u^{**})_{j}&=&\frac{1}{\tau\log(1+\tau^{-1})}\Bigg[\frac{\tau\sqrt{q_{k_{j}}^{*}}v_{j}^{*}}{\tau}-\frac{\tau\sqrt{q_{l_{j}}^{**}}v_{j}^{**}}{\tau}\Bigg]\nonumber\\
&=&\frac{\big(\sqrt{q_{k_{j}}^{*}}-\sqrt{q_{l_{j}}^{**}}\big)}{\log(1+\tau^{-1})}.
\end{eqnarray}

Now consider case (ii). If (ii) holds, then by definition, $v_{j}^{*}=\beta_{j}^{\prime}/||\beta_{G_{k_{j}}^{*}}||_{2}$. In addition, since $||\beta_{G_{l_{j}}^{**}}^{\prime}||_{2}=0$, therefore $v_{j}^{**}$ is an arbitrary quantity such that $0\leq (v_{j}^{**})^{2}\leq \sum_{j\in G_{l_{j}}^{*}}(v_{j}^{**})^{2}\leq 1$. For practical purposes, we choose $v_{j}^{**}=\tau$. Moreover, $||\beta_{G_{l_{j}}^{**}}^{\prime}||_{2}=0$ implies that all coefficients with indices in $G_{l_{j}}^{**}$ are zero. Therefore $\beta_{j}^{\prime} = 0$ and $v_{j}^{*}=\beta_{j}^{\prime}/||\beta_{G_{k_{j}}^{*}}^{\prime}||_{2}=0$. Then under case (ii), 
\begin{eqnarray}
\label{as_results072}
(\sqrt{q^{*}}u^{*} - \sqrt{q^{**}}u^{**})_{j}&=&\frac{1}{\tau\log(1+\tau^{-1})}\Bigg[\frac{\tau\sqrt{q_{k_{j}}^{*}}v_{j}^{*}}{\tau+||\beta_{G_{k_{j}}^{*}}^{\prime}||_{2}}-\frac{(\tau + ||\beta_{G_{k_{j}}^{*}}^{\prime}||_{2})\sqrt{q_{l_{j}}^{**}}v_{j}^{**}}{\tau+||\beta_{G_{k_{j}}^{*}}^{\prime}||_{2}}\Bigg]\nonumber\\
&=&-\frac{\sqrt{q_{l_{j}}^{**}}}{\log(1+\tau^{-1})}.
\end{eqnarray}
Now consider case (iii). Under case (iii), since $||\beta_{G_{l_{j}}^{**}}^{\prime}||_{2} > 0$, therefore $v_{j}^{**}=\beta_{j}^{\prime}/||\beta_{G_{l_{j}}^{**}}^{\prime}||_{2}$. In addition, $||\beta_{G_{k_{j}}^{*}}^{\prime}||_{2}=0$ implies that all coefficients with indices in $G_{k_{j}}^{*}$ are zero. Therefore $\beta_{j}^{\prime} = 0$ and $v_{j}^{**}=\beta_{j}^{\prime}/||\beta_{G_{l_{j}}^{**}}^{\prime}||_{2}=0$. In addition, $v_{j}^{*}$ is an arbitrary quantity such that $0\leq (v_{j}^{*})^{2}\leq \sum_{j\in G_{k_{j}}^{*}}(v_{j}^{*})^{2}\leq 1$. Here we let $v_{j}^{*}=\tau$. Therefore under case (iii),
\begin{eqnarray}
\label{as_results073}
(\sqrt{q^{*}}u^{*} - \sqrt{q^{**}}u^{**})_{j}&=&\frac{1}{\tau\log(1+\tau^{-1})}\Bigg[\frac{(\tau+||\beta_{G_{l_{j}}^{**}}^{\prime}||_{2})\sqrt{q_{k_{j}}^{*}}v_{j}^{*}}{\tau+||\beta_{G_{l_{j}}^{**}}^{\prime}||_{2}}-\frac{\tau\sqrt{q_{l_{j}}^{**}}v_{j}^{**}}{\tau+||\beta_{G_{l_{j}}^{**}}^{\prime}||_{2}}\Bigg]\nonumber\\
&=&\frac{\sqrt{q_{k_{j}}^{*}}}{\log(1+\tau^{-1})}.
\end{eqnarray} 

Finally we consider case (iv). Under case (iv), $v_{j}^{*}=\beta_{j}^{\prime}/||\beta_{G_{k_{j}}^{*}}^{\prime}||_{2}$ and $v_{j}^{**}=\beta_{j}^{\prime}/||\beta_{G_{l_{j}}^{**}}^{\prime}||_{2}$. Further by direct calculation, we have
\begin{eqnarray}
\label{as_results074}
(\sqrt{q^{*}}u^{*} - \sqrt{q^{**}}u^{**})_{j}&=&\frac{1}{\log(1+\tau^{-1})}\Bigg[\frac{\sqrt{q_{k_{j}}^{*}}v_{j}^{*}}{(\tau+||\beta_{G_{k_{j}}^{*}}^{\prime}||_{2})}-\frac{\sqrt{q_{l_{j}}^{**}}v_{j}^{**}}{(\tau+||\beta_{G_{l_{j}}^{**}}^{\prime}||_{2})}\Bigg]\nonumber\\
&=&\frac{1}{\log(1+\tau^{-1})}\Bigg[\frac{\sqrt{q_{k_{j}}^{*}}\beta_{j}^{\prime}}{(\tau+||\beta_{G_{k_{j}}^{*}}^{\prime}||_{2})||\beta_{G_{k_{j}}^{*}}^{\prime}||_{2}}\nonumber\\
& &-\frac{\sqrt{q_{l_{j}}^{**}}\beta_{j}^{\prime}}{(\tau+||\beta_{G_{l_{j}}^{**}}^{\prime}||_{2})||\beta_{G_{l_{j}}^{**}}^{\prime}||_{2}}\Bigg].\nonumber\\
\end{eqnarray}

Now with $\tau = n^{-1}$ and the results from (\ref{as_results071}), (\ref{as_results072}), (\ref{as_results073}), and (\ref{as_results074}), we can see that $|U_{\tau}(\beta^{\prime},G^{*},G^{**})|=O(\rho_{2}\max_{k}\sqrt{q_{k}}[\log(n)]^{-1})$. Therefore if $\rho_{2}\max_{k}\sqrt{q_{k}}=o(\log(n))$,
$U_{\tau}(\beta^{\prime},G^{*},G^{**})$ will approach to zero when $n\rightarrow\infty$. It further implies that the right hand side of (\ref{as_results05}) will become zero when $n\rightarrow \infty$. On the other hand, the left hand side of (\ref{as_results05}) is just the subgradient vector of the objective function $V_{\tau}(0,\beta^{\prime},G^{**})$. Therefore when $\tau\rightarrow 0$, (\ref{as_results05}) becomes the subgradient equations of $\lim_{\tau\rightarrow 0}V_{\tau}(0,\beta^{\prime},G^{**})$. Since $\widehat{\beta}_{\text{gvsnss}}^{*}$ is the solution of the subgradient equations (\ref{as_results04}) when $\tau\rightarrow 0$ and (\ref{as_results05}) is just a rearrangement of (\ref{as_results04}), therefore $\widehat{\beta}_{\text{gvsnss}}^{*}$ is also the solution to (\ref{as_results05}) when $\tau\rightarrow 0$. Since (\ref{as_results05}) becomes the subgraident equations of $\lim_{\tau\rightarrow0}V_{\tau}(0,\beta^{\prime},G^{**})$ when $\tau\rightarrow 0$, and the solution of (\ref{as_results05}) at $\tau\rightarrow 0$ is the minimizer of $\lim_{\tau\rightarrow0}V_{\tau}(0,\beta^{\prime},G^{**})$, there we conclude that $\widehat{\beta}_{\text{gvsnss}}^{*}$ is the minimizer of $\lim_{\tau\rightarrow0}V_{\tau}(0,\beta^{\prime},G^{**})$, which completes the proof.\qed
\section{Proof of Theorem \ref{theorem_sign_consistency}}\label{appendix_c}
\emph{Proof of Theorem \ref{theorem_sign_consistency}.} Define $w=\widehat{\beta}^{\tau}-\beta$. It can be shown that given $\beta$ and $G$ are fixed, $w$ is the minimizer of $V_{\tau}(w^{*},\beta,G)$, therefore $w$ is also the solution to the following subgradient equations:
\begin{eqnarray}
\label{msc021}
2X^{T}Xw -2X^{T}\epsilon + 2\lambda(\beta+w) + \rho_{1}g + \rho_{2}\sqrt{q}u= 0,
\end{eqnarray}
where
\begin{eqnarray}
\label{msc03}
(g)_{j} = \frac{h_{j}}{[\tau\log(1+\tau^{-1})](1+\tau^{-1}|w_{j}+\beta_{j}|)},\nonumber
\end{eqnarray} 
with $h_{j} =$ sign$(w_{j}+\beta_{j})$ if $w_{j}+\beta_{j}\neq 0$ and $-1\leq h_{j}\leq 1$ if $w_{j}+\beta_{j} = 0$, and
\begin{eqnarray}
\label{msc04}
(u)_{j} = \frac{v_{j}}{[\tau\log(1+\tau^{-1})](1+\tau^{-1}||w_{G_{k_{j}}}+\beta_{G_{k_{j}}}||_{2})}\nonumber
\end{eqnarray}
with $v_{j} = (w_{j}+\beta_{j})/||w_{G_{k_{j}}} + \beta_{G_{k_{j}}}||_{2}$ if $||w_{G_{k_{j}}}+\beta_{G_{k_{j}}}||_{2}> 0$ and $\sum_{j\in G_{k_{j}}}(v_{j})^{2}\leq 1$ if $||w_{G_{k_{j}}}+\beta_{G_{k_{j}}}||_{2}=0$, where $k_{j}$ is the index for the group that $j$ belongs to. 

Let $S_{1}^{c}= S^{c}\cap G_{R}$ and $S_{2}^{c}=S^{c}\cap G_{R^{c}}$. Here $S^{c}$ is the set of indices for redundant covariates, i.e. the covariates with zero coefficients. In addition, $S_{1}^{c}$ is the set of indices for the redundant covariates covered by $G_{R}$, and $S_{2}^{c}$ is the set of indices for the redundant covariates covered by $G_{R^{c}}$. By definition, $G_{R^{c}}\subseteq S^{c}$, therefore we have $S_{2}^{c}= G_{R^{c}}$. In addition, $S$, $S_{1}^{c}$ and $G_{R^{c}}$ are three disjoint index sets and $S\cup S_{1}^{c}\cup G_{R^{c}}=\{1,2,\cdots,p\}$. With the results given above, we can re-express (\ref{msc021}) as
\begin{eqnarray}
\label{msc02}
& &2\begin{pmatrix}
X_{S}^{T}X_{S} & X_{S}^{T}X_{S_{1}^{c}} & X_{S}X_{G_{R^{c}}}\\ 
X_{S_{1}^{c}}^{T}X_{S} & X_{S_{1}^{c}}^{T}X_{S_{1}^{c}} & X_{S_{1}^{c}}^{T}X_{G_{R^{c}}}\\ 
X_{G_{R^{c}}}^{T}X_{S} & X_{G_{R^{c}}}^{T}X_{S_{1}^{c}} & X_{G_{R^{c}}}^{T}X_{G_{R^{c}}}\end{pmatrix}
\begin{pmatrix}w_{S} \\ w_{S_{1}^{c}} \\ w_{G_{R^{c}}}\end{pmatrix}-2\begin{pmatrix}X_{S}^{T}\epsilon \\ X_{S_{1}^{c}}^{T}\epsilon \\ X_{G_{R^{c}}}^{T}\epsilon\end{pmatrix}\nonumber\\
& & + 2\lambda \begin{pmatrix}w_{S}+\beta_{S} \\ w_{S_{1}^{c}} +\beta_{S_{1}^{c}}\\ w_{G_{R^{c}}}+\beta_{G_{R^{c}}}\end{pmatrix}+\rho_{1}\begin{pmatrix}g_{S} \\ g_{S_{1}^{c}}\\ g_{G_{R^{c}}}\end{pmatrix} + \rho_{2}\begin{pmatrix}\sqrt{q_{S}}u_{S} \\ \sqrt{q_{S_{1}^{c}}}u_{S_{1}^{c}}\\ \sqrt{q_{G_{R^{c}}}}u_{G_{R^{c}}}\end{pmatrix}=0.
\end{eqnarray}
For practical purposes, we define $\vartheta_{j}^{S}$ as the position of index $j$ in the set $S$. It is equivalent to say that index $j$ is the $\vartheta_{j}^{S}$th element in $S$. If $j\notin S$, then we just leave $\vartheta_{j}^{S}$ undefined. Similar definitions are applied to $\vartheta_{j}^{S_{1}^{c}}$ and $\vartheta_{j}^{G_{R^{c}}}$. 

To make the sign consistency hold, we must have $w_{j}=\widehat{\beta}_{j}^{\tau}-\beta_{j}=0$ for all $j\in S_{1}^{c}\cup G_{R^{c}}$, and sign$(\widehat{\beta}_{j})=$ sign$(\beta_{j})$ for all $j\in S$. Given that $w$ is the solution to (\ref{msc02}), then with the arguments given above, we obtain the following conditions:
\begin{eqnarray}
\label{cond1}
\big(X_{S}^{T}X_{S}w_{S}-X_{S}^{T}\epsilon + \lambda(w_{S}+\beta_{S})+\frac{\rho_{2}}{2}\sqrt{q_{S}}u_{S}\big)_{\vartheta_{j}^{S}}= \bigg(-\frac{\rho_{1}}{2}g_{S}\bigg)_{\vartheta_{j}^{S}},
\end{eqnarray}
for $j\in S$, and
\begin{eqnarray}
\label{cond3}
-\frac{\rho_{1}}{2\tau\log(1+\tau^{-1})}&<&\bigg(X_{S_{1}^{c}}^{T}X_{S}w_{S} - X_{S_{1}^{c}}^{T}\epsilon + \frac{\rho_{2}}{2}\sqrt{q_{S_{1}^{c}}}u_{S_{1}^{c}}\bigg)_{\vartheta_{j}^{S_{1}^{c}}}\nonumber\\
&<&\frac{\rho_{1}}{2\tau\log(1+\tau^{-1})}
\end{eqnarray}
for $j\in S_{1}^{c}$, and
\begin{eqnarray}
\label{cond4}
||2X_{G_{k}}^{T}X_{S}w_{S}-2X_{G_{k}}^{T}\epsilon + \rho_{1}g_{G_{k}}||_{2}<\frac{\rho_{2}\sqrt{q_{k}}}{\tau\log(1+\tau^{-1})}
\end{eqnarray}
for $k\in R^{c}$. 

The subgradient equations (\ref{cond1}) are a result from the KKT conditions and the inequalities (\ref{cond3}) and (\ref{cond4}) are used to ensure that estimated coefficients with indices in $S_{1}^{c}$ and $G_{R^{c}}$ are zero. 

Now by solving equations in (\ref{cond1}) for $w_{S}$, we have
\begin{eqnarray}
\label{msc05}
w_{S}&=&(X_{S}^{T}X_{S}+\lambda I_{s\times s})^{-1}X_{S}^{T}\epsilon\nonumber\\
& &-(X_{S}^{T}X_{S}+\lambda I_{s\times s})^{-1}\bigg(\frac{\rho_{1}}{2}g_{S}+\frac{\rho_{2}}{2}\sqrt{q_{S}}u_{S}+\lambda\beta_{S}\bigg).
\end{eqnarray}

Note that the $\vartheta_{j}^{S}$th element in the last term on the right hand side of (\ref{msc05}) can be expressed as
\begin{eqnarray}
\label{msc051}
\bigg(\frac{\rho_{1}}{2}g_{S}+\frac{\rho_{2}}{2}\sqrt{q_{S}}u_{S}+\lambda\beta_{S}\bigg)_{\vartheta_{j}^{S}}&=&\frac{1}{2[\tau\log(1+\tau^{-1})]}\bigg[\frac{\tau\rho_{1}h_{j}}{(\tau+|w_{j}+\beta_{j}|)}\nonumber\\
& &+\frac{\tau\rho_{2}\sqrt{q_{k_{j}}}v_{j}}{(\tau+||w_{G_{k_{j}}}+\beta_{G_{k_{j}}}||_{2})}+2\lambda\tau\log(1+\tau^{-1})\beta_{j}\bigg].\nonumber\\
\end{eqnarray}

Here we define $B_{S,\tau}$ by
\begin{eqnarray}
\label{msc053}
(B_{S,\tau})_{\vartheta_{j}^{S}} = \frac{\tau\rho_{1}h_{j}}{(\tau+|w_{j}+\beta_{j}|)}+\frac{\tau\rho_{2}\sqrt{q_{k_{j}}}v_{j}}{(\tau+||w_{G_{k_{j}}}+\beta_{G_{k_{j}}}||_{2})}+2\lambda\tau\log(1+\tau^{-1})\beta_{j}.
\end{eqnarray}
By Assumption 2, $C_{SS}=n^{-1}(X_{S}^{T}X_{S}+\lambda I)$. Practically we can express $w_{S}$ as
\begin{eqnarray}
\label{msc052}
w_{S} = n^{-1}C_{SS}^{-1}X_{S}^{T}\epsilon - \frac{1}{2n\tau\log(1+\tau^{-1})}C_{SS}^{-1}B_{S,\tau}.
\end{eqnarray}

\textbf{Sign consistency for estimated coefficients with indices in $S$.} Now in order to ensure the sign consistency for estimated coefficients with indices in $S$, we impose some constraint on each entry of $w_{S}$. We focus on the following inequality:
\begin{eqnarray}
\label{cond2}
|w_{j}|<|\beta_{j}|.
\end{eqnarray}
Inequality (\ref{cond2}) implies that for $j\in S$, sign$(\widehat{\beta}_{j}^{\tau})=$ sign$(\beta_{j})$. To see why it is, let us consider the case when $\beta_{j} > 0$. If $\beta_{j}> 0$, then $|w_{j}|<|\beta_{j}|$ means that either  $-\beta_{j}<w_{j}=\widehat{\beta}_{j}^{\tau}-\beta_{j}<\beta_{j}$ or $-\beta_{j}<-w_{j}=\beta_{j}-\widehat{\beta}_{j}^{\tau}<\beta_{j}$, which jointly imply that $0<\widehat{\beta}_{j}^{\tau}<2\beta_{j}$. A similar argument can be applied to the case when $\beta_{j}<0$. Therefore given that (\ref{cond2}) holds, sign consistency holds for estimated coefficients with indices in $S$. 

With representation (\ref{msc052}), for $j\in S$, we can bound $|w_{j}|$ in a way such that 
\begin{eqnarray}
\label{msc07}
|w_{j}|&\leq&n^{-1}\Big|\big(C_{SS}^{-1}X_{S}^{T}\epsilon\big)_{\vartheta_{j}^{S}}\Big|+\frac{1}{2n\tau\log(1+\tau^{-1})}\Big|\big(C_{SS}^{-1}B_{S,\tau}\big)_{\vartheta_{j}^{S}}\Big|.
\end{eqnarray}
By plugging the right hand side of (\ref{msc07}) into the left hand side of (\ref{cond2}) and doing some rearrangements, we obtain the following inequality:
\begin{eqnarray}
\label{msc081}
n^{-1}\Big|\big(C_{SS}^{-1}X_{S}^{T}\epsilon\big)_{\vartheta_{j}^{S}}\Big|\leq |\beta_{j}|-\frac{1}{2n\tau\log(1+\tau^{-1})}\Big|\big(C_{SS}^{-1}B_{S,\tau}\big)_{\vartheta_{j}^{S}}\Big|.
\end{eqnarray}
Further note that for any $j\in S$, $|(C_{SS}^{-1}X_{S}^{T}\epsilon)_{\vartheta_{j}^{S}}|\leq||C_{SS}^{-1}X_{S}^{T}\epsilon||_{\infty}\leq \kappa_{\min}^{-1}||X_{S}^{T}\epsilon||_{\infty}$, where $\kappa_{\min}$ is the minimum eigenvalue of $C_{SS}$. Now with the results given above, we construct the following event:
\begin{eqnarray}
\label{msc082}
E_{1}=\Bigg\{\epsilon: n^{-1}\kappa_{\min}^{-1}||X_{S}^{T}\epsilon||_{\infty}<\min_{j\in S}|\beta_{j}| - \frac{1}{2n\tau\log(1+\tau^{-1})}||C_{SS}^{-1}B_{S,\tau}||_{\infty}\Bigg\}.
\end{eqnarray}
Since the left hand side of the inequality stated in $E_{1}$ is larger than the left hand side of (\ref{msc081}), and the right hand side of the inequality stated in $E_{1}$ is smaller than the right hand side of (\ref{msc081}), therefore if the inequality stated in $E_{1}$ hold, then (\ref{msc081}) will hold. In turn, (\ref{cond1}) and (\ref{cond2}) will hold, and the sign consistency for estimated coefficients with indices in $S$ can be established. 

We go on to derive an estimate for the tail probability of $E_{1}$. Define $\psi_{1,n}$ by 
\begin{eqnarray}
\label{rho1}
\psi_{1,n} = \min_{j\in S}|\beta_{j}| - \frac{1}{2n\tau\log(1+\tau^{-1})}||C_{SS}^{-1}B_{S,\tau}||_{\infty}.
\end{eqnarray}
Note that $E_{1}$ is equivalent to the event $\cap_{j\in S}\{n^{-1}|\sum_{i=1}^{n}x_{ij}\epsilon_{i}|< \psi_{1,n}\kappa_{\min}\}$. On the other hand, by the assumptions on $\epsilon_{i}$'s and $\sum_{i=1}^{n}x_{ij}^{2}$, one can show that $\sum_{i=1}^{n}x_{ij}\epsilon_{i}$ is a normal variable with mean zero and variance $\sigma^{2}\sum_{i}x_{ij}^{2}=n\sigma^{2}$. Therefore, we can bound the probability of $E_{1}^{c}$ in a way such that
\begin{eqnarray}
\label{msc083}
\mathbb{P}(E_{1}^{c})\leq \sum_{j\in S}\mathbb{P}\bigg(\frac{1}{n}\bigg|\sum_{i=1}^{n}x_{ij}\epsilon_{i}\bigg|\geq \psi_{1,n}\kappa_{\min}\bigg)\leq s\mathbb{P}\bigg(|Z|\geq \frac{\sqrt{n}\psi_{1,n}\kappa_{\min}}{\sigma}\bigg),
\end{eqnarray}
where $Z$ is a standard normal variable. By applying a Chernoff bound argument to the right hand side of (\ref{msc083}), we further obtain
\begin{eqnarray}
\label{msc084}
\mathbb{P}(E_{1}^{c})&\leq&\exp\bigg(-\frac{n\psi_{1,n}^{2}\kappa_{\min}^{2}}{2\sigma^{2}}+\log s\bigg)=\exp\bigg\{-n\bigg(\frac{\psi_{1,n}^{2}\kappa_{\min}^{2}}{2\sigma^{2}}-\frac{\log s}{n}\bigg)\bigg\}.\nonumber\\
\end{eqnarray}

\textbf{Sign consistency for estimated coefficients with indices in $S_{1}^{c}$.} Now by plugging (\ref{msc052}) in the middle term of (\ref{cond3}) and then taking absolute value on the quantity, for $j\in S_{1}^{c}$, we have
\begin{eqnarray}
\label{msc09}
& &\Bigg|\bigg(n^{-1}X_{S_{1}^{c}}^{T}X_{S}C_{SS}^{-1}X_{S}^{T}\epsilon-\frac{1}{2n\tau\log(1+\tau^{-1})}X_{S_{1}^{c}}^{T}X_{S}C_{SS}^{-1}B_{S,\tau}\nonumber\\
& &- X_{S_{1}^{c}}^{T}\epsilon + \frac{\rho_{2}}{2}\sqrt{q_{S_{1}^{c}}}u_{S_{1}^{c}}\bigg)_{\vartheta_{j}^{S_{1}^{c}}}\Bigg|\nonumber\\
&\leq&n^{-1}\Big|(X_{S_{1}^{c}}^{T}X_{S}C_{SS}^{-1}X_{S}^{T}\epsilon)_{\vartheta_{j}^{S_{1}^{c}}}\Big|+ \Big|(X_{S_{1}^{c}}^{T}\epsilon)_{\vartheta_{j}^{S_{1}^{c}}}\Big|\nonumber\\
& &+\frac{1}{2\tau\log(1+\tau^{-1})}\bigg|\frac{\rho_{2}\sqrt{q_{k_{j}}}v_{j}}{(1+\tau^{-1}||w_{G_{k_{j}}}+\beta_{G_{k_{j}}}||_{2})}\bigg|\nonumber\\
& &+\frac{1}{2n\tau\log(1+\tau^{-1})}\Big|\big(X_{S_{1}^{c}}^{T}X_{S}C_{SS}^{-1}B_{S,\tau}\big)_{\vartheta_{j}^{S_{1}^{c}}}\Big| 
\end{eqnarray}
By plugging the right hand side of (\ref{msc09}) into the left hand side of (\ref{cond3}) and doing some rearrangements, we obtain the following inequality:
\begin{eqnarray}
\label{msc091}
& &n^{-1}\Big|(X_{S_{1}^{c}}^{T}X_{S}C_{SS}^{-1}X_{S}^{T}\epsilon)_{\vartheta_{j}^{S_{1}^{c}}}\Big|+ \Big|(X_{S_{1}^{c}}^{T}\epsilon)_{\vartheta_{j}^{S_{1}^{c}}}\Big|\nonumber\\
&<&\frac{1}{2\tau\log(1+\tau^{-1})}\bigg(\rho_{1}-n^{-1}\Big|\big(X_{S_{1}^{c}}^{T}X_{S}C_{SS}^{-1}B_{S,\tau}\big)_{\vartheta_{j}^{S_{1}^{c}}}\Big|-\bigg|\frac{\rho_{2}\sqrt{q_{k_{j}}}v_{j}}{(1+\tau^{-1}||w_{G_{k_{j}}}+\beta_{G_{k_{j}}}||_{2})}\bigg|\bigg)\nonumber\\
\end{eqnarray}
Note that by Assumption 3, the maximum eigenvalue value of the matrix $X_{S}X_{S}^{T}$ is $n\varsigma_{\max}$. Therefore,
\begin{eqnarray}
\label{msc092}
|(X_{S_{1}^{c}}^{T}X_{S}C_{SS}^{-1}X_{S}^{T}\epsilon)_{\vartheta_{j}^{S_{1}^{c}}}|\leq ||X_{S_{1}^{c}}^{T}X_{S}C_{SS}^{-1}X_{S}^{T}\epsilon||_{\infty}\leq n\varsigma_{\max}\kappa_{\min}^{-1}||X_{S_{1}^{c}}^{T}\epsilon||_{\infty}.
\end{eqnarray}
Further define 
\begin{eqnarray}
\label{msc093}
(B_{S_{1}^{c},\tau})_{\vartheta_{j}^{S_{1}^{c}}} = \frac{\rho_{2}\sqrt{q_{k_{j}}}v_{j}}{(1+\tau^{-1}||w_{G_{k_{j}}}+\beta_{G_{k_{j}}}||_{2})}.
\end{eqnarray}
With (\ref{msc092}) and (\ref{msc093}), we construct the following event:
\begin{eqnarray}
\label{msc094}
E_{2}=\Bigg\{\epsilon&:&2\bigg(\frac{\varsigma_{\max}}{\kappa_{\min}}+1\bigg)||X_{S_{1}^{c}}^{T}\epsilon||_{\infty}\nonumber\\
&<&\frac{1}{\tau\log(1+\tau^{-1})}\Big(\rho_{1}-n^{-1}||X_{S_{1}^{c}}^{T}X_{S}C_{SS}^{-1}B_{S,\tau}||_{\infty}-||B_{S_{1}^{c},\tau}||_{\infty}\Big)\Bigg\}.\nonumber\\
\end{eqnarray}
Since the left hand side of the inequality stated in $E_{2}$ is larger than the left hand side of (\ref{msc091}), and the right hand side of the inequality stated in $E_{2}$ is smaller than the right hand side of (\ref{msc091}), therefore if the inequality stated in $E_{2}$ holds, then (\ref{msc091}) will hold. In turn both (\ref{cond1}) and (\ref{cond3}) will hold, and the sign consistency for estimated coefficients with indices in $S_{1}^{c}$ can be established.

Now define $\psi_{2,n}$ by
\begin{eqnarray}
\label{rho2}
\psi_{2,n} =\frac{1}{\tau\log(1+\tau^{-1})}\Big(\rho_{1}-n^{-1}||X_{S_{1}^{c}}^{T}X_{S}C_{SS}^{-1}B_{S,\tau}||_{\infty}-||B_{S_{1}^{c},\tau}||_{\infty}\Big).
\end{eqnarray}
Then following the technique similar to the one used in deriving (\ref{msc083}) and (\ref{msc084}), We can bound the probability of $E_{2}^{c}$ in a way such that
\begin{eqnarray}
\label{msc095}
\mathbb{P}(E_{2}^{c})&\leq&\exp\bigg(-\frac{\psi_{2,n}^{2}\kappa_{\min}^{2}}{8(\varsigma_{\max} + \kappa_{\min})^{2}\sigma^{2}}+\log s_{1}^{c}\bigg)\nonumber\\
&=&\exp\bigg\{-n\bigg[\frac{\psi_{2,n}^{2}\kappa_{\min}^{2}}{8n(\varsigma_{\max} + \kappa_{\min})^{2}\sigma^{2}}-\frac{\log s_{1}^{c}}{n}\bigg]\bigg\}.
\end{eqnarray} 

\textbf{Sign consistency for estimated coefficients with indices in $G_{R^{c}}$.}  Now by plugging (\ref{msc052}) into the left hand side of (\ref{cond4}), we have
\begin{eqnarray}
\label{msc10}
& &\Bigg|\Bigg|2n^{-1}X_{G_{k}}^{T}X_{S}C_{SS}^{-1}X_{S}^{T}\epsilon\nonumber\\
& &-\frac{1}{n\tau\log(1+\tau^{-1})}X_{G_{k}}^{T}X_{S}C_{SS}^{-1}B_{S,\tau} - 2X_{G_{k}}^{T}\epsilon + \rho_{1}g_{G_{k}}\Bigg|\Bigg|_{2}\nonumber\\
&\leq&2\Big|\Big|n^{-1}X_{G_{k}}^{T}X_{S}C_{SS}^{-1}X_{S}^{T}\epsilon\Big|\Big|_{2} + 2\big|\big|X_{G_{k}}^{T}\epsilon\big|\big|_{2}\nonumber\\
& &+\frac{\rho_{1}}{\tau\log(1+\tau^{-1})}\bigg|\bigg|\frac{h_{G_{k}}}{(1+\tau^{-1}|w_{G_{k}}+\beta_{G_{k}}|)}\bigg|\bigg|_{2}\nonumber\\
& &+\frac{1}{n\tau\log(1+\tau^{-1})}\big|\big|X_{G_{k}}^{T}X_{S}C_{SS}^{-1}B_{S,\tau}\big|\big|_{2}
\end{eqnarray}
for $k\in R^{c}$. Further by plugging the right hand side of (\ref{msc10}) into the left hand side of (\ref{cond4}) and doing some rearrangements, we can obtain the following inequality:
\begin{eqnarray}
\label{msc101}
& &n^{-1}\Big|\Big|X_{G_{k}}^{T}X_{S}C_{SS}^{-1}X_{S}^{T}\epsilon\Big|\Big|_{2} + \big|\big|X_{G_{k}}^{T}\epsilon\big|\big|_{2}\nonumber\\
&<&\frac{1}{2\tau\log(1+\tau^{-1})}\bigg(\rho_{2}\sqrt{q_{k}}-n^{-1}\big|\big|X_{G_{k}}^{T}X_{S}C_{SS}^{-1}B_{S,\tau}\big|\big|_{2}\nonumber\\
& &-\rho_{1}\bigg|\bigg|\frac{h_{G_{k}}}{(1+\tau^{-1}|w_{G_{k}}+\beta_{G_{k}}|)}\bigg|\bigg|_{2}\bigg).
\end{eqnarray}
By Assumption 3, the maximum eigenvalue of the matrix $X_{G_{k}}X_{G_{k}}^{T}$ is $n\nu_{k,\max}$. Further note that
\begin{eqnarray}
\label{msc102}
||X_{G_{k}}^{T}X_{S}C_{SS}^{-1}X_{S}^{T}\epsilon||_{2}\leq n\varsigma_{\max}\kappa_{\min}^{-1}||X_{G_{k}}^{T}\epsilon||_{2}\leq n\varsigma_{\max}\kappa_{\min}^{-1}\sqrt{n\nu_{k,\max}}||\epsilon||_{2}. 
\end{eqnarray}
With (\ref{msc102}), we construct the following event:
\begin{eqnarray}
\label{msc103}
E_{3}=\Bigg\{\epsilon&:&2\bigg(\frac{\varsigma_{\max}}{\kappa_{\min}}+1\bigg)\sqrt{n\nu_{k,\max}}||\epsilon||_{2}\nonumber\\
& &<\frac{1}{\tau\log(1+\tau^{-1})}\bigg(\rho_{2}-n^{-1}\max_{k\in R^{c}}\big|\big|X_{G_{k}}^{T}X_{S}C_{SS}^{-1}B_{S,\tau}\big|\big|_{2}\nonumber\\
& &-\max_{k\in R^{c}}\rho_{1}\bigg|\bigg|\frac{h_{G_{k}}}{(1+\tau^{-1}|w_{G_{k}}+\beta_{G_{k}}|)}\bigg|\bigg|_{2}\bigg),\text{ for all }k\in R^{c}\Bigg\}.
\end{eqnarray}
Since the left hand side of the inequality stated in $E_{3}$ is larger than the left hand side of (\ref{msc101}), and the right hand side of the inequality stated in $E_{3}$ is smaller than the right hand side of (\ref{msc101}), therefore if the inequality stated in $E_{3}$ holds, then (\ref{msc101}) will also hold. In turn, if (\ref{msc101}) holds for all $k\in R^{c}$, then both (\ref{cond1}) and (\ref{cond4}) will hold, and the sign consistency for estimated coefficients with indices in $G_{R^{c}}$ can be established.

We follow a strategy similar to those given above to derive an estimate for the tail probability of $E_{3}$. Define $\psi_{3,n}$ by 
\begin{eqnarray}
\label{rho3}
\psi_{3,n}&=&\frac{1}{\tau\log(1+\tau^{-1})}\bigg(\rho_{2}-n^{-1}\max_{k\in R^{c}}\big|\big|X_{G_{k}}^{T}X_{S}C_{SS}^{-1}B_{S,\tau}\big|\big|_{2}\nonumber\\
& &-\max_{k\in R^{c}}\rho_{1}\bigg|\bigg|\frac{h_{G_{k}}}{(1+\tau^{-1}|w_{G_{k}}+\beta_{G_{k}}|)}\bigg|\bigg|_{2}\bigg).
\end{eqnarray}
Note that $E_{3}$ is equivalent to the event $\cap_{k\in R^{c}}\{4n\nu_{k,\max}\kappa_{\min}^{-2}(\varsigma_{\max}+\kappa_{\min})^{2}||\epsilon||_{2}^{2}<\psi_{3,n}^{2}\}$. Therefore the probability of $E_{3}^{c}$ can be bounded in a way such that
\begin{eqnarray}
\label{msc104}
\mathbb{P}(E_{3}^{c})&\leq&\sum_{k\in R^{c}}\mathbb{P}\bigg\{\frac{||\epsilon||_{2}^{2}}{\sigma^{2}}\geq \frac{\kappa_{\min}^{2}\psi_{3,n}^{2}}{4n\nu_{k,\max}(\varsigma_{\max}+\kappa_{\min})^{2}\sigma^{2}}\bigg\}\nonumber\\
&\leq&r^{c}\mathbb{P}\bigg\{\frac{||\epsilon||_{2}^{2}}{\sigma^{2}}\geq \frac{\kappa_{\min}^{2}\psi_{3,n}^{2}}{4n\nu_{\max}(\varsigma_{\max}+\kappa_{\min})^{2}\sigma^{2}}\bigg\}.
\end{eqnarray}
In addition, since $\epsilon$'s are i.i.d. normal variables with mean zero and variance $\sigma^{2}$, therefore $||\epsilon||_{2}^{2}/\sigma^{2}$ is a Chi-square variable with $n$ degrees of freedom. It can be shown that $\mathbb{E}[\exp(a||\epsilon||_{2}^{2}/\sigma^{2})] = (1-2a)^{-n/2}$ for $a<1/2$. We let $a=1/4$, then $\mathbb{E}[\exp(4^{-1}||\epsilon||_{2}^{2}/\sigma^{2})] = 2^{n/2}$. Wit the arguments given above, the probability of $E_{3}^{c}$ can be further bounded in a way such that
\begin{eqnarray}
\label{msc105}
\mathbb{P}(E_{3}^{c})&\leq& \exp\bigg(-\frac{\kappa_{\min}^{2}\psi_{3,n}^{2}}{16n\nu_{\max}(\varsigma_{\max}+\kappa_{\min})^{2}\sigma^{2}}+\frac{n}{2}\log2 + \log r^{c}\bigg)\nonumber\\
&\leq&\exp\bigg\{-n\bigg[\frac{\kappa_{\min}^{2}\psi_{3,n}^{2}}{16n^{2}\nu_{\max}(\varsigma_{\max}+\kappa_{\min})^{2}\sigma^{2}}-0.35 - \frac{\log r^{c}}{n}\bigg]\bigg\}.
\end{eqnarray}
Since $E_{1}$, $E_{2}$ and $E_{3}$ jointly implies conditions (\ref{cond1}), (\ref{cond2}), (\ref{cond3}) and (\ref{cond4}), which further implies the sign consistency sign$(\widehat{\beta}^{\tau})=$ sign$(\beta)$, therefore
\begin{eqnarray}
\label{msc111}
\mathbb{P}\big\{\text{sign}(\widehat{\beta}^{\tau})=\text{sign}(\beta)\big\}\geq \mathbb{P}(E_{1}\cap E_{2}\cap E_{3})= 1 - \mathbb{P}\{(E_{1}\cap E_{2}\cap E_{3})^{c}\}.\nonumber
\end{eqnarray}
Further note that $\mathbb{P}\{(E_{1}\cap E_{2}\cap E_{3})^{c}\}=\mathbb{P}(E_{1}^{c}\cup E_{2}^{c}\cup E_{3}^{c})\leq \mathbb{P}(E_{1}^{c})+\mathbb{P}(E_{2}^{c})+\mathbb{P}(E_{3}^{c})$. Therefore we have 
\begin{eqnarray}
\label{msc112}
\mathbb{P}\{\text{sign}(\widehat{\beta}^{\tau})=\text{sign}(\beta)\}\geq 1 - \mathbb{P}(E_{1}^{c})-\mathbb{P}(E_{2}^{c})-\mathbb{P}(E_{3}^{c}).
\end{eqnarray}
Then by applying the tail probability results (\ref{msc084}), (\ref{msc095}) and (\ref{msc105}) to construct a lower bound for the quantity on the right hand side of (\ref{msc112}), we recover the inequality (\ref{sign01}).

\textbf{Asymptotic behavior of $\psi_{1,n}$, $\psi_{2,n}$ and $\psi_{3,n}$.} Now we go on to show that as $n\rightarrow\infty$, $\psi_{1,n}$, $\psi_{2,n}$  and $\psi_{3,n}$, defined in (\ref{rho1}), (\ref{rho2}) and (\ref{rho3}), respectively, can satisfy the requirements stated in Theorem \ref{theorem_sign_consistency}. We first consider the asymptotic behavior of $(B_{S,\tau})_{\vartheta_{j}^{S}}$, which is defined in (\ref{msc053}). Note that by assumptions, if $|w_{j}+\beta_{j}|\neq 0$, then $h_{j}=1$ or $h_{j}=-1$. Therefore given that $\tau= n^{-1}$ and $\rho_{1}=O(n^{1/2})$, the first term on the right hand side of (\ref{msc053}) will be $O(n^{-1/2})$. In addition, if $|w_{j}+\beta_{j}|=0$, then $h_{j}$ is an arbitrary quantity in $[-1,1]$. In this situation we may let $h_{j}$ be proportional to $n^{-1}$, then the first term on the right hand side of (\ref{msc053}) will be $O(n^{-1/2})$. An argument similar to the one given above can be applied to the second term on the right hand side of (\ref{msc053}). Further note that given $\lambda=O(n^{1/2})$, the third term on the right hand side of (\ref{msc053}) will be $O(n^{-1/2}\log(1+n))$. With the arguments given above, we conclude that 
\begin{eqnarray}
\label{psi1_01}
(B_{S,\tau})_{\vartheta_{j}^{S}}= O(n^{-1/2}\log(1+n))
\end{eqnarray} 
for all $j\in S$. An argument similar to the one given above can be applied to $(B_{S_{1}^{c},\tau})_{\vartheta_{j}^{{S_{1}^{c}}}}$ in (\ref{msc093}) and the term $\rho_{1}||h_{G_{k}}(1+\tau^{-1}|w_{G_{k}}+\beta_{G_{k}}|)^{-1}||_{2}$ in $E_{3}$, which leads to
\begin{eqnarray}
\label{psi2_01}
(B_{S_{1}^{c},\tau})_{\vartheta_{j}^{{S_{1}^{c}}}}= O(q_{k_{j}}^{1/2}n^{-1/2})
\end{eqnarray}
for all $j\in S_{1}^{c}$ and 
\begin{eqnarray}
\label{psi3_01}
\rho_{1}\bigg|\bigg|\frac{h_{G_{k}}}{1+\tau^{-1}|w_{G_{k}}+\beta_{G_{k}}|}\bigg|\bigg|_{2}=O(q_{k}^{1/2}n^{-1/2})
\end{eqnarray}
for all $k\in R^{c}$. 

Next we go on to deal with the $l_{\infty}$-norm terms involved in $\psi_{1,n}$, $\psi_{2,n}$ and $\psi_{3,n}$. First note that for a $p$ dimensional vector $b$, we can bound $||b||_{\infty}$ in a way such that $||b||_{\infty}=\sqrt{\max_{j}|b_{j}|^{2}}\leq \sqrt{\sum_{j=1}^{p}b_{j}^{2}}=\sqrt{b^{T}b}$. Therefore for $\psi_{1,n}$ defined in (\ref{rho1}), we can bound the term $||C_{SS}^{-1}B_{S,\tau}||_{\infty}$ in a way such that
\begin{eqnarray}
\label{psi1_l_infty01}
||C_{SS}^{-1}B_{S,\tau}||_{\infty}\leq \frac{\sqrt{B_{S,\tau}^{T}B_{S,\tau}}}{\kappa_{\min}} = O\bigg(\frac{s^{1/2}\log(n+1)}{n^{1/2}\kappa_{\min}}\bigg).
\end{eqnarray}   
Now consider $\psi_{2,n}$ defined in (\ref{rho2}). First note that since $S_{1}^{c}\subseteq G_{R^{c}}$, therefore we can bound the term $n^{-1}||X_{S_{1}^{c}}^{T}X_{S}C_{SS}^{-1}B_{S,\tau}||_{\infty}$ in a way such that
\begin{eqnarray}
\label{psi2_l_infty01}
n^{-1}||X_{S_{1}^{c}}^{T}X_{S}C_{SS}^{-1}B_{S,\tau}||_{\infty}&\leq&n^{-1}||X_{G_{R^{c}}}^{T}X_{S}C_{SS}^{-1}B_{S,\tau}||_{\infty}\nonumber\\
&\leq&\max_{k\in R^{c}}n^{-1}||X_{G_{k}}^{T}X_{S}C_{SS}^{-1}B_{S,\tau}||_{\infty}.
\end{eqnarray}
The right hand side of (\ref{psi2_l_infty01}) can be further bounded in a way such that
\begin{eqnarray}
\label{psi2_l_infty02}
\max_{k\in R^{c}}n^{-1}||X_{G_{k}}^{T}X_{S}C_{SS}^{-1}B_{S,\tau}||_{\infty}&\leq&\max_{k\in R^{c}}n^{-1}\frac{\sqrt{n^{2}\nu_{k,\max}\varsigma_{\max}}}{\kappa_{\min}}\sqrt{B_{S,\tau}^{T}B_{S,\tau}}\nonumber\\
&=&O\bigg(\frac{s^{1/2}\sqrt{\nu_{\max}\varsigma_{\max}}\log(1+n)}{n^{1/2}\kappa_{\min}}\bigg).
\end{eqnarray}
A similar argument can be applied to the term $\max_{k\in R^{c}}n^{-1}||X_{G_{k}}^{T}X_{S}C_{SS}^{-1}B_{S,\tau}||_{2}$ in $\psi_{3,n}$ defined in (\ref{rho3}), which leads to
\begin{eqnarray}
\label{psi3_l_infty01}
\max_{k\in R^{c}}n^{-1}||X_{G_{k}}^{T}X_{S}C_{SS}^{-1}B_{S,\tau}||_{2}= O\bigg(\frac{s^{1/2}\sqrt{\nu_{\max}\varsigma_{\max}}\log(1+n)}{n^{1/2}\kappa_{\min}}\bigg).
\end{eqnarray}
Note that we have assumed $p = o(n(\log(n+1))^{-2})$ and since $s\leq p$ and $q_{k}\leq p$ for $k=1,2,\cdots,m$, therefore we have $s^{1/2}=o(n^{1/2}(\log(n+1))^{-1})$ and $q_{k}^{1/2}=o(n^{1/2}(\log(1+n))^{-1})$ for $k=1,2,\cdots,m$. Then with (\ref{psi1_l_infty01}), the second term on the right hand side of (\ref{rho1}) will approach to zero as $n\rightarrow\infty$, therefore we have $\psi_{1,n}=O(1)$ as $n\rightarrow \infty$. In addition, with results in (\ref{psi2_01}), (\ref{psi2_l_infty01}) and (\ref{psi2_l_infty02}), the second and third terms on the right hand side of (\ref{rho2}) will approach to zero as $n\rightarrow\infty$, therefore we have $\psi_{2,n}=O(n^{3/2}(\log n)^{-1})$ as $n\rightarrow\infty$. Moreover, with results in (\ref{psi3_01}) and (\ref{psi3_l_infty01}), the second and third terms on the right hand side of (\ref{rho3}) will approach to zero as $n\rightarrow\infty$, therefore we have $\psi_{3,n}=O(n^{3/2}(\log n)^{-1})$ as $n\rightarrow\infty$, which completes the proof.

\clearpage
\begin{table}[t]
\begin{center}
\caption{Out-of-sample mean squared error. Each value is an average over the 415 time blocks. The value in the bracket is the standard error. }
\begin{tabular}{lccc}
\hline\hline
Method & Model 1  & Model 2 \\ \hline
{\small gvsnss} & {\small 16.99 (1.83)}  & {\small 16.67 (1.81)} \\ 
{\small lasso} & {\small 21.87 (1.74)}  & {\small 22.48 (1.81)} \\
{\small gvsnss-PC} & {\small 17.50 (1.90)}  & {\small 17.03 (1.86)} \\  
{\small lasso-PC} & {\small 17.66 (1.83)}  & {\small 18.39 (1.88)} \\ 
{\small PC} & {\small 16.75 (1.88)}  & {\small 17.61 (1.92)} \\ 
{\small AR} & - & {\small 18.68 (2.03)} \\ 
\hline\hline
\end{tabular} \label{table2}
\end{center}
\end{table}

\begin{table}[h]
\begin{center}
\caption{Estimation results based on 100 sub-sampling simulations. Each value is an average over 100 sub-sampling simulations and the value in the bracket is the standard error. PMSE: Predictive mean squared error; $\widehat{s}$: The number of covariates with non-zero estimated coefficients; $\widehat{s}_{\text{SSE}}^{+}$: The number of covariates with positive estimated coefficients in the Susan Shepard Effect group.}
\begin{tabular}{lccc}
\hline\hline
& gvsnss 5CV& gvsnss BF & lasso 10CV\\
\hline
$\text{PMSE}_{\text{test}}$ & 0.43 (0.01) &0.38 (0.01) &0.41 (0.01)\\
$\widehat{s}$& 2.81 (0.25) &1.03 (0.02) &3.89 (0.22)\\
$\widehat{s}_{\text{SSE}}^{+}$&0.35 (0.13) &0.00 (0.00)& 1.33 (0.09)\\    
\hline\hline
\end{tabular}\label{table_real_02}
\end{center}
\end{table}

\clearpage

\clearpage
\bibliographystyle{plain}

\end{document}